\documentclass[acmsmall]{acmart}
\renewcommand\footnotetextcopyrightpermission[1]{} % removes footnote with conference information in first column
\usepackage[T1]{fontenc}% optional T1 font encoding
\usepackage[utf8]{inputenc}% this is added to correct the apostrophe errors
\usepackage{amsmath}
\usepackage{subfigure}

\usepackage{algorithmic}
\usepackage{array}
\usepackage{fixltx2e}
\usepackage{stfloats}
\usepackage{url}
\usepackage{mathtools}
\usepackage{float}

\newcolumntype{P}[1]{>{\centering\arraybackslash}p{#1}}
\usepackage{verbatim}
\usepackage{ltablex}
\usepackage{lscape}
\usepackage[ruled]{algorithm2e}
\usepackage{wrapfig}
\usepackage{color,soul}
\usepackage{longtable}
\usepackage{amssymb,amsmath}
\usepackage{multirow}
\usepackage{varwidth}
\usepackage{color, colortbl}

\usepackage{graphicx}
\usepackage{stfloats}
\usepackage{tikz}
\usetikzlibrary{shapes,snakes,patterns}
\usetikzlibrary{shapes.gates.logic.US,trees,positioning,arrows}
\usepackage{pgfplots}
\pgfplotsset{width=18cm, height=6cm}
\usepackage{pgf-pie}
\SetAlFnt{\small}
\SetAlCapFnt{\small}
\SetAlCapNameFnt{\small}
\SetAlCapHSkip{0pt}
\IncMargin{-\parindent}

\usepackage{enumitem}
\setlist[itemize]{leftmargin=*}
\DeclareMathOperator*{\argmax}{argmax}

\definecolor{Gray}{gray}{0.9}

\newcommand{\cb}{\cellcolor{black!20}}
\usepackage{comment}
\usepackage{bm}

\AtBeginDocument{%
  \providecommand\BibTeX{{%
    \normalfont B\kern-0.5em{\scshape i\kern-0.25em b}\kern-0.8em\TeX}}}

\usepackage{xr}
\usepackage[title]{appendix}

\begin{document}
%\tableofcontents
%\newpage
\title{A Survey on Centrality Metrics and Their Implications in Network Resilience}

\author{Zelin Wan}
\email{zelin@vt.edu}
\orcid{0001-5293-0363}
\affiliation{%
  \department{The Department of Computer Science}
  \institution{Virginia Tech}
  \streetaddress{7054 Haycock Rd}
  \city{Falls Church}
  \state{VA}
  \country{USA}
  \postcode{22043}
}

\author{Yash Mahajan}
\email{yashmahajan@vt.edu}
\affiliation{%
  \department{The Department of Computer Science}
  \institution{Virginia Tech}
  \streetaddress{7054 Haycock Rd}
  \city{Falls Church}
  \state{VA}
  \country{USA}
  \postcode{22043}
}

\author{Beom Woo Kang}
\email{beomwookang@gmail.com}
\affiliation{%
  \department{The Department of Electronic Engineering}
  \institution{Hanyang University}
  \streetaddress{Seongdong-gu}
  \city{Seoul}
  \country{Republic of Korea}
}

\author{Terrence J. Moore}
\email{terrence.j.moore.civ@mail.mil}
\affiliation{%
  \institution{US Army Research Laboratory}
  \city{Adelphi}
  \state{MD}
  \country{USA}
}

\author{Jin-Hee Cho}
\email{jicho@vt.edu}
\affiliation{%
  \department{The Department of Computer Science}
  \institution{Virginia Tech}
  \streetaddress{7054 Haycock Rd}
  \city{Falls Church}
  \state{VA}
  \country{USA}
  \postcode{22043}
}

\renewcommand{\shortauthors}{Z. Wan, et al.}

\begin{abstract}
Centrality metrics have been used in various networks, such as communication, social, biological, geographic, or contact networks. In particular, they have been used in order to study and analyze targeted attack behaviors and investigated their effect on network resilience. Although a rich volume of centrality metrics has been developed for decades, a limited set of centrality metrics have been commonly in use.  This paper aims to introduce various existing centrality metrics and discuss their applicabilities and performance based on the results obtained from extensive simulation experiments to encourage their use in solving various computing and engineering problems in networks.
\end{abstract}





\maketitle

\section{Introduction} \label{sec:intro}

\subsection{Motivation} \label{subsec:motivation}
Identifying central nodes in a network is critical to designing a network that is resilient against faults or attacks. However, identifying which nodes are vital in a network is a nontrivial task. Centrality metrics have been studied since the 1940s and began being more formally incorporated into graph theory in the 1970s~\cite{Freeman78}. Although many of these early studies had particular applications and language in the social sciences, a more interdisciplinary approach emerged in the late 1990s and the early 2000s in the nomenclature of {\em Network Science}~\cite{NAP11516}. In the resilience context, there is an extensive literature studying the effect of targeted attacks, or attacks on nodes that have high centrality~\cite{barabasi16-book, Newman10}. A typical scenario includes an intelligent attacker that selects a target node or nodes to disrupt or compromise the network. Since the 2000s, centrality metrics have grown in significance in communication networks as network resilience and cybersecurity concerns have become more prominent. The most common centrality metrics used in this area of research are degree and betweenness~\cite{holme2002attack, Yoon17}, but they are often used because they are popular and without justification of their relevance to the particular scenario. Other studies have used other metrics~\cite{albert2000error, Kim19, Newman02-virus}, such as eigenvector, closeness, pagerank, and so forth. However, given the rich volume of existing centrality metrics that have been studied in other scientific fields for decades, their merits and relevant usages have been insufficiently appreciated and leveraged in various communication and network domains. In this survey, we aim to present this rich volume of centrality metrics and how they can be used and useful in various network and communication research. In addition, we demonstrate the performance of each centrality metric in terms of the effect of targeted attack based on the metrics on the resilience of several example real-world networks. We hope this work can open a door for researchers in engineering fields to fully leverage the existing centrality metrics and their relevancy for network system design and attack modeling.

\subsection{Comparison of Our Survey Paper and Existing Centrality Metrics Survey Papers} \label{subsec:comparision-other-surveys}
The study of centrality metrics has a long history. However, comprehensive surveys only appear in recent work. In 2002, \citet{Dhyani02-web-metrics} conducted a survey on metrics only used in Web information networks to measure graph properties, page importance, page similarity, search, retrieval, the characteristics of usage, and information theoretic properties.  Other fairly recent efforts surveyed centrality applicable in multiple domains. For example, \citet{Guille13} surveyed a small set of centrality metrics and tested their impact on information diffusion in terms of topic propagation originating at those central nodes. This work is limited to the application of information diffusion with only 14 well-known centrality metrics.  \citet{Lu16} conducted a more comprehensive survey on centrality metrics and demonstrated their performance in various network types. The authors considered biological networks, financial networks, social networks, and software networks; they also studied different types of networks, such as directed, undirected, bi-partite, and weighted networks. Their performance analyses of applications included the effect of centrality on information diffusion, identification of scientific influence, detecting financial risks, predicting essential proteins, and so forth.  However, network resilience has not been considered as the application performance metric, which is a central theme of our paper.  

%We could find fairly recent efforts on the survey of centrality measures in multiple domains.  \citet{Guille13} surveyed a small set of centrality metrics and tested them to see their impact on information diffusion in terms of how popular topics can be propagated via those central nodes.  This work is limited to the application of information diffusion with only 14 well-known centrality metrics.  \citet{Lu16} has conducted a most comprehensive survey on centrality metrics and demonstrated their performance in various network types. The authors considered biological networks, financial networks, social networks, and software networks where a network can be directed, undirected, bi-partite, and weighted networks. Their performance analyses of applications included the effect of centrality metrics in information diffusion, identification of scientific influence, detecting financial risks, predicting essential proteins, and so forth.  However, network resilience has not been considered as the application performance metric, which is a central theme in our paper.  

More recently, two survey papers on centrality measures~\cite{Das18, Ashtiani18} have been published.  \citet{Das18} surveyed only 14 centrality metrics from 1948 to 2017, capturing the evolution of centrality concepts. %However, this survey is still limited to only 14 centrality metrics and did not show any experimental results demonstrating their effect on achieving any application goals.  
\citet{Ashtiani18} conducted a comprehensive survey on centrality metrics to investigate protein-protein interaction networks.  They examined node centrality in yeast protein-protein interaction networks (PPINs) for detection or prediction of influential proteins. However, this work is also limited to applying the centrality metrics in biological networks.  Unlike the above survey papers~\cite{Ashtiani18, Das18, Dhyani02-web-metrics, Guille13, Lu16}, our survey paper primarily focuses on the investigation of node centrality, graph centrality, and group-selection centrality in the context of the impact of centrality on network resilience under targeted attacks.

%Very recently, two survey papers on centrality measures~\cite{Das18, Ashtiani18} have been published.  \citet{Das18} surveyed 14 centrality metrics and provided how to compute them. In addition, the authors discussed centrality metrics from 1948 to 2017 which gives useful insights to capture the evolution of centrality metrics. However, this survey is limited to only 14 centrality metrics and did not show any experimental results showing their effect on achieving application goals.  \citet{Ashtiani18} conducted a comprehensive survey on centrality metrics.  The authors mainly looked at centrality metrics to investigate protein-protein interaction networks.  They examined nodes' centrality in yeast protein-protein interaction networks (PPINs) for detecting in which a centrality metric is used as a feature to predict influential proteins.  However, this work is only limited to applying the centrality metrics in biological networks.  Unlike the above survey papers~\cite{Ashtiani18, Das18, Dhyani02-web-metrics, Guille13, Lu16}, our survey paper mainly focuses on investigating a node's centrality, a graph's centrality, and group selection centrality as targeted attacks and examined their impact on network resilience.  

\subsection{Key Contributions \& Scope} \label{subsec:scope_contribution}
Unlike the other state-of-the-art survey papers above, this survey paper makes the following {\bf key contributions}: 
\begin{itemize}
\item We discussed multidisciplinary concepts of centrality and its historical evolution in the research literature. This provides insights on how centrality metrics have been applied in various kinds of networks, in particular their applicability in communication and social networks of interest to many engineers.
\item We conducted an extensive survey on three types of centrality metrics, consisting of point centrality metrics, graph centrality metrics, and group selection metrics, covering over 60 centrality metrics in total. We also described how each metric is computed and its computational complexity. Due the space constraint, we included tables summarizing asymptotic complexity of centrality metrics surveyed in the supplement document. This may inform other researchers into what metric will be more relevant for a particular network or system design of interest. 
\item Unlike other conventional survey papers, we conducted an extensive simulation study to demonstrate the performance of the surveyed centrality metrics in terms of network resilience based on the size of the giant component where each centrality metric is used to pick targets to model either non-infectious or infectious. This will provide a clear and in-depth understanding on how one metric is more relevant than others based on a comparative performance analysis using four different real network topologies. Due to the space constraint, we placed these experimental results in the supplement document.
\item Based on the extensive survey and experimental performance comparison of the centrality metrics, we share what we have learned, providing both insights, limitations as well as promising future research directions.
\end{itemize}

\section{Multidimensional Concepts of Centrality and Its Applications in Diverse Domains} \label{sec:concept-centrality}

The multi-disciplinary development of concepts of node or network centrality has generated multifaceted interpretations of the subject. In this section, we discuss how centrality has been described and applied in several different disciplines.

%Due to the nature of multifaceted concepts of centrality, the concept of an individual's or a network's centrality has been studied in multiple disciplines with different purposes.  In this section, we discuss how the concept of centrality has been defined in different disciplines.  In addition, we discuss how the centrality is differently defined by different researchers or domains. 

\subsection{Multidimensional Concepts of Centrality} \label{subsec:multidimensional-concept-centrality}
A fundamental motivation for the study of centrality is the belief that one's position in the network impacts their access to information \cite{Leavitt51,Stephenson89}, status \cite{Katz53}, power \cite{Bonacich87}, prestige \cite{Rusinowska2011}, and influence \cite{Friedkin91}. We categorize these concepts into three classes as follows: (1) {\em communication activity} based on individual characteristics; (2) {\em influence} based on both individual and network characteristics; and (3) {\em communication control} based primarily on network characteristics. Individual characteristics refer to the way an individual node (i.e., user) interacts with other nodes such as the frequency of interactions (e.g., posting or sending information in online social networks, OSNs, or sending signals or packets in communication networks), the degree of information sharing with others, and the quality of the signals (e.g., posted comments). Network characteristics predominantly indicate the manner in which the node is connected with other nodes; it is these characteristics which can be captured by centrality.

%The fundamental motivation of centrality studies has started from the view that a person with more people in his/her network will have more chances to access more information~\cite{Leavitt51, Stephenson89}, be in higher status~\cite{Katz53}, be more powerful~\cite{Bonacich87}, have more prestige~\cite{Rusinowska2011}, and be more influential~\cite{Friedkin91} than those who are not~\cite{Nicosia12}.  We categorize these centrality concepts into three classes as follows: (1) {\em communication activity} based on individual characteristics; (2) {\em influence} affected by both individual and network characteristics; and (3) {\em communication control} mainly impacted by network characteristics.  The individual characteristics refer to the way an individual node (i.e., user) interacts with other nodes such as a frequency of interactions (e.g., posting or sending information in online social networks, OSNs, or sending signals or packets in communication networks), the degree of information sharing with others, and/or the influence/quality of signals (i.e., posted comments).  The network characteristics mainly indicate the way a node is connected with other nodes, in which many network centrality captures this characteristic.

{\bf Communication Activity.}  This aspect of centrality covers the amount and type of activity an individual node participates in as part of its communications with other nodes. The relative activity, compared with other nodes, can ultimately affect its power or influence. \citet{Klein15} demonstrated a connection between the communication activity and the influence of a user in an OSN. In OSNs, influential users tend to more easily spread information they choose to communicate. However, such well-connected users are less likely to disseminate information received from their extensive network. Hence, this characteristic in terms of frequency or type of interactions of information sharing is a critical factor related to centrality \cite{Bakshy12}.

%This centrality aspect is mainly involved with how actively an individual node communicates with other nodes which can ultimately affect its power or influence. \citet{Klein15} used the communication activity of a user in an OSN to consider the user's influence because the active interactions with others can naturally lead to have more of other users which increase a chance to boost its influence. In OSNs, influential information spreaders tend to frequently spread information they receive.  Even if a person is well connected in OSNs, it is not necessarily for him/her to disseminate received information to other nodes in his/her network.  Hence, this individual characteristics in terms of the frequent interactions or sharing information with other nodes are a critical factor representing a centrality~\cite{Bakshy12}.     

{\bf Influence.}  The term {\em influence} has been used to interpret what centrality may represent in networks. In addition, a number of terms are used to characterize and study the `influence' of a node as follows:  
\begin{itemize}
\item {\em Power}: \citet{Friedkin91} examined the relationships between network centrality and the mutual influence of members in a group.  An individual member's centrality affects other members' opinions and informs a dynamic process of updating their opinions.

\item {\em Status}: \citet{Katz53} proposed the idea that a member's centrality within a network depends upon not only the number of adjacent neighbors but also the {\it status} of each neighbor, i.e., the highest-status member who obtains the majority of choices in a network becomes the most influential. Katz introduced an advanced metric to calculate the status of each member in a network based on the total number of choices, implying the edges in a directed graph, toward each member via a single step up to multiple steps that entail attenuation in a connection of a series~\cite{sade72}.

\item {\em Prestige}: \citet{Bonacich87} and \citet{Katz53} defined a vertex's prestige in a network based on its neighbors. For example, eigenvector centrality is used to derive the prestige of each vertex~\cite{Rusinowska2011}.

%\citet{Cook83} listed the requisites of an exchange network~\cite{emerson62}: (1) a set of members; (2) a dispersion of resources among the members; (3) a set of exchange opportunities between each pair of members; (4) {\it exchange relations} which is a set of historically evolved and employed exchange opportunities; (5) a set of network linkages that connects and combines exchange relations into a network structure. %A significant feature proposed is that each members in the network is unaware of the members above its own set of exchange opportunities; this feature enables the analysis of the distribution of power as related to the position in the network.
\item {\em Resources}:  How much resource one can obtain from their network has been discussed within the context of an exchange network~\cite{Cook83}.  In an exchange network, consisting of a set of members exchanging opportunities, each member needs to decide whether to connect with others to increase their opportunities or resources even when unaware of members outside of its own set of exchange opportunities~\cite{emerson62}.  This feature facilitates the analysis of the power distribution as related to the position in the network~\cite{Cook83, emerson62}.  In exchange networks, a node's power is not necessarily aligned with the number of connections~\cite{Cook83} while most centrality metrics that are more relevant to quick spreading or mitigating influence (e.g., information diffusion or disease transmission) are more reliant on the number of direct or indirect connections with other nodes.  \citet{Bonacich87} reflected this belief in his eigenvector-type centrality where a node's power is measured based on the power of its neighbors.  \citet{Laumann73} discussed {\em community elite}, a set of necessary members in exchange networks in which their position and other attributes determine the structure of influence. 
\item {\em Bridging}: \citet{Saito10} introduced the concept of {\it super-mediators} as the set of nodes that transfer information between nodes.  The capability of a certain node to receive information from numerous nodes and propagate this information to others indicates the their influence~\cite{Leavitt51, Stephenson89}.  Betweenness metrics~\cite{Newman10, Freeman77} is an example representing a bridging role in a network where the node with high betweenness can connect other nodes as a key mediator.  This concept of a broker in sociology is commonly described as a node with high betweenness that can play a key role in bridging two separate groups~\cite{Newman10}.
\end{itemize}

{\bf Communication Control.}  A node's communication control describes how the node can control communications with others, which can naturally affect the node's centrality.  The common two factors affecting this communication control are~\cite{campbell86, Leavitt51}:
\begin{itemize}
\item {\em Commnicability}: With respect to group performance and individual behavioral patterns, \citet{Leavitt51} stressed the importance of a network topology because it determines information accessibility that can affect successful task executions.
\item {\em Network size}: A network can be viewed as resources as each individual gathers information via connections within networks~\cite{campbell86}.  A node's network size is a typical measure of the node's centrality in terms of the resources available to it, including both the quality and the quantity of information in its network~\cite{mcpherson82potential}.
\end{itemize}

\subsection{Centrality Metrics Research in Multidisciplinary Domains} \label{subsec:multidisciplinary-applications-centrality}

\begin{figure*}
    \centering \includegraphics[width=\textwidth]{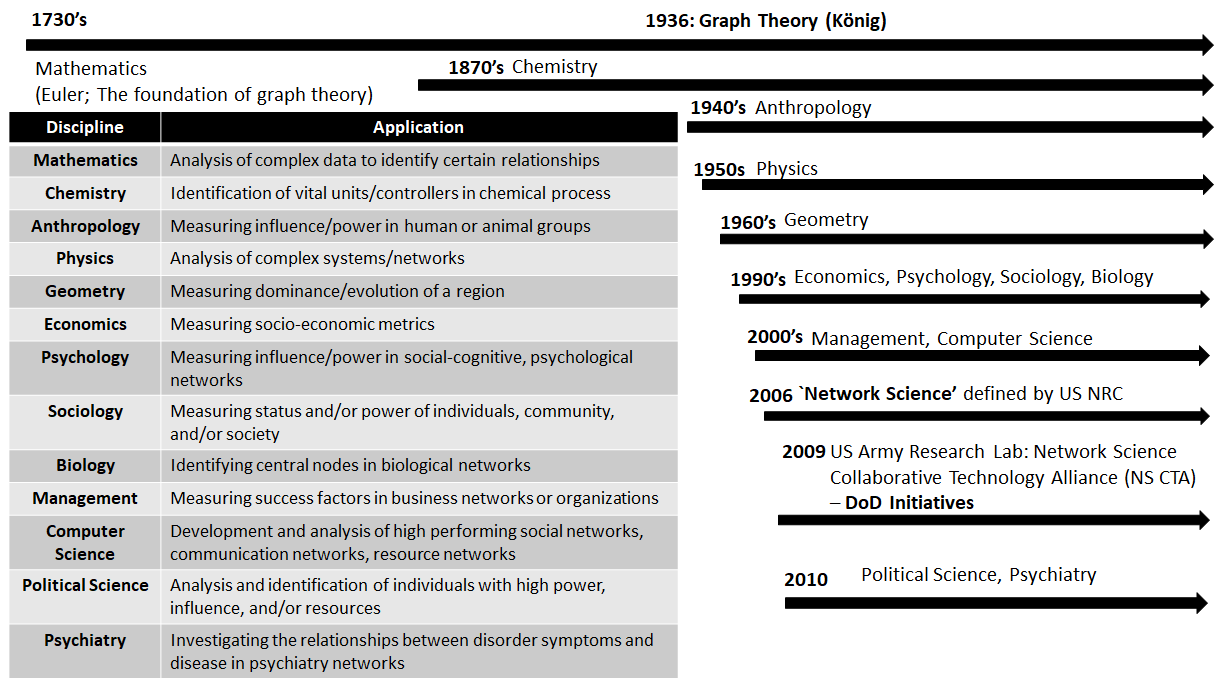}
    \caption{Development of network centrality metrics across multiple disciplines, along with the evolution of the Network Science.}
    \label{fig:network-science-history}
\end{figure*}

Centrality metrics have been studied since the 1940s.  Even in the late 1970s, there exists a rich volume of studies discussing and experimenting with centrality~\cite{Freeman78}. Hence, in the late 1970s, \citet{Freeman78} tried to clarify the concepts and utility of existing metrics. In this section, we have surveyed how centrality has been studied in various disciplines, including mathematics, chemistry, anthropology, geography, economics, psychology, sociology, biology, management, computer science, political science, and psychiatry.  Due to the space constraint, our discussions on this are placed in Section 1 of the supplement document.  As a highly multidisciplinary academic field, we discuss how `Network Science' has studied centrality from a graph theoretical perspective. 

Fig.~\ref{fig:network-science-history} summarizes the evolution of centrality across diverse disciplines along with the emergence of the {\em Network Science} discipline.  The origin of developing centrality metrics is linked with the birth of graph theory~\cite{Koenig90}. Although many fields have used centrality metrics for a variety of purposes, high visibility of the usefulness of these metrics has been much increased as the Network Science field has officially formed in 2000s.  In particular, in 2006, US National Research Council (NRC) defined {\em Network Science} as an academic field~\cite{NAP11516}.  In 2009, The Department of Defense (DoD) initiated a research effort on Network Science for developing battlefield platforms with advanced technology reflecting the theme of {\em network-centric warfare}.  The US Army Research Laboratory (US-ARL) initiated a collaborative research program, the {\em Network Science Collaborative Technology Alliance} (NS CTA), in order to encourage the development of advanced network science-based technologies to support ground soldiers in network-centric warfare~\cite{nscta}, which has further triggered the advancement and maturity of network science research. 

Now we discuss a variety of centrality metrics in various disciplines.  We categorize the types of centrality metrics into three classes: point centrality metrics, graph centrality metrics, and group selection centrality metrics.  The following three sections address these three classes of centrality metrics.

\section{Point Centrality Metrics} \label{sec:point-centrality}

In this section, we introduce various types of centrality metrics using the following common notations. A network is represented by a graph $\mathcal{G}$ with a set of vertices, $\mathcal{V}=\{v_1, v_2, \dots, v_n\}$ representing nodes and a set of edges, $\mathcal{E} = \{e_1, e_2, \dots, e_m\}$ representing connections (links or relationhships) between pairs of nodes, where $n$ is the number of nodes in the network and $m$ is the number of edges~\cite{Borgatti06}. An adjacency matrix $\mathbf{A}$ captures the links between nodes by the value in its entries, e.g., $A_{ij} \neq 0$ only when an edge exists between nodes $v_i$ and $v_j$ with the value $1$ in a simple, undirected graph. We classify point centrality metrics in terms of three classes: local centrality metrics, iterative centrality metrics, and global centrality metrics.
%NOTE FROM TERRY: I got rid of some notation that wasn't used, e.g., e_{i,j} for edges (the only time e_{ij} was used was in excess degree in the assortativity definition), N for nodes (never used), etc.

\subsection{Local Centrality Metrics}
{\em Local centrality metrics} measure the centrality of a node based on its local neighborhood topology. Each of these metrics are variations of the degree of a node, sometimes in combinations with the degree of nodes in the local neighborhood.

%Hence, the local centrality metrics relatively tend to incur less cost to estimate a node's centrality than the global centrality metrics.

\subsubsection*{\bf Degree centrality}\label{sssec:degree} 
The simplest and most well-known centrality metric is the node degree or the number of links or edges incident to the node. 
%This is occasionally also referred to as the valency of the vertex (node) due to the use of the term in chemical bonding (measuring the combining power of an element with other atoms) and linguistics (the number of words associated with a verb). 
The degree of vertex $v_i$ is defined, mathematically, by:
\begin{equation}\label{eq:degree-centrality}
C_{\text{deg}}(v_i) = \# \text{ of edges incident to } v_i = \sum_{j=1}^n A_{ij} = \sum_{j=1}^n A_{ji}.
\end{equation}
In the social network context, the degree indicates amount of activity of the actor \cite{wasserman1994social, Freeman78}. \citet{Hanneman05} describe the degree as measuring the opportunity and alternatives for the actor. In social, communication, and computer networks, degree represents a measure of the number of channels for information exchange (i.e., sending and receiving data) \cite{Brandes03, Newman10}. A standardization or normalization of degree is given by $C_{\mbox{deg}}(v_i) / (n-1)$. This form is useful for comparison across networks~\cite{Freeman78}. Nodes with high degree are called {\em hubs}. 
In a directed network, the in-degree and the out-degree of the node may be unequal, so the adjacency matrix is not symmetric. 
For in-degree, $C_{\text{in-deg}}(v_i) = \# \mbox{~edges directed toward } v_i = \sum_{j=1}^n A_{ji}$, with the out-degree defined analogously based on nonzero entries in the $i$th row of the adjacency matrix. A node with significantly higher in-degree than out-degree or higher in-degree on average compared with other nodes is considered to have prestige \cite{wasserman1994social}. A popular example exists in citation networks where the directed edges correspond to one document citing another. Documents with many citations have high in-degree. Other modified examples corresponding to in-degree in citations include the number of citations of a given author and journal impact factor~\cite{Garfield72}.

\subsubsection*{\bf Semi-Local centrality}\label{sssec:local} While a hub node has immediate access to a large number of neighbors, the hub may exist on the periphery of the network where most of those neighbors have little to no access to the rest of the network. Hence, hubs may not be the ideal nodes for measuring influence, the capability of spreading (information or disease) with efficacy. 
Seeking a middle ground between hub nodes and nodes that have high betweenness (see Eq.~\eqref{eq:betweenness}), \citet{chen2012identifying} developed {\em semi-local centrality}, sometimes called {\em local centrality}, as a low-complexity approach that takes into account neighbor degrees of the node. This semi-local centrality of a node $v_i$ is defined as:
\begin{equation}\label{eq:semilocal-centrality}
C_{\text{semi-local}}(v_i)  = \sum_{u\in N(v)} Q(u) \mbox{ where } Q(u) = \sum_{w\in N(u)} d_2(w),
\end{equation}
where $N(u)$ is the set of neighbors of $u$ and $d_2(w)$ is the number of nearest and next nearest neighbors of $w$. This metric compares favorably to ranks generated from an SIR process.

\subsubsection*{\bf Hybrid degree centrality}
In the context of spreading processes, whether it be for information sharing or disease transmission, the spreading probability $p$ can determine the difference between the influence of the local and near-local neighborhood topology. A small $p$ would intuitively favor a measure like degree centrality, while a larger $p$ would favor a more global measure. \citet{ma2017identifying} incorporated the influence of the scale of $p$ into centrality by adapting degree centrality and semi-local centrality~\cite{chen2012identifying} to create the {\em hybrid degree centrality} of node $v$, defined mathematically as:
\begin{equation}
\label{eq:hybriddegree-centrality}
C_{\text{hybrid}}(v) = (\beta - p) \cdot \alpha \cdot C_{\text{deg}}(v) + p \cdot C_{\text{m-local}}(v),
\end{equation}
where $p$ is the spreading probability, $C_{\text{deg}}$ is the degree centrality, $C_{\text{m-local}}$ is the modified local centrality, $\alpha$ is a normalizing factor to scale the degree centrality to the magnitude of the modified local centrality, and $\beta$ is an optimization parameter.\footnote{For the datasets considered in \cite{ma2017identifying}, $\alpha = 1000$ and $\beta$ near $0.1$ seemed to return favorable results.} The modified local centrality is defined as $C_{\text{m-local}}(v) = \sum_{u \in N(v)} \sum_{w \in N(u)} d_2(w) - 2 \sum_{u \in N(v)} d(u)$, where $N(v)$ is the set of neighbors of $v$, $d_2(w)$ is the number of nearest and next nearest neighbors of node $w$, and $d(u) = C_{\text{deg}}(u)$ is the number of neighbors of $u$. 

\subsubsection*{\bf Volume centrality} If the spreading process dies out, has a limited reach from its initial source, or has a time out component, then it makes sense that this might be entirely captured by the topology in the local neighborhood of the source node.  Let $N_h(v)$ denote the set of neighbors within a distance $h$ of $v$. Then, the volume centrality of the node for a given $h$ is defined as~\cite{Kim12-volume}:
\begin{equation}\label{eq:volume-centrality}
C_{\text{volume}}(v) = \sum_{u \in N_h(v)} d(u).
\end{equation}
This is actually a slight modification of the original definition \cite{wehmuth2012distributed} that uses the set $\tilde{N}_h(v) = N_h(v) \cup \{v\}$. With this latter definition, then when $h=0$, volume centrality is degree centrality. However, this is already captured when calculating the degrees of nodes in $N_1(v)$. \citet{Kim12-volume} showed that larger $h$ correlates well with closeness centrality (see Eq.~\eqref{eq:closeness}).  However, as $h$ increases, the complexity of the method will increase. Hence, \citet{wehmuth2012distributed} demonstrated that $h=2$ results in a good trade-off between identifying nodes that diffuse information well and the cost of this identification.

%\subsubsection*{\bf Volume centrality} A node's volume centrality is measured by the sum of the degrees of the node's all neighbors.  When node $v_i$'s neighbors are denoted by $v_j$'s where the neighbors are defined as nodes within $h$-hop distance from node $i$, node $i$'s volume centrality is estimated by~\cite{Kim12-volume}:
%\begin{equation}
%V_C (v_i) = \sum_{j=0}^h \kappa^j,
%\label{eq:volume-centrality}
%\end{equation}
%where $\kappa^j$ refers to the degree of node $v_j$ which is a neighbor of node $i$ within $h$-hop distance from node $i$.

\subsubsection*{\bf Clustering coefficient} One of the characterizations of small-world networks is the increased likelihood of neighbors of a node to be connected. Social networks tend to exhibit this property and an early characterization of this high clustering property is the density of an ego network (i.e., as described by \citet{Burt95}, the network of the neighbors of a given node excluding that node). \citet{Watts98} proposed the same metric independently as a way to quantify the clustering of nodes in a given graph and characterize the position of the graph within the spectrum of random to small-world graphs. Their definition has proven incredibly popular. It is expressed by:
\begin{equation} \label{eq:clusteringcoefficient}
C_{\text{clustering}}(v) = \frac{1}{|N(v)|\cdot(|N(v)|-1)} \sum_{r,s \in N(v)} A_{rs}.
\end{equation}
Note that each edge will be counted twice in an undirected graph in the summation and the number of such unique edges is normalized by ${|N(v)| \choose 2}$, which is the number of possible edges between the neighbors of $v$. For a directed network, there are twice as many  possible directed edges as the undirected case since the adjacency matrix is no longer symmetric, i.e., $A_{rs}$ may not equal $A_{sr}$ and the set $N^{\text{out}}(v)$ of neighbors $v$ links to is used. This measure is often called the {\em local clustering coefficient} to distinguish it from a global measure of transitivity.

%\subsubsection*{\bf Clustering coefficient} \citet{Watts98} proposed this metric to prove a given graph is a small world network. This metric is mainly to measure how well a node's neighbors are also connected to each other. The clustering coefficient of node $i$, $CC_i$, can be obtained by:
%\begin{equation} \label{eq:lcc}
%CC_i = \frac{|\{e_{ij}| v_j, v_k \in \mathcal{N}_i, e_{jk} \in \mathcal{E}\}|}{\tau_{\mathcal{G}}}
%\end{equation} 
%where $\mathcal{N}_i$ is a set of $i$'s neighbors and $\mathcal{E}$ is a set of edges for a given graph $\mathcal{G}$. $\tau_{\mathcal{G}}$ refers to the number of subgraphs with two edges and three vertices among $i$'s neighbors where the number of $i$'s neighbor is $|\mathcal{N}_i|$, denoted by $k_i$. For a directed network, $\tau_{\mathcal{G}}$ is $k_i(k_i-1)$. For an undirected network, $\tau_{\mathcal{G}}$ is $k_i(k_i-1)/2$ as the graph is symmetric. This metric is often called the {\em local clustering coefficient}.

\subsubsection*{\bf Redundancy} \citet{Burt95} introduced the notion of {\em redundancy} in social networks to describe the concept of neighborhood overlap of a node and its neighbors within the node's ego network. Burt demonstrated redundancy's detriment to {\em social capital} within socio-economic networks. This is defined as:
\begin{equation} \label{eq:redundancy}
C_{\text{redundancy}}(v) = \sum_{r \in N(v)} \sum_{s \in N(r) \cap N(v)} p_{vs} m_{rs},
\end{equation}
where $p_{vs} = \frac{A_{vs} + A_{sv}}{\sum_{r \in N(v)} A_{vr} + A_{rv}}$ and $m_{rs} = \frac{A_{rs} + A_{sr}}{\max_{t \in N(r) \cap N(v)} A_{rt} + A_{tr}}$.
Burt uses redundancy to calculate the {\em effective size (or degree)} of a node's ego (or neighborhood) network taking redundancy into account as $n - C_{\text{redundancy}}(v)$. \citet{Borgatti97} reformulated these expressions to show that for a simple undirected graph, the redundancy is simply $C_{\text{redundancy}}(v) = 2e/C_{\text{degree}}(v)$, where $e$ is the number of links between the neighbors of $v$, and the effective size of $v$ is $C_{\text{degree}}(v)-2e/ C_{\text{degree}}(v)$.

\subsubsection*{\bf Entropy-based measures}
In the thermodynamics context, entropy is a measure of the order of systems. In the information theory context, entropy measures the amount of information absent in a given process. These concepts of entropy have been used in networks, either in characterizing systems or processes~\cite{anand2009entropy}. \citet{nie2016using} adapted the concept of entropy to centrality. They constructed two variants to measure the entropy, {\em local entropy} as the node's contribution to network entropy and {\em mapping entropy} to incorporate a consideration of the neighbors of the node, defined by:
\begin{equation} \label{eq:entropy}
C_{\text{local-entropy}}(v) = - \sum_{u \in N(v)} d(u) \log d(u) \text{ ~\&~ } C_{\text{mapping-entropy}}(v) = - d(v) \sum_{u \in N(v)} \log d(u) .
\end{equation}

%\subsubsection*{\bf Entropy-based measures}
%These measures incorporate entropy into the determination of the centrality ranks.  \citet{nie2016using} estimated {\em local entropy} based on:
%\begin{equation}
%\label{eq:localentropy}
%C_{\text{LE}}(v) = - \sum_{u \in N(v)} d(u) \cdot \log d(u),
%\end{equation}
%and defined {\em mapping entropy} by:
%\begin{equation}
%\label{eq:mappingentropy}
%C_{\text{ME}}(v) = - d(v) \sum_{u \in N(v)} \log d(u).
%\end{equation}

\subsubsection*{\bf ClusterRank} As noted with the redundancy measure, high clustering can have an adverse effect on information propagation or spreading. With this insight, \citet{chen13cluster} proposed {\em ClusterRank}, incorporating both the degree as well as the interactions among the neighbors via the clustering coefficient \cite{Watts98}. The ClusterRank of node $v$ is defined as:
\begin{equation} \label{eq:clusterrank}
C_{\text{clusterrank}}(v) = f(C_{\text{clustering}}(v)) \sum_{u \in N^{\text{out}}(v)} (C_{\text{out-deg}}(u) + 1) ,
\end{equation}
where Chen et al. choose $f(C_{\text{clustering}}(v)) = 10^{-C_{\text{clustering}}(v)}$, $N^{\text{out}}(v)$ is the set of directed edges emanating from $v$ (i.e., the ``followers'' of $v$), and $C_{\text{clustering}}(v)$ is the local clustering coefficient defined for directed networks. The summation also adds the degree of the node $v$ in the unity term. The coefficient acts as a damping weight where higher clustering is penalized for having fewer unique links to different parts of the network. This damping weight is mitigated if many of the neighbors of $v$ have large numbers of additional neighbors.

%\subsubsection*{\bf ClusterRank} Based on the interactions among the neighbors in order to identify influential nodes in networks \citet{chen13cluster} proposed a neighborhood centrality called {\it ClusterRank}. Extended from LocalRank~\cite{chen2012identifying} \jhc{(check which one is more prevalent (clusterrank vs. localrank)}, this metric considers the degree as well as the interactions among the neighbors by adopting clustering coefficient~\cite{Watts98} in directed networks.  Although the clustering coefficient is commonly interpreted as an adverse indicator in information propagation, the information from a node with high clustering coefficient is disseminated rapidly if the node's neighbors are connected to many other nodes in networks. The ClusterRank value of node $i$ is defined as:
%\begin{equation}
%    C_{CR}(i) = f(c_i) \sum_{j\in {\Gamma}_i} (k_j^{out} + 1),
%\end{equation}
%where $f(c_i)$ is defined as $f(c_i) = 10^{-c_i}$, ${\Gamma}_i$ is the set of neighbors directed from node $i$, i.e., followers, $k_j^{out}$ is the out-degree of node $j$, $c_i$ is the clustering coefficient of node $i$ defined as:
%\begin{equation}
%    c_i = \frac{|\left\{ e_{jk}|j,k\in {\Gamma}_i \right\}|}{k_i^{out}(k_i^{out}-1)},
%\end{equation}
%where $|{e_{jk}|j,k\in {\Gamma}_i}|$ is the number of directed edges between a pair of node $i$'s followers.  If this metric is estimated via multiple iterations, it is an iterative metric. However, as we consider the first iteration, we treat this metric as a local metric.

\subsubsection*{\bf H-index} \citet{Hirsch05hindex} introduced the {\em h-index} to measure the impact of the scientific output of a researcher.  A researcher has index $h$ if $h$ is the largest integer $\ell$ such that the researcher has at least $\ell$ papers each having at least $\ell$ citations. \citet{korn09lindex} adapted $h$-index (calling it the {\em lobby index}) to discover important nodes in networks. A node has index $h$ if the node has at least $h$ neighbors, each having at least degree $h$, with the rest of the neighbors having at most degree $h$. Extending this concept, \citet{lu2016hindex} defined the $\mathcal{H}$ operator that, for any node $v$, takes the degrees of the set of its neighbors as an input and returns the maximum number $h$ such that $h$ inputs have value at least $h$. This can be expressed as:
\begin{equation} \label{eq:hindex}
C_{\text{h-index}}(v) = h(v) = \mathcal{H}\left(d(u_{1}), d(u_{2}), \ldots, d(u_{d(v)})\right) ,
\end{equation}
where $u_{1}, u_{2}, \ldots, u_{d(v)}$ are the neighbors of $v$. If the zero-order $h$-index of node $v$ is its degree, i.e., $h^{(0)}(v) = d(v)$, then the value in Eq.~\eqref{eq:hindex} can be called the first-order $h$-index. Then the $k$-order $h$-index is defined as $h^{(k)}(v) = \mathcal{H}\left(h^{(k-1)}(u_{1}), h^{(k-1)}(u_{2}), \ldots, h^{(k-1)}(u_{d(v)})\right)$; this sequence converges to the coreness as the order increases, i.e., $c_i = lim_{k \rightarrow \infty} h^{(k)}(v)$.

%\subsubsection*{\bf H-index} {\it H-index} was first introduced by \citet{Hirsch05hindex} in order to measure the academic impact of a researcher where H-index score $h$ of the researcher indicates that one has in maximum $h$ papers with at least $h$ citations for each paper. \citet{korn09lindex} adopted H-index to find important nodes in networks; The H-index $h$ of node $i$ signifies that node $i$ has $h$ neighbors with at least degree of $h$ for each of them.  Based upon these concepts, \citet{lu2016hindex} defined $\mathcal{H}$ operator that, for any node $i$, takes the set of neighbors of node $i$ as an input and returns the maximum number $h$ of node $i$'s neighbors of which the value is at least $h$. Thus, the H-index of node $i$ can be represented as:
%\begin{equation}
%h_i = \mathcal{H} (k_{j_1}, k_{j_2}, \cdots, k_{j_{k_i}}),
%\end{equation}
%where $k_i$ is the degree of node $i$, $j_1$ is one of the neighbors of $i$. This value is also named as the first-order H-index, i.e., $h_i^{(1)}$ = $h_i$ where the zero-order H-index of node $i$, $h_i^{(0)}$ equals $k_i$. Based on this, where $n$ is bigger than 0, the $n$-order H-index is defined as:
%\begin{equation}
%h_i^{(n)} = \mathcal{H} (h_{j_1}^{(n-1)}, h_{j_2}^{(n-1)}, \cdots, h_{j_{k_i}}^{(n-1)}).
%\end{equation}
%The H-index of node $i$ eventually converges to the coreness as the order increases,
%\begin{equation}
%c_i = \lim_{n\rightarrow \infty} h_i^{(n)}.
%\end{equation}

\subsubsection*{\bf Curvature}  The success of hyperbolic models for networks~\cite{krioukov2010hyperbolic} in reproducing observations from real networks has spurred some interest in measuring the intrinsic geometry of complex networks.  Curvature in networks is a particularly interesting aspect to measure since the models typically presume a constant curvature but the reality (and data) is rarely that convenient.  There are several competing approaches for curvature. One early measure by \citet{Eckmann02} derives a curvature that is identical to the local clustering coefficient of \citet{Watts98} and is used to reveal a connection between high curvature and common topics in the World Wide Web. A popular approach is derived from a Gaussian curvature on planar graphs~\cite{keller2011curvature, higuchi2001combinatorial}, that has been generalized for complex networks~\cite{knill2012index} as:
\begin{equation} \label{eq:curvature-gaussian}
C_{\text{Gauss-curv}}(v) = \sum_{k \ge 0} (-1)^k \frac{s_{v}^{k+1}}{k+1},
\end{equation}
where $s_{v}^{k}$ is the number of $k$-cliques incident to $v$. A truncated version of this is used in \cite{wu2015emergent} to compare a network model with data. A third approach of recent interest adapts a notion of Ricci curvature to networks via the transfer of a mass distribution from one vertex to another, and hence can be defined on an edge~\cite{ollivier2007ricci,jost2014ollivier}. The curvature at a vertex is then a weighted sum of the curvature of the incident edges, $C_{\text{Ricci-curv}}(v) =  \frac{1}{k_v} \sum_{u \in N(v)} \kappa(u,v)$, where $\kappa(u,v) = 1 - W(m_u,m_v)$ and $W(m_u,m_v)$ is the optimal mass transport cost and the mass is typically a unit weight distributed proportionally by an edge weight to the neighbors of the vertices. Curvature has been shown to have relevance to network fragility~\cite{sandhu2016ricci} and network congestion~\cite{wang2016interference}. An alternate adaptation of Ricci curvature~\cite{forman2003bochner,sreejith2016forman} has also received some interest.

\subsection{Iterative Centrality Metrics}

{\em Iterative centrality metrics} rely on iterative processes to calculate. In some cases, the number of iterations is fixed and determined by a characteristic of the network (e.g., maximum degree), and these metrics still incorporate mostly local information of the network. However, in most cases, the number of iterations depends on the convergence rate of values at each node. Global information is incorporated into the metric at the node via these iterative processes.  

%{\em Iterative centrality metrics} require multiple iterations to converge to a centrality and naturally lead to higher cost in computation. But even if centrality is measured locally but it can reflect a global centrality to some extent because each node's centrality is computed through multiple iterations until it reaches a convergence.

\subsubsection*{\bf $k$-shell index or coreness}  The most efficient spreaders have been found to reside in the {\em core} of the network~\cite{kitsak2010identification}, which can be determined by the process of assigning each node an index (or a positive integer) value derived from the $k$-shell decomposition. The decomposition and assignment are as follows: Nodes with degree $k=1$ are successively removed from the network until all remaining nodes have degree strictly greater than $1$.  All the removed nodes at this stage are assigned to be part of the $k$-shell of the network with index $k_{\text{S}}=1$ or the $1$-shell.  This is repeated with the increment of $k$ to assign each node to distinct $k$-shells. The $k$-shell of node $v$ is:
\begin{equation} \label{eq:kshell}
C_{\text{k-shell}}(v) = \max\{k|v \in H_{k} \subset G\} , %\text{ is in the maximal subgraph of G with nodes having degree at least } k\}
\end{equation}
where $H_{k}$ is the maximal subgraph of $G$ with all nodes having degree at least $k$ in $H$. The coreness and $k$-shell of networks have been used to characterize network structure, determine network degeneracy, and identify clusters \cite{wasserman1994social}.

%\subsubsection*{\bf $k$-shell index or coreness}  The most efficient spreaders have been found to reside in the core of the network~\cite{kitsak2010identification}, which can be determined by the process of assigning each node an index (or a positive integer) value derived from the $k$-shell decomposition. The assignment is as follows: Nodes with degree $k=1$ are successively removed from the network until all remaining nodes have degree strictly greater than $k=1$.  All the removed nodes (and edges) at this stage are assigned to be part of the $k$-shell of the network with index $k_{\text{S}}=1$.  This is repeated with the increment of $k$ to assign each node (and some of its edges) to distinct $k$-shells.

\subsubsection*{\bf Mixed degree decomposition}  $k$-shell decomposition methods ignore differences in the degree of nodes within the same shell. \citet{zeng2013ranking} developed a {\em mixed degree decomposition} that retains elements of the degree mixed with the $k$-shell index; for node $v$, this is given by:
\begin{equation}
\label{eq:mixeddegree}
C_{\text{mixed-deg}}(v) = k^{(r)}(v) + \lambda \cdot k^{(e)}(v), %k^{(m)} = k^{(r)} + \lambda * k^{(e)}_i,
\end{equation}
where each node starts with mixed degree equal to the residual degree $k^{(r)}(v)$ (i.e., the $k$-shell index) and the nodes with smallest mixed degrees ($M$) are removed and assigned to the $M$-shell.  Via Eq.~\eqref{eq:mixeddegree}, the mixed degrees of the remaining nodes are updated by the current residual degree $k^{(r)}(v)$ and the exhausted degree $k^{(e)}(v)$ (i.e., removed edges from $v$ due to the nodes in the $M$-shell) and nodes with updated mixed degree not larger than $M$ are also removed and assigned to the $M$-shell. This is repeated iteratively for the next smallest remaining mixed degree to determine each node's mixed degree.  When $\lambda=0$, then mixed degree is simply the $k$-shell index; on the other hand, when $\lambda=1$, then mixed degree is simply the degree.

\subsubsection*{\bf Neighborhood coreness}
This metric adapts the notion of $k$-shell (or $k$-core) of vertices that, although linked to efficient spreaders in networks~\cite{kitsak2010identification}, lacks sufficient diversification for ranking. The $k$-shell is a maximal connected subgraph where all vertex degrees are at least $k$. 
%If node $v$ has a $k$-shell index of $C_{\text{k-shell}}(v)$, then it belongs to the $C_{\text{k-shell}}(v)$-shell, but not the $(C_{\text{k-shell}}(v)+1)$-shell.  
The core of a network consists of nodes with high $k$-shell index. Unfortunately, many nodes have the same $k$-shell index.  \citet{bae2014identifying} introduce more diversity by considering the $k$-shell of neighbors. The {\em neighborhood coreness} and the {\em extended neighborhood coreness} are defined as: 
\begin{equation}\label{eq:nbhdcoreness}
C_{\text{nc}}(v) = \sum_{u\in N(v)} C_{\text{k-shell}}(u) \text{ ~\&~ } C_{\text{nc+}}(v) = \sum_{u\in N(v)} C_{nc}(u)
\end{equation}
These metrics introduce a more distinguishable monotonicity than using the $k$-shell.  
%\citet{bae2014identifying} compare neighborhood coreness to the common centrality metrics (e.g., degree, betweenness, and closeness) as well as several other metrics (e.g., extended $k$-shell, mixed degree~\cite{zeng2013ranking},  an improved method of $k$-shell~\cite{liu2013ranking}) based on $k$-shell against simulations generating a {\em spreading influence} of each node, determined by the number of recovered nodes from the {\em Susceptible-Infected-Recovered} (SIR) process initiated on a node.

\subsubsection*{\bf Eigenvector centrality} \label{subsubsec:eigenvector-centrality} This metric is occasionally called Bonacich's degree centrality~\cite{Bonacich72, Bonacich87, Hanneman05}.  Bonacich supported a claim of Cook~\cite{Cook83} that centrality is not the same as power and a node with high centrality (e.g., degree) is not necessarily powerful or influential. Accordingly, Bonacich developed an {\em eigenvector centrality}, which incorporates notions of both centrality and power, where a node's centrality is determined from its direct connections with other nodes and its power is from the centralities of these neighbors directly and other nodes in the network indirectly. The eigenvector centrality of node $v$ is defined as~\cite{Bonacich72}:
\begin{equation} \label{eq:eigenvector}
C_{\text{eigenvector}}(v) = \frac{1}{\lambda} \sum_{u \in N(v)} C_{\text{eigenvector}}(u) = \frac{1}{\lambda} \sum_{u \in \mathcal{G}} a_{uv} C_{\text{eigenvector}}(u), 
\end{equation}
where $a_{uv}$ is an entry of the adjacency matrix $\mathbf{A}$ and $\lambda$ is an eigenvalue associated with the eigenvector.\footnote{Note that the iterative approach to attain this centrality requires positive values at initialization to guarantee convergence to the eigenvector corresponding to the maximum eigenvalue, which has non-negative values.}. Note, the second equality makes clear that the ranking of centralities is determined by the eigenvector of the adjacency matrix. 

\subsubsection*{\bf Katz centrality} \citet{Katz53} proposed a new status measure by considering the number of direct connections to a node and the statuses of nodes connected to the node. Katz centrality is well-defined in vector notation~\cite{Newman10} as:
\begin{equation} \label{eq:katz-c}
\mathbf{C_{\text{katz}}}(\alpha, \beta) = \alpha \mathbf{A} \mathbf{C_{\text{katz}}}(\alpha, \beta) + \beta \mathbf{1},
\end{equation}
where $\alpha$ is a weight that determines the relative influence of the centrality of the node's neighbors to other nodes in the network by their distances and $\beta$ is a `free part' representing a constant extra credit all nodes receive. This can be reformulated with $\beta = 1$ as $\mathbf{C_{\text{katz}}}(\alpha) = (\mathbf{I}-\alpha \mathbf{A})^{-1} \mathbf{1}$. \citet{Newman10} indicates that Katz centrality resolves a problem of zero-valued eigenvector centrality of nodes not in strongly connected components of directed graphs.

\subsubsection*{\bf Authority and Hub centralities}  For a directed network, the in-degree of node $v$ alone does not provide any notion of the relevant nodes to node $v$.  Moreover, the out-degree of node $v$ does not provide any notion of the important nodes to node $v$.  \citet{Kleinberg99} introduced an iterative process in the context of hyperlinked web pages to determine which pages are authoritative and which pages are hubs to authoritative pages to assist in web search queries. In this process, each page $v$ is assigned two non-negative weights, one corresponding to its relevance as an authority $x_v$ and another corresponding to its relevance as a hub $y_v$.  Each set of weights are normalized so that the sum of their squares is unity, i.e., $\sum x_v^2 = 1$ and $\sum y_v^2 = 1$.  The update process is given by $x_v \leftarrow \sum_{u:(u,v) \in E} y_u$ and $y_v  \leftarrow \sum_{u:(v,u) \in E} x_u$ subject to the normalization invariance. A page's authority depends on the hub weights of the pages linking to it. Similarly, a page's hub weight is determined by the authority weights of the pages it links to. In matrix terms, where $\mathbf{x}$ and $\mathbf{y}$ are vector collections of the authority and hub weights of the nodes, respectively, then the update equations can be expressed as $\bm{x} \leftarrow \bm{A} \bm{y} / (\bm{y}^T\bm{A}\bm{A}^T\bm{y})$ and $\bm{y} \leftarrow \bm{A}^T \bm{x}/ (\bm{x}^T \bm{A}^T\bm{A}\bm{x})$. Some simple linear algebra can be used to show that these converge to the principle eigenvectors of the matrices $\bm{A}^T \bm{A}$ and $\bm{A} \bm{A}^T$, respectively, provided the initial weights in the process are not orthogonal to the principle eigenvectors. Thus, the authority and hub centrality of the node $v$ is given by:
\begin{equation}
\label{eq:authority-centrality}
C_{\text{auth}}(v) = [\bm{e}_1(\bm{A}^T \bm{A})]_v ~~,~~ C_{\text{hub}}(v) = [\bm{e}_1(\bm{A} \bm{A}^T)]_v,
\end{equation}
where $\bm{e}_1(\cdot)$ denotes the principle eigenvector. Kleinberg proposed stopping the process after $10,000$ iterations, as convergence may be slow for large networks.

%\tjm{Connect to eigenvector-type centralities...}

\subsubsection*{\bf PageRank} PageRank is a modern-day variant of Katz centrality that was developed by \citet{Brin98theanatomy}, the founders of Google. PageRank measures the importance of websites by the number of links to the website, and is defined by \cite{Newman10}
\begin{equation} \label{eq:pagerank}
C_{\text{pagerank}}(v,\alpha,\beta) = \alpha \sum_{u \in \mathcal{G}, u \neq v} A_{uv} \frac{C_{\text{pagerank}}(u,\alpha,\beta)}{\max(C_{\text{out-deg}}(u),1)} + \beta,
\end{equation}
where $C_{\text{out-deg}}(u)$ refers to the out-degree of node $u$. The interpretations of $\alpha$ and $\beta$ are similar to the ones described for the Katz centrality in that $\alpha$ is a weight damping the influence of nodes further away from $v$,  while $\beta$ represents a weight for free part or credit that each node receives. The key difference is the relative weighting of links to $v$ by the out degree of the nodes linking to $v$. In vector form, page rank can be expressed, with $\beta = 1$, as $\mathbf{C_{\text{pagerank}}}(\alpha,\beta) = (\mathbf{I}-\alpha \mathbf{A} \mathbf{D}^{-1})^{-1} \mathbf{1} = \mathbf{D}(\mathbf{D}-\alpha \mathbf{A})^{-1} \mathbf{1}$, where $\mathbf{D}$ is a diagonal matrix with entries $D_{uu} = \max(C_{\text{out-deg}}(u),1)$.

%\subsubsection*{\bf PageRank} PageRank complements Katz centrality, aiming to show higher centrality with higher out-degree~\cite{Newman10}.  \citet{Brin98theanatomy}, the founders of the Google, developed the pagerank metric in order to measure the importance of website pages. \citet{Newman10} provides a well-defined mathematical expression for pagerank:
%\begin{equation} \label{eq:pagerank}
%C_{\text{pagerank}}(v,\alpha,\beta) = \alpha \sum_{u \in \mathcal{G}, u \neq v} A_{uv} \frac{C_{\text{pagerank}}(u,\alpha,\beta)}{deg^{out}(u)} + \beta,
%\end{equation}
%where $deg^{out}(u)$ refers to the out-degree of node $u$ for $deg^{out}(u)> 0$ or $deg^{out}(u)=1$ when $deg^{out}(u)=0$ to avoid zero in the denominator.  The interpretations of $\alpha$ and $\beta$ are similar to the ones described for the Katz centrality in that $\alpha$ is a weight damping the influence of nodes further away from $v$  while $\beta$ represents a weight for free part. The key difference is the relative weighting of links to $v$ by their out degree. In vector form, page rank can be expressed, with $\beta = 1$, as 
%\begin{eqnarray} \label{eq:pagerank-matrix}
%\mathbf{C_{\text{pagerank}}}(\alpha,\beta) & = & (\mathbf{I}-\alpha \mathbf{A} \mathbf{D}^{-1})^{-1} \mathbf{1} \nonumber \\
%& = &\mathbf{D}(\mathbf{D}-\alpha \mathbf{A})^{-1} \mathbf{1},
%\end{eqnarray} 
%where $\mathbf{D}$ is a diagonal matrix with entries $D_{uu} = max(deg^{out}(u),1)$.

\subsubsection*{\bf Contribution centrality} \citet{Alvarez-Socorro15} refined the eigenvector centrality to account the similarity of the neighbors that link to a node. The concept presumes that nodes with greater dissimilarity, in the sense of Jaccard~\cite{Bank08}, should have a greater contribution weight than more similar nodes. Dissimilar nodes may provide different information than similar nodes. This {\em contribution centrality} is given by:
\begin{equation} 
\label{eq:contribution-centrality}
C_{\text{contribution}}(v) = \frac{1}{\lambda} \sum_{u \in N(v)} W_{u,v} C_{\text{contribution}}(u) ,
\end{equation}
where $W_{u,v} = A_{u,v} D_{u,v}$ is the contribution of node $u$ to node $v$, $\mathbf{A}$ is the adjacency matrix, $D_{u, v} = 1- \frac{|N(v) \cap N(u)|}{|N(v) \cup N(u)|}$ is a dissimilarity coefficient, and $N(v)$ refers to a set of $v$'s neighbors. This measure can also be considered as the eigenvector centrality of a weighted network, where the weights are informed by the structural dissimilarity coefficient. The weighted adjacency matrix can be expressed as $\mathbf{W} = \mathbf{A} \bigodot \mathbf{D}$, where $\bigodot$ is the Hadamard or element-wise product. The $\lambda$ in Eq.~\eqref{eq:contribution-centrality} is the maximum eigenvalue of $\mathbf{W}$.

\subsubsection*{\bf Diffusion centrality} This metric approximates communication centrality (i.e., a fraction of the number of participating nodes, such as in buyers of a product, after being informed over the total number of informed nodes), and is given in vector form by~\cite{Banerjee13}:
\begin{eqnarray} \label{eq:diffusion}
\mathbf{C}_{\text{diffusion}}(\mathbf{A}; q, T) := \sum_{t=1}^T {(q \mathbf{A})}^t \mathbf{1},
\end{eqnarray}
where $\mathbf{A}$ is the adjacency matrix, $\mathbf{1}$ is a vector of ones, $q$ is the passing probability, and $T$ is the number of iterations. The diffusion centrality of $i$th node is the $i$th entry. This centrality actually captures a number of different measures depending on the value of $T$ or the number of iterations of passing. When $T=1$, then diffusion centrality will be proportional to degree centrality. When $T \to \infty$, $\mathbf{A}$ is diagonalizable (this is always true for real symmetric matrices, thus true for undirected network adjacency matrices), and $q \ge \frac{1}{\lambda}$ (where $\lambda$ is the maximum eigenvalue of $\mathbf{A}$), then diffusion centrality is proportional to eigenvector centrality. But when $q < \frac{1}{\lambda}$, this is a type of Katz-Bonacich centrality.

\subsubsection*{\bf Subgraph centrality} {\em Subgraph centrality} measures the weighted sum of the closed paths starting and ending at $v$ in the network, including both cyclic and acyclic paths, where the contribution or weight if each path in the sum decreases as the path length increases~\cite{Estrada05}. Thus, this metric measures the inclusion of the node in all connected subgraphs of the network but is characterized significantly by the inclusion of the node in motifs. Subgraph centrality is given by:
\begin{equation}\label{eq:subgraphcentrality}
C_{\text{subgraph}}(v) = \sum_{k=0}^\infty \frac{\mu_k(v)}{k!} = \sum_{j=1}^N (u_j^v)^2 e^{\lambda_j},
\end{equation}
where $\mu_k(v) = (\mathbf{A}^k)_{ii}$, $\lambda_j$ is the $j$th eigenvalue of $\mathbf{A}$ and $u_j$ is its corresponding eigenvector ($u_j^v$ is the $v$th element of this vector).  Inclusion in smaller subgraphs (closed walks) is given more significance due to the scaling, which is also necessary for convergence of the sum. The measure is useful to distinguish between nodes with equivalent values of degree centrality, betweenness, closeness, or eigenvector centrality. The authors conjecture that if the subgraph centrality is identical for all nodes, then these other measures will also be identical. Note that the average centrality of all the nodes is trivial to determine to be $\langle C_{\text{subgraph}} \rangle = \frac{1}{N}\sum_{i=1}^N e^{\lambda_j}$.

\subsubsection*{\bf LeaderRank} \citet{Lu11} proposed {\em LeaderRank} to find prominent members, or {\em leaders}, and thereby rank them in terms of their influence, particularly in a social network context. Given a {\em leadership network} or a directed graph with {\em leaders} and {\em fans}, where a directed edge existing signifies the subscription from a fan to a leader, LeaderRank generates a supplemental network, created via the addition of {\em ground node} $g$ with bidirectional edges between all the nodes in the leadership network. This ensures a strongly connected graph with $n+1$ nodes and $m+2n$ directed edges containing the subgraph of the original leadership network of $n$ nodes and $m$ directed edges. Each node, except the ground node, is assigned an initial unit score. In each unit of time or iteration, the current score of each node is equi-distributed to the neighbors the node is linked to, until equilibrium. The proportion of score allocated from node $u$ to node $v$ in one unit of time is $A_{uv}/C_{\text{out-deg}}(u)$, where $\mathbf{A}$ is the adjacency matrix so $A_{uv} = 1$ if $u$ points to (is a fan of) $v$ and $A_{uv} = 0$ otherwise. At time $t$, the amount of score allocated at node $v$ is $s_v(t+1) = \sum_{u \in N^{\text{in}}(v)}^{n+1} \frac{1}{C_{\text{out-deg}}(u)}s_u(t)$, where $s_v(0) = 1$ for all non-ground nodes and $s_g(0) = 0$. At the equilibrium time $t_e$, the score of the ground node is equi-distributed to the other nodes, which ensures no loss of value in the distribution scheme for the leadership network. Hence, the final LeaderRank score of node $v$ is:
\begin{equation}\label{eq:leaderrank}
C_{\text{LeaderRank}}(v) = s_v(t_e) + \frac{s_g(t_e)}{n}.
\end{equation}

\subsubsection*{\bf Dynamical influence} \citet{Klemm12} proposed the concept of dynamic influence as a centrality measure that can quantify the influence of a node's dynamic state on the collective system behavior based on the interplay between dynamics and structure in complex networks.  Given systems with $n$ time-dependent real variables, $\mathbf{x}=[x_1, \cdots, x_N]$ associated with linear dynamics denoted by $n \times n$ real matrix, $\mathbf{M}$, we have the update function $\dot{\mathbf{x}} = \mathbf{M} \mathbf{x}$. The largest eigenvalue $\mu_{\max}$ for $\mathbf{M}$ is considered to obtain a first classification of dynamics. When $\mu_{\max}$ is negative, $\mathbf{x}(t)$ converges to a null vector as a stable, fixed solution. When $\mu_{\max}$ is positive, $\mathbf{x}(t)$ will grow indefinitely from the initial state $\mathbf{x}(0)$. Assuming that there exists a non-degenerate $\mu_{\max}$ for $\mathbf{M}$, we define a scalar product $\mathbf{\phi}_c=\mathbf{c} \cdot \mathbf{x}$ as a conserved quality where $\mathbf{c}$ is a left eigenvector of $\mathbf{M}$ for $\mu_{\max}$ governed by $\frac{d \mathbf{\phi}_c}{d{\it t}} = \mathbf{c} \cdot \mathbf{\dot{x}} (t) = [\mathbf{c}\mathbf{M}]\cdot \mathbf{\dot{x}} (t) =0$. When the conserved quality exists, the final state can be calculated from the initial state $\mathbf{x}(0)$ by:
\begin{equation} \label{eq:dynamic-influence-x-infini}
\mathbf{C}_{\text{dynamic-influence}} := \mathbf{x}(\infty) = \lim_{t \rightarrow \infty} \mathbf{x}(t) = \frac{\mathbf{c} \cdot \mathbf{x}(0)}{\mathbf{c} \cdot \mathbf{e}} \mathbf{e},
\end{equation}
where $\mathbf{e}$ refers to a right eigenvector of $\mathbf{M}$ for $\mu_{\max}$.  The above equation means that $\mathbf{C}_{\text{dynamic-influence}}$ is projected based on $\mathbf{x}(0)$ where $c_i$ represents the effect of $\mathbf{x}(0)$ on the final state $\mathbf{x}(\infty)$.

\subsubsection*{\bf Cumulative nomination} \citet{poulin2000nomination} introduced {\em cumulative nomination} whereby the reputation of a node is derived from the nominations of its neighbors and, hence, a node located at the center of the network is nominated more frequently than a node located on the periphery. Initially, a unit of nomination is provided to each node in the network. Then for each nomination round or iteration, the nomination value of each node is updated as the sum of the nominations from its neighbors, i.e., for node $v$, $p^{n\prime}_v = p^{n-1\prime}_v + \sum_{u \in N(v)} p^{n-1\prime}_u$, where $p_v^{0\prime}$ = 1. It is convenient to normalize this process at each step: $p^n_v = \frac{p^{n-1}_v + \sum_{u \in N(v)} p^{n-1}_u}{\sum_{w \in \mathcal{G}}\left[ p^{n-1}_w + \sum_{u \in N(w)} p^{n-1}_u \right]}$. At equilibrium, the cumulative nomination of node $v$ is given by:
\begin{equation}
C_{\text{cumulative-nomination}}(v) = \lim_{n \to \infty} p^n_v.
\end{equation}
This metric is analogous to the one proposed by~\citet{Bonacich72}, but it is empirically proven to be faster in convergence to the steady state~\cite{poulin2000nomination}.

%\subsubsection*{\bf Cumulative nomination} \citet{poulin2000nomination} introduced a centrality called {\it cumulative nomination} assuming that a node's reputation is derived from the nomination by neighbors and the node located at the center of the network is nominated relatively more than those located on the periphery.  First of all, an unit of the nomination is given to every nodes in the network.  Then, after each step, the nomination value of each node is updated where the new value of a node is the sum of the neighbor's nomination value added to the previous value of the node. This process is repeated until it converges to the steady state and the accumulated nomination of node $i$ at $n$-th step can be represented by:
%\begin{equation}
%    p^{n\prime}_i = p^{n-1\prime}_i + \sum_j a_{ij} p^{n-1\prime}_j,
%\end{equation}
%where $a_{ij}$ = 1 if $j$ is adjacent to $i$, 0 otherwise, and $p_i^{0\prime}$ = 1. The normalized version of this value at $n$-th step is given by:
%\begin{equation}
%p^n_i = \frac{p^{n\prime}_i}{\sum_j p^{n\prime}_j},
%\end{equation}
%where it can be finalized as:
%\begin{equation} \label{eq:cummulativenomination}
%p^n_i = \frac{p^{n-1}_i + \sum_j a_{ij} p^{n-1}_j}{\sum_k\left[ p^{n-1}_k + \sum_j a_{kj} p^{n-1}_j \right]}.
%\end{equation}
%While this metric is analogous to the one proposed by~\citet{Bonacich72}, it is empirically proven to be faster in the convergence to the steady state~\cite{poulin2000nomination}.

\subsubsection*{\bf SALSA} \citet{lempel2000salsa} developed a {\em Stochastic Approach for Link Structure Analysis} (or {\em SALSA}) as an alternative to the hubs and authorities approach of \citet{Kleinberg99} for web links. The given directed graph $\mathcal{G}$ is converted into an undirected bipartite graph $\tilde{G}$ between a hub side $V_h$ and an authority side $V_a$. Each node $v$ in $\mathcal{G}$ is represented by two nodes, one on the hub side $v_h$ and one on the authority side $v_a$. Each directed edge from $v$ to $u$ in $\mathcal{G}$ is represented by an undirected edge between $v_h$ and $u_a$ in $\tilde{G}$. Two random walks, starting from either side of $\tilde{G}$, of path length two, construct Markov chains that reveal a ranking of nodes as hubs and authorities in the network. The transition matrices of these Markov chains can be defined by a hub matrix $\tilde{\mathbf{H}}$, with element entries $\tilde{H}_{u,v} = \sum_{x \in \mathcal{G}|(u_h,x_a),(v_h,x_a)\in \tilde{G}} \frac{1}{\mathrm{deg}(u_h)} \cdot \frac{1}{\mathrm{deg}(x_a)}$, and an authority matrix $\tilde{\mathbf{A}}$, with entries $\tilde{A}_{u,v} = \sum_{x \in \mathcal{G}|(x_h,u_a),(x_h,v_a)\in \tilde{G}} \frac{1}{\mathrm{deg}(u_a)} \cdot \frac{1}{\mathrm{deg}(x_h)}$, where the degree is in $\tilde{G}$. The updates for these transition matrices are $\mathbf{h}^{n} = \tilde{\mathbf{H}} \mathbf{h}^{n-1}$ and $\mathbf{a}^{n} = \tilde{\mathbf{A}} \mathbf{a}^{n-1}$, where the initial value assigned for each node is 1. As with the mutual reinforcement approach of Kleinberg's hubs and authorities, the principal eigenvectors of the transition matrices are the convergent points of the iterations, i.e.,
\begin{equation} \label{eq:salsa}
C_{\text{SALSA-hub}}(v) = [e_1(\tilde{\mathbf{H}})]_v ~\text{\&}~ C_{\text{SALSA-auth}}(v) = [e_1(\tilde{\mathbf{A}})]_v ,
\end{equation}
where $e_1(\cdot)$ denotes the principle eigenvector.

\subsection{Global Centrality Metrics} \label{subsec:global-point-centrality}

{\em Global centrality metrics} require a measurement using possibly the entire network topology. These approaches involve the measurement of path lengths between nodes that are separated (non-adjacent) in the network. The calculations of shortest paths often do not scale well with network size; hence, these metrics are generally more computationally expensive.

%{\em Global centrality metrics} mean that a centrality is measured based on an entire network topology.  Thus, this type of centrality metrics requires a node to have a global view on the entire network topology.  Therefore, the global centrality metrics often assume that each node know the entire network topology in advance under a static network or is periodically informed of the changed network topology, which incurs cost as well.  The part of using a distance between two nodes (e.g., $d_{ij}$) mainly represents a given centrality is globally estimated based on an entire network topology.

\subsubsection*{\bf Improved method}
As observed in the prior subsection, the $k$-shell method~\cite{kitsak2010identification} does not discriminate between nodes within the same $k$-shell, leading to approaches like mixed degree decomposition and neighborhood coreness. \citet{liu2013ranking} introduced an {\em improved method} as an alternative approach to distinguish these intra-$k$-shell nodes, whereby each node in the $k_s$ core is further ranked by $\theta(v|k_s) = (k_s^{\max}-k_s+1)\sum_{u\in J} d(v,u)$, where $k_s^{\max}$ is the largest $k$-shell index in the network, $J$ is the network core (nodes in the subset with the largest $k$-shell index), and $d(u,v)$ is the length of the geodesic (shortest path) between nodes $v$ and $u$. This centrality can be considered as a two element vector:
\begin{equation}\label{eq:improvedmethod}
C_{\text{improved-method}}(v) = (k_s, \theta(v|k_s)),
\end{equation}
where nodes are sorted first by large $k_s$ and then, for the same $k_s$, by small $\theta(v|k_s)$. Essentially, nodes within the same $k$-shell are distinguished by how close the nodes are to all other nodes in the network core.

\subsubsection*{\bf Betweenness centrality} 
One of the earliest concepts of centrality, learned from studies on human interactions in a laboratory setting~\cite{Bavelas48,Leavitt51}, was developed from the observation of certain nodes having control on the communication between a pair of other nodes based on their position in the network. The ability of a node to control this communication grants it a position of influence as a broker or enabler. Locally, a node with high degree has potential for fulfilling such a role, depending on the level of clustering (links) between the neighbors of the node, but this would be true only for its immediate neighbors. It does not capture the control the node has on the communication between a pair of nodes that are distant from each other. A centrality that encapsulates this concept was formally described by \citet{Freeman77} as {\em betweenness centrality} and is mathematically defined for node $v$ by:
\begin{equation} \label{eq:betweenness}
C_\text{bet} (v) = \sum_{s,t|s \neq v \neq t} \frac{\sigma_{st}(v)}{\sigma_{st}},
\end{equation} 
where $\sigma_{st}$ is the number of the shortest paths between $s$ and $t$ and $\sigma_{st}(v)$ is the number of the shortest paths between $s$ and $t$ that include $v$ in the paths.  For comparing the relative betweenness between nodes in different networks, the centrality can be scaled or normalized by $\binom{n-1}{k}$~\cite{Freeman78}, the number of possible pairs of shortest paths node $v$ can be between. This extreme example only occurs for the center node in a star network.  
Betweenness centrality has received significant interest in applications in information flow~\cite{yan2006efficient}, network resilience~\cite{holme2002attack}, or network classification~\cite{goh2002classification}. A variant of this centrality adapted for edges is popularly used to detect community structure~\cite{Girvan02}.  This interest has led to a number of algorithms for faster computation~\cite{brandes2001faster}, although for large and dense networks, the measure can become computationally prohibitive.

\subsubsection*{\bf $L$-betweenness centrality} Betweenness centrality is often an expensive calculation, especially for large networks. \citet{ercsey2010centrality} formalized a notion of betweenness, originally described by~\citet{Borgatti06}, considering shortest paths of length at most $L$, i.e., 
\begin{equation} 
\label{eq:l-betweenness}
C_{\text{L-bet}} (v) = \sum_{s,t|s \neq v \neq t, d(s,t) \le L} \frac{\sigma_{st}(v)}{\sigma_{st}}.
\end{equation}
If $L$ is at least the diameter of the network, then $L$-betweenness is equivalent to betweenness centrality.  \citet{ercsey2010centrality} explicitly express this quantity in terms of the summation of betweenness centralities at each vertex for shortest paths of fixed length $\ell$ over the range $\ell = 1,\ldots, L$. That construction is particularly useful for their analysis demonstrating a scaling factor with respect to $L$ and that for relatively small values of $L$, the $L$-betweenness centrality is a good indicator of the true betweenness centrality in terms of ranking the nodes with highest centrality. For small $L$, this metric straddles the boundary between the classes of global and local centrality metrics. %\tjm{Stress centrality was mentioned here, but should be mentioned with betweenness centrality generally.}

%\subsubsection*{\bf $L$-betweenness centrality}
%Ercsey-Ravasz and Toroczkai \cite{ercsey2010centrality} show that (shortest-path) betweenness can be predicted by only considering contributions from geodesics at most length $L$ on large networks, i.e.,
%\begin{equation}
%B_{\text{L}}(v) = \sum_{l=1}^L b_l(v),
%\end{equation}
%where $b_l(v)$ is the betweenness centrality of node $v$ for all shortest paths of length $l$. \tjm{(Check if this is correct. The equation looks wrong.)}This is also defined for the stress centrality, which simply counts the number of times a node is on a shortest path between a pair of nodes without normalization. This process works because of a scaling behavior of the metric as a function of $L$. The process is similar to other betweenness centrality approaches that start at a root node and count contributions while building shells outward\footnote{It would be interesting to introduce a decay component here if it hasn't been done already.}, except stopping at a distance $L$ instead of the root's eccentricity (or the diameter).

\subsubsection*{\bf Flow betweenness centrality} \citet{Freeman91} proposed a variant of betweenness to capture the capacity of information that can flow in a valued or weighted graph. The concept borrows from maximum flow-minimum cut theory~\cite{Ford87}. Given the maximum flow $m_{rs}$ between vertices $r$ and $s$, denote by $m_{rs}(v)$ the portion of this flow that passes through node $v$. Then the {\em flow betweenness} for node $v$ is given by:
\begin{equation}\label{eq:flow-betweenness}
C_{\text{flow-bet}}(v) = \sum_{s,t|s \neq v \neq t} m_{st}(v).
\end{equation}
This expression can be normalized by replacing each summand $m_{st}(v)$ with $\frac{m_{st}(v)}{m_{st}}$. This metric can be used to estimate the mean difference between the highest centrality and the centralities of the other nodes as a graph centrality metric, as discussed in Section~\ref{subsubsec:gfbc}.

%\subsubsection*{\bf Flow betweenness centrality} \citet{Freeman91} proposed a variant of his betweenness metric to capture the degree of capacity that information can flow based on the idea of {\em maximum flow}. Based on the maximum flow theory, the maximum flow, $f_{ij}^{\max}$ between $i$ and $j$ is bounded by the minimum capacity between them, $c_{ij}^{\min}$, which is denoted by $f_{ij}^{\max} \leq c_{ij}^{\min}$ as no flow can exceed the minimum cut capacity based on minimum cut, maximum flow theory~\cite{Ford87}. Given $m_{jk}$ as the maximum flow between $v_j$ to $v_k$ and $m_{jk}(v_i)$ as the maximum flow between $v_j$ and $v_k$ via $v_i$ where $j \neq k \neq i$, the degree of maximum flow passing through $v_i$ between $v_j$ and $v_k$ is defined in a weighted (or valued) network by:
%\begin{equation} 
%\label{eq:fbc}
%C_F (v_i) = \sum_{i \neq j \neq k} m_{jk}(v_i).
%\end{equation}
%This flow betweenness can be normalized as:
%\begin{equation} 
%\label{eq:fbc-norm}
%C'_F (v_i) = \sum_{i \neq j \neq k} \frac{m_{jk}(v_i)}{m_{jk}}
%\end{equation}
%This flow betweenness can also be used to estimate the mean difference between the highest centrality and the centralities of other nodes as a graph centrality metric, which is discussed in Section~\ref{subsubsec:gfbc}.

\subsubsection*{\bf Random-walk betweenness centrality}  Like flow betweenness, this also captures a notion of betweenness beyond shortest paths. \citet{newman2005measure} introduced {\em random-walk betweenness} to incorporate the contribution from all paths (short and long) with more weights given to shorter paths. Actually, Newman first defined the measure via a current flow analogy and showed it to be equivalent to random walks. Formally, this measure is defined by:
\begin{equation}\label{eq:randomwalk}
C_{\text{random-walk-bet}}(v) = \frac{\sum_{s,t|s<t} I^{(st)}_v}{\frac{1}{2}n(n-1)},
\end{equation}
where $I^{(st)}_v = \frac{1}{2} \sum_u A_{vu}|T_{vs} - T_{vt} - T_{us} + T_{ut}|$ and $T$ is the matrix $(D_w - A_w)^{-1}$ where $D_w-A_w$ is the Laplacian with the $w$-th row and column removed (e.g., the last column and row). Note $I^{(st)}_s = I^{(st)}_t = 1$.

%\subsubsection*{\bf Random-walk betweenness centrality}  This measure captures a notion of betweenness beyond shortest paths since the spread of information likely does not follow such paths.  \citet{newman2005measure} introduced a measure that captures the contribution from all paths (short and long) with more weights given to shorter paths. Newman first defined {\em random-walk betweenness} via a current flow analogy and showed it to be equivalent to random walks. Formally, it is defined as:
%\begin{equation}
%\label{eq:randomwalk}
%b(v) = \frac{\sum_{s<t} I^{(st)}_v}{\frac{1}{2}n(n-1)},
%\end{equation}
%where $I^{(st)}_v = \frac{1}{2} \sum_u A_{vu}|T_{vs} - T_{vt} - T_{us} + T_{ut}|$ and $T$ is the matrix $(D_w - A_w)^{-1}$ where $D_w-A_w$ is the Laplacian with the $w$-th row and column removed (e.g., the last column and row). Note $I^{(st)}_s = I^{(st)}_t = 1$.

\subsubsection*{\bf Load centrality} 
In the context of the transportation of data over a network, high centrality nodes encounter a heavy load in terms of the data packets that may be transmitted over shortest paths. \citet{Goh01} defined the load centrality of node $v$ as the total quantity of data packets traversing over node $v$ after every node in the network sends a single packet to every other node along the shortest path. For the scenario where more than one shortest path exists between two nodes, the quantity is divided at each branching point evenly. Explicitly, 
\begin{equation}\label{eq:load}
C_{\text{load}}(v) = \sum_{s,t|s \neq v \neq t} \theta_{st}(v),
\end{equation}
where $\theta_{st}(v)$ is the amount of the unit quantity that passed through node $v$ from node $s$ to node $t$ such that the quantity is split uniformly at each branch encountered in the shortest paths from $s$ to $t$. There has been some confusion that this load centrality is equivalent to the betweenness centrality (even in the original paper by~\citet{Goh01}). However, the quantity in betweenness is split evenly along each shortest path and not at the branching points. 
%For example, when there exists, say, three shortest paths and a branching point is encountered with two options to the terminal vertex, the quantity is split differently. For betweenness, two-thirds of the unit quantity is diverted to the path where two shortest paths must overlap and one-third to the other path; whereas for load, one-half of the quantity is diverted to each path at that point. 
For this reason, it is often the case that even in simple graphs the load due to a pair of vertices is not symmetric at every vertex, i.e., $\theta_{st}(v) \neq \theta_{ts}(v)$. A simple algorithm for the calculation of load is provided in~\cite{brandes2008variants}.

%In the context of Internet, high centrality nodes are heavily loaded and congested with heavy data packets transmitted over the shortest paths~\cite{Goh01}. \citet{Goh01} defined each node's load $\ell_i$ in a scale-free network where the index $i$ is associated with the probability that two nodes $i$ and $j$ are being connected by $p_i = p_i/\sum_{k} p_k$ or $p_j = p_j/\sum_{k} p_k$ with $p_i = i^{-\alpha}$ for a controlling factor $\alpha$ ranged in $(0, 1]$ as a real number.  Each node's load $\ell_i$ follows:
%\begin{equation} \label{eq:load-cov1}
%\frac{\ell_i}{\sum_j \ell_j} \sim \frac{1}{N^{(1-\beta)}i^{ \beta}},
%\end{equation}
%where $\beta$ is a controlling parameter in $[0, 1)$. Since $\sum_j \ell_i$ converges to $N \log N$, %$\ell_i$ converges to:
%\begin{equation} \label{eq:load-cov2}
%\ell_i \sim (N \log N) \Big(\frac{N}{i}\Big)^{\beta}.
%\end{equation}

\subsubsection*{\bf Routing betweenness centrality} Considering the traffic load on the network like load centrality~\cite{Goh01}, \citet{Dolev10} defined a variant of betweenness based on the routing strategy. This {\em routing betwenness centrality} measures the expected number of packets passing through a given vertex. For the vertex $v$, the routing betweenness is calculated by:
\begin{equation} \label{eq:routingbetweenness}
C_{\text{routing-bet}}(v) = \sum_{s, t \in \mathcal{V}} \sigma_{st}(v) \cdot T(s, t),
\end{equation}
where $\sigma_{st}(v)$ is the probability that a packet will go through $v$ when it is sent from $s$ to $t$, and $T(s, t)$ is the total number of paths from $s$ to $t$. This probability is dependent on the particular routing protocol.

%\subsubsection*{\bf Routing betweenness centrality} Considering the traffic load like the load centrality~\cite{Goh01}, \citet{Dolev10} defined a betweenness centrality based routing strategy, called {\em routing betwenness centrality} (RBC). This centrality measures an expected number of packets passing through a given vertex. The RBC of vertex $v$ is given by: %\tjm{(independent paths?)}
%\begin{equation} \label{eq:rbc}
%\sigma (v) = \sum_{s, t \in \mathcal{V}} \sigma_{st}(v) \; T(s, t),
%\end{equation}
%where $\sigma_{st}(v)$ is the probability that a packet will go through $v$ when it is sent from $s$ to $t$, and $T(s, t)$ is the total number of paths from $s$ to $t$.

\subsubsection*{\bf Closeness centrality} \citet{bavelas1950communication} was interested in distinguishing between different positions in small group networks. One approach was {\em closeness centrality}, defined as the reciprocal of {\em farness}, or the inverse proportion of the average distance to all other nodes in the network.
Formally, this can be expressed as:
\begin{equation}\label{eq:closeness}
C_{\text{closeness}}(v) = \frac{1}{\sum_{u \in \mathcal{V}} d(v,u)}.
\end{equation}
Often, this quantity is normalized for comparisons across networks by multiplying by $n-1$ (or $n$ for large networks). Another approach to compare the relative position of nodes with the same farness in different structure groups is given by~\cite{bavelas1950communication}, $C_{\text{bavelas}}(v) = \frac{\sum_{s,t \in \mathcal{V}} d(s,t)}{\sum_{u \in \mathcal{V}} d(v,u)}$, which is equivalent to $C_{\text{closeness}}(v)/\sum_{u \in \mathcal{V}} C_{\text{closeness}}(u)$.

\subsubsection*{\bf Information centrality} \citet{Stephenson89} developed a centrality measure that uses all paths between pairs of nodes to incorporate the notion of the potential transmission of information. This {\em information centrality} borrows from the statistical estimation perspective that there is noise from a signal transmission captured by the variance of the signal passing through a path so that the information decreases as the distance between nodes grows. Treating this variance as unity for each link, the information for node $v$ is then defined as the harmonic mean of the information between $v$ and every other node, that is,
\begin{equation}\label{eq:information-centrality}
C_{\text{information}}(v) = \frac{n}{\sum_{u \in \mathcal{V}} \frac{1}{I_{uv}}},
\end{equation}
where $I_{uv}$ is the information along all paths from $u$ to $v$, weighted by the length of each path. This quantity is ultimately given by $I_{uv} = 1/(C_{uu} + C_{vv} - 2 C_{uv})$, where $\mathbf{C} = \mathbf{D} - \mathbf{A} + \mathbf{1}\mathbf{1}^T$, $\mathbf{D}$ is a diagonal matrix of node degrees and $\mathbf{1}$ is a vector of ones. Hence, the information centrality can be rewritten as $C_{\text{information}}^{-1}(v) = C_{vv} + \frac{\text{tr}(\mathbf{C})}{n} - \frac{2}{n^2}$.

%\subsubsection*{\bf Information centrality} This metric is developed to identify a node's importance based on all possible paths that information can flow between two nodes~\cite{Stephenson89}.  Taking the perspective of statistical estimation in that the noise from transmitting a signal is captured by the variance of the signal passing the links between nodes, this centrality decreases as more links (i.e., a path consisting of multiple hops between the two nodes) exists between two nodes. Assuming that the distance between two nodes that are directly connected are all same with 1 unit, the higher distance, the lower centrality.  Given $n$ is the number of nodes in a given network, the information centrality is simply calibrated by:
%\begin{equation} 
%\label{eq:information-c}
%I_i = \frac{n}{\sum_{j=1}^{n} \frac{1}{I_{ij}}},
%\end{equation}
%where $I_{ij}$ is given by:
%\begin{equation} \label{eq:information-c-ij}
%I_{i, j} = \sum_{r=1}^{n} \sum_{s=1}^{n} D_{i,j}^{-1}(r, s),
%\end{equation}
%where $D_{i,j}^{-1}(r, s)$ is an element of the inverse of the matrix $\mathbf{D}_{ij}$ whose element refers to the number of links in common between $P_{ij}(r)$ and $P_{ij}(s)$ that refer to the $r$-th and $s$-th paths between node $i$ and node $j$, respectively~\cite{Stephenson89}.

%A direct calculation of information centrality can be simply represented by:
%\begin{equation}
%\label{eq:information-c2}
%I_i^{-1} = C^{I}_{ii} + \frac{\text{tr}(C^{I})}{n} - \frac{2}{n^2},
%\end{equation}
%where $C^I = (L + \mathbf{1}\mathbf{1}^T)^{-1}$.

\subsubsection*{\bf Current-flow betweenness and closeness}
An alternative notion of flow, similar to the max-flow-min-cut approach for flow betweenness, is to model information spread over a network as an electric current~\cite{brandes2005centrality}. {\em Current-flow betweenness} is defined as:
\begin{equation}\label{eq:currentbetweenness}
C_{\text{current-bet}}(v) = \frac{1}{(n-1)(n-2)} \sum_{s,t\in \mathcal{V}} \tau_{st}(v),
\end{equation}
where $\tau_st(v)$ is the electrical current that passes through node $v$ given a supply entering the source node $s$ and exiting the terminus node $t$. More formally, $\tau_{st}(v) = \frac{1}{2} \begin{pmatrix} -|b(v)| + \sum_{e:v\in e} |x(\overrightarrow{e})| \end{pmatrix}$, where $b(s)=1$, $b(t) = -1$, and $b$ is zero elsewhere and $x$ satisfies Kirchhoff's Current and Potential Laws. This is equivalent to {\em random-walk betweenness}~\cite{newman2005measure}. This approach with current can be extended to other path-based centralities. For example, {\em current-flow closeness} is defined as:
\begin{equation}\label{eq:currentcloseness}
C_{\text{current-closeness}}(v) = \frac{n-1}{\sum_{w \neq v} p_{vw}(w) - p_{vw}(w)},
\end{equation}
where $p(\overrightarrow{e}) = x(\overrightarrow{e})/c(e)$ by Ohm's Law, and where the conductance $c(e)$ is the inverse of the resistance $r(e)$ or length of an edge. This variant of closeness has been shown to be equivalent to {\em information centrality}~\cite{Stephenson89}.

%\subsubsection*{\bf Current-flow betweenness and closeness}
%This metric seeks to model information spread over a network as an electric current~\cite{brandes2005centrality}.  The {\em current-flow betweenness} is defined as:
%\begin{equation}
%\label{eq:currentbetweenness}
%C_{\text{cb}}(v) = \frac{1}{(n-1)(n-2)} \sum_{s,t\in V} \tau_{st}(v),
%\end{equation}
%where $\tau_st(v)$ is the electrical current that passes through node $v$ given a supply entering node $s$ and exiting node $t$. More formally,
%\begin{equation}
%\label{eq:currentthroughput}
%\tau_{st}(v) = \frac{1}{2} \begin{pmatrix} -|b(v) + \sum_{e:v\in e} |x(\overrightarrow{e})| \end{pmatrix},
%\end{equation}
%where $b(s)=1$, $b(t) = -1$, and $b$ is zero elsewhere and $x$ satisfies Kirchhoff's Current and Potential Laws. This is equivalent to {\em random-walk betweenness}~\cite{newman2005measure}. The {\em current-flow closeness} is defined as:
%\begin{equation}
%\label{eq:currentcloseness}
%C_{\text{cc}}(v) = \frac{n-1}{\sum_{w \neq v} p_{vw}(w) - p_{vw}(w)},
%\end{equation}
%where $p(\overrightarrow{e}) = x(\overrightarrow{e})/c(e)$ by Ohm's Law, and where the conductance $c(e)$ is the inverse of the resistance $r(e)$ or length of an edge. This is equivalent to {\em information centrality}~\cite{Stephenson89}.

\subsubsection*{\bf Residual closeness}
\citet{dangalchev2006residual} developed {\em residual closeness} to determine the vulneratiblity in the graph using a variation of closeness. This is defined by:
\begin{equation}\label{eq:residual}
C_{\text{residual-closeness}}(v) = \sum_{u \neq v} \left(\frac{1}{2}\right)^{d(v,u)}.
\end{equation}
Rather than taking the reciprocal of the sum of distances, residual closeness uses a weighting scheme. A generalization of this idea already exists in the literature~\cite{jackson1996strategic}, although it was not explicitly expressed as a centrality metric until later~\cite{jackson2010social}. \citet{jackson2010social} calls this metric {\em decay centrality}, expressed as $C_{\text{decay}}(v) = \sum_{u \neq v} \delta^{d(v,u)}$. Recently, \citet{tsakas2016decay} has shown that the maximum decay centrality often coincides with the maximum degree centrality when $\delta > \frac{1}{2}$ and with the maximum closeness centrality when $\delta < \frac{1}{2}$, at least on Erd{\"o}s-R{\'e}nyi graphs.

%\subsubsection*{\bf Residual closeness}
%To determine vulnerability in the graph, \citet{dangalchev2006residual} defined the {\em residual closeness} of a node as:
%\begin{equation}
%\label{eq:residual}
%C_{\text{r}}(v) = \sum_{u \neq v} (\frac{1}{2})^{d(v,u)}.
%\end{equation}
%This is a variation on closeness. Instead of taking the reciprocal of the sum of distances, this approach uses a weighting scheme. A generalization of this idea already existed in the literature~\cite{jackson1996strategic}, although not explicitly as a centrality metric until after \cite{jackson2010social}. Jackson calls this metric {\em decay centrality} and defines it as:
%\begin{equation}
%\label{eq:decay}
%C_{\text{decay}}(v) = \sum_{u \neq v} \delta^{d(v,u)}.
%\end{equation}
%Recently, \citet{tsakas2016decay} has shown that the maximum decay centrality often coincides with the maximum degree centrality when $\delta > \frac{1}{2}$ and with the maximum closeness centrality when $\delta < \frac{1}{2}$, at least on Erd{\:o}s-R{\'e}nyi graphs.

\subsubsection*{\bf Spatial centrality}
In spatial networks, the distance between neighbors is not uniform (or unweighted).  \citet{crucitti2006centrality} applied and developed generalizations of some common metrics that account for the network's embedding in space. Closeness and betweenness centralities are identical to their weighted distance versions~\cite{wasserman1994social}, i.e., the distance between two nodes is the true distance (or weight) from one node to the other. The new metric developed by \citet{crucitti2006centrality} is {\em straightness centrality}, which is given for node $v$ by:
\begin{equation}\label{eq:straightness}
C_{\text{straightness}}(v) = \frac{1}{n-1} \sum_{u \in \mathcal{V}, u \neq v} \frac{d_{\text{Euclidean}}(u,v)}{d(u,v)},
\end{equation}
where $d_{\text{Euclidean}}(u,v)$ is the Euclidean distance in the real or embedded space. Straightness centrality measures the efficiency of the route between two nodes using node $v$.

\begin{comment}
Another metric is {\em information centrality} developed by~\citet{latora2007measure} (and a different one from that of~\citet{Stephenson89}), which for node $v$ is given by: 
\begin{equation}
\label{eq:spaceinformation}
C_{\text{information2}}(v) = \frac{\text{Eff}(G) - \text{Eff}(G_v)}{\text{Eff}(G)},
\end{equation}
where $\text{Eff}(\cdot)$ is the efficiency of the network (i.e., the average of the straightness centralities in the network) and $G_v$ is the graph without node $v$.
\end{comment}

\subsubsection*{\bf AHP-based centrality} \citet{bian2017identifying} developed the {\em Analytic Hierarchy Process (AHP)} as a decision making process to identify influential nodes. The steps to process are as follows: 
\begin{enumerate}[leftmargin=*]
\item Calculate centrality values (e.g., degree, betweenness, closeness) for each node and combine in an $n \times 3$ matrix.
\item Calculate weights. \citet{bian2017identifying} appended another vector to the above matrix derived from results of SI (Susceptible-Infected) processes run on the nodes, i.e., $\mathbf{D} = [\mathbf{C_D}, \mathbf{B_D}, \mathbf{C_C}, \mathbf{F(t)}]$, where $\mathbf{D}$ is $n \times 4$ matrix, $\mathbf{C_D}$ is degree centrality, $\mathbf{B_D}$ is betweenness centrality, $\mathbf{C_C}$ is closeness centrality, and $\mathbf{F(t)}$ is results of SI model~\cite{hu2016modified}. The matrix is normalized and weights are determined by matching the attributes to the SI column, i.e., $r_{ij} = \frac{D_{ij}}{\sum_{i=1}^{n} D_{ij}}$, for $i=1, \ldots,n; j = 1, \ldots, 4$, $v_{ij} = \frac{1}{|r_{ij} - r_{i4}|}$ for $i=1, \ldots, n; j = 1,2,3$, $e_j = \sum_{i=1}^{n} v_{ij}$, and finally $w_j = \frac{e_j}{\sum_{j=1}^{3} e_j}$, for $j = 1,2,3$. $\mathbf{w}$ is $3 \times 1$ vector which represent the weight for three metrics.
\item Calculate the matrix of option scores using the AHP, i.e., $B^{(j)}_{ik} = \frac{D_{ij}}{D_{kj}}$ for $i = 1, \ldots, n; k = 1, \ldots, n; j = 1,2,3$, where $\mathbf{B}^{(j)}$ is an $n \times n$ matrix. Then the option scores are $\mathbf{s_j} = \max_{\mathrm{eigen}}\mathbf{B}^{(j)} \times \mathbf{B}^{(j)}$, for $j = 1,2,3$, where $\max_{\mathrm{eigen}}\mathbf{B}^{(j)}$ is the largest eigenvalue of matrix $\mathbf{B}^{(j)}$.
\item The nodes are then ranks using
\begin{equation}
\label{eq:ahp}
\mathbf{C}_{\text{AHP}} = \mathbf{s} \times \mathbf{w}^T,
\end{equation}
where $\mathbf{s}$ is $n \times 3$ matrix with columns $\mathbf{s}_j$ for $j = 1, 2, 3$ and $\mathbf{w}^T$ is a transpose vector of $\mathbf{w}$, which is a vector of weights $w_j$ for $j = 1, 2, 3$, respectively.
\end{enumerate}
The presumption is that the SI scores in the above process are based on short time horizons, whereas the results of the AHP may have value for longer time horizons. Thus, AHP combines three classic centrality metrics and weights them via a short-run epidemic compartmental model process.

\subsubsection*{\bf Generalized degree and shortest paths}
For weighted networks, extensions to the usual centrality measures already exist for degree~\cite{barrat2004architecture}, closeness~\cite{newman2001scientific}, and betweenness~\cite{brandes2001faster}. In incorporating weights, the measures ignore the number of ties or intermediaries.  \citet{opsahl2010node} sought to remedy this with the creation of generalized measures that also encompass both the traditional measures and the weighted versions: \begin{equation}
\label{eq:gen-degree-paths}
\begin{gathered}
C_{\text{gen-deg}}^{w}(v,\alpha) = C_{\text{deg}}(v)^{(1-\alpha)} \cdot C_{\text{deg}}^{w}(v)^{\alpha} , \\
C_{\text{gen-closeness}}^{w}(v,\alpha) = \left[ \sum_{u} d^{w}(v,u,\alpha) \right]^{-1}, \text{ ~\&~ } C_{\text{gen-bet}}^{w}(v,\alpha) = \sum_{s,t} \frac{\sigma_{st}^{w}(v,\alpha)}{\sigma_{st}^{w}}
\end{gathered}
\end{equation}
where the shortest path weighted distances given by $d^{w}(u,v) = \min \left(\frac{1}{w_{ui_1}} + \cdots + \frac{1}{w_{i_k v}}\right)$ are replaced with $d^{w}(u,v,\alpha) = \min \left(\frac{1}{(w_{ui_1})^{\alpha}} + \cdots + \frac{1}{(w_{i_k v})^{\alpha}}\right)$. For each generalization, when $\alpha=0$, the measures are the usual (unweighted) centrality measures; when $\alpha=1$, the measures are the common weighted measures. When $\alpha \in (0,1)$, having many weak ties correlates with higher generalized centrality; and when $\alpha > 1$, having fewer weak ties correlates with higher generalized centrality.

\subsubsection*{\bf Weight neighborhood centrality}  \citet{wang2017novel} included a notion of the diffusion importance of links based on the power-law property found in the distribution of many measures (e.g., degree, betweenness) in real networks. Their {\em weight neighborhood centrality} is defined as:
\begin{equation}
\label{eq:weightneighborhood}
C_{\text{weight-nbhd}}(v,\phi) = \phi_v + \sum_{u \in N(v)} \frac{w_{uv}}{\langle w \rangle} \cdot \phi_u,
\end{equation}
where the weights are given by $w_{uv} = (C_{\text{deg}}(u) \cdot C_{\text{deg}}(v))^\alpha$ and $\phi$ is the benchmark centrality (e.g., degree, betweenness, $k$-shell). $N(v)$ is the neighbors of node $v$, $\alpha$ is a tunable parameter between 0 and 1, and $\langle w \rangle$ is average weight for edges. 
%The computational complexity is $O(m)$ plus the complexity needed to obtain the benchmark centrality $\phi$. 
This metric can be classified as a local or iterative centrality metric provided $\ell$ is small and the benchmark centrality is also local or iterative; otherwise it is a global centrality. 

\subsubsection*{\bf Percolation centrality} \citet{Piraveenan13} developed {\em percolation centrality} to capture the dynamic changes of a network topology based on the percolation process. Typically, the percolation state of a node $v$ at time $t$ might be denoted by $x_v(t)$ and has discrete values, where a $0$ value indicates $v$ is not percolated (e.g., infected) at time $t$ and a value of $1$ indicates it is percolated. When $0<x_v(t)<1$, then $v$ might be said to be is in the process (or probability) of being percolated. Hence, a higher value of $x_v(t)$ implies that $v$ is closer to (has a greater chance of) being percolated. Piraveenan et al. defined this percolation centrality as the proportion of percolated paths passing through a node, which for node $v$ is measured by:
\begin{equation} \label{eq:percolation-c}
C_{\text{percolation}}(v,t) = \frac{1}{n-2} \sum_{r \neq v \neq s} \frac{\sigma_{rs} (v)}{\sigma_{rs}} \frac{x_r(t)}{[\sum_{u \in \mathcal{G}} x_u(t)] - x_v(t)}, 
\end{equation}
where $\sigma_{rs}$ is the total number of shortest paths between $r$ and $s$ and $\sigma_{rs} (v)$ is the total number of shortest paths between $r$ and $s$ passing through $v$. When only a single source node is (partially) percolated, then the average of the percolation centrality for every node over all possible sources (excluding itself) is proportional to betweenness centrality (see Eq.~\eqref{eq:betweenness}) as $x_r(t)/([\sum_{u \in \mathcal{G}} x_u(v)] - x_v(t)) = 1$ when only when $r$ is the source, thereby contributing a $1/(n-1)$ factor.  If all nodes are (partially) percolatied at the same level, all shortest paths are percolated paths, leading to the state that percolation centrality is proportional to betweenness centrality.

\subsubsection*{\bf Eccentricity} Based on the idea that the centrality of a node depends on the distance, i.e., the shortest path, between other nodes in networks, \citet{hage1995eccentricity} introduced the concept of {\it eccentricity}, which is the maximum distance between a node and any other node in the network. Lower eccentricity indicates higher centrality. Eccentricity centrality can be mathematically expressed as:
\begin{equation}
\label{eq:eccentricity}
    C_{\text{eccentricity}}(v) = \frac{1}{\max \left\{d(v,u)|u\in V\right\}},
\end{equation}
where $d(v,u)$ is the distance between the nodes $v$ and $u$.

\section{Graph Centrality Metrics} \label{sec:graph-centrality-metrics}
In Section~\ref{sec:point-centrality}, we surveyed an individual node's centrality. Now we look into the centrality of a given graph, which represents the degree of centrality in an entire network, not just points (or vertices).  We discuss the existing 14 graph centrality (GC) metrics as below.

\subsubsection*{\bf Distance-based GC} This measures the distances between all pairs of vertices in order to measure the {\em compactness} of a network. The distance-based GC is defined by~\cite{Freeman78, Shimbel53}:
\begin{equation} \label{eq:distance-based-gc}
C_{\text{distance-GC}}(\mathcal{G}) = \sum_{u \in \mathcal{V}} \sum_{v \in \mathcal{V}} d(u, v),
\end{equation}
where $d(u, v)$ refers to the distance between vertices $u$ and $v$. \citet{Shimbel53} used this same metric but called it {\em dispersion} as this metric is interpreted as vertex's {\em accessibility} to $\mathcal{G}$. The average shortest path \cite{Watts98} is a similar metric in order to compare the breadth of a network at different scales.

\subsubsection*{\bf Degree-based GC} This metric measures the relative dominance of a single vertex in a network.  \citet{Nieminen74} measured this metric by:
\begin{equation} \label{eq:dgc-1}
C_{\text{deg-GC}} (\mathcal{G}) = \sum_{i=1}^n \binom{1+d^* - d_i}{2},
\end{equation}
where $\mathcal{G}$ has the degree set \{$d_1,d_2, \ldots, d_n$\} and $d^*$ denotes the maximum degree in the graph $\mathcal{G}$. The maximum sum of the differences between the largest centrality and all other centralities can be derived as follows: The maximum degree of a vertex, $C_{\text{deg-GC}}(v^*)$, is $n-1$. If the graph is a star or wheel, other vertices have only one neighbor and $C_{\text{deg}}(v)=1$ for all $v \neq v^*$, resulting in the difference $(n-1)-1=n-2$.  Since $n-1$ comparisons would be considered, the sum of these maximum difference is $(n-2)(n-1)=n^2-3n+2$. Therefore, the normalized $C_{\text{deg-GC}} (\mathcal{G})$ can be expressed as $C_{\text{norm-deg-GC}}(\mathcal{G}) = \frac{\sum_{i =1}^n [C_{\text{deg-GC}}(v^*) - C_{\text{deg-GC}}(v_i)]}{n^2-3n+2}$.

\subsubsection*{\bf Betweenness-based GC} This metric is calculated by the mean difference between the maximum betweenness and all other betweennesses~\cite{Freeman77}, as below:
\begin{equation} \label{eq:bgc}
C_{\text{bet-GC}} (\mathcal{G}) 
= \frac{\sum_{i=1}^n [C_{\text{bet}}'(v^*)-C_{\text{bet}}'(v_i)]}{n-1} 
%= \frac{\sum_{i=1}^n \Big[\frac{C_{\text{bet}}(v^*)}{n^2-3n+2}-\frac{C_{\text{bet}}(v_i)}{n^2-3n+2}\Big]}{n-1} 
= \frac{\sum_{i=1}^n [C_{\text{bet}}(v^*)-C_{\text{bet}}(v_i)]}{n^3-4n^2+5n-2},
\end{equation} 
where $C_{\text{bet}}'(v_i)$ and $C_{\text{bet}}'(v^*)$ are determined based on the normalized betweenness~\cite{Freeman78}.

\subsubsection*{\bf Flow betweenness-based GC} \label{subsubsec:gfbc} This metric determines the centrality of a weighted (or valued) graph based on the difference between the highest maximum flow of a node with the highest betweenness and the maximum flow of other nodes. This is computed by~\cite{Freeman91}:
\begin{equation}
\label{eq:flow-betweenness-gc}
C_{\text{flow-bet-GC}}(\mathcal{G})=\frac{\sum_{i=1}^n [C'_{\text{flow-bet-GC}}(v^*)-C'_{\text{flow-bet-GC}}(v_i)]}{n-1},
\end{equation} 
where $C'_{\text{flow-bet-GC}}(v^*)$ refers to the normalized flow centrality of the most central node and $C'_{\text{flow-bet-GC}}(v_i)$ is the normalized flow centrality of node $i$ based on Eq.~\eqref{eq:flow-betweenness}.

\subsubsection*{\bf Closeness-based GC} \citet{Freeman78} generalized the closeness-based graph centrality measure based on the previous trials~\cite{Leavitt51, Sabidussi66}. This metric can be simply derived based on the normalized closeness metric, $(n-1) C_{\text{closeness}}(v)$, from Eq.~\eqref{eq:closeness} by:
\begin{equation} \label{eq:cgc}
C_{\text{close-GC}}(\mathcal{G}) 
= \frac{\sum_{i=1}^n [C_{\text{closeness}}'(v^*)-C_{\text{closeness}}'(v_i)]}{\mathrm{max} \sum_{i=1}^n [C_{\text{closeness}}'(v^*)-C_{\text{closeness}}'(v_i)]} 
= \frac{\sum_{i=1}^n [C_{\text{closeness}}'(v^*)-C_{\text{closeness}}'(v_i)]}{(n^2-3n+2)/(2n-3)},
\end{equation}
where $C_{\text{closeness}}'(v^*)$ is the largest closeness metric among $v \in \mathcal{G}$ and $C_{\text{closeness}}'(v_i)$ is the closeness metric of $v_i$. 
\begin{comment}
$\mathrm{max} \sum_{i=1}^n [C_{\text{closeness}}'(v^*)-C_{\text{closeness}}'(v_i)]$ was derived as $(n^2-3n+2)/(2n-3)$ because the maximum graph centrality can be when a network follows a star or wheel graph where a center node is connected to all other nodes with one direct edge. The sum of all edges, $\sum_{i=1}^n \sum_{j=1}^n d(i, j)$ in the star network, is obtained by:
\begin{eqnarray} \label{eq:start_sum_edges_cgc}
\sum_{i=1}^n \sum_{j=1}^n d(i, j) = 1+2(n-2)=2n-3.
\end{eqnarray} 
$2n-3$ is obtained where a center point is accessed by one edge while all other points (i.e., $n-2$ except itself and the center point) are accessible by two edges. The closeness in the star network, $C_{\text{closeness}}'(v_i)$, is obtained by:
\begin{eqnarray} \label{eq:denom_cgc}
C_{\text{closeness}}'(v_i) = \frac{n-1}{2n-3}.
\end{eqnarray} 
The difference between $C_{\text{closeness}}'(v^*)$ and $C_{\text{closeness}}'(v_i)$ is:
\begin{eqnarray} \label{eq:max_cc}
C_{\text{closeness}}'(v^*)-C_{\text{closeness}}'(v_i) = 1-\frac{n-1}{2n-3} = \frac{n-2}{2n-3}.
\end{eqnarray} 
Since there are $(n-1)$ differences, the maximum possible closeness metric, $\mathrm{max} \sum_{i=1}^n [C_{\text{closeness}}'(v^*)-C_{\text{closeness}}'(v_i)]$, is obtained by:
\begin{eqnarray} \label{eq:max_cgc}
\begin{aligned}
(n-1)\Big(\frac{n-2}{2n-3}\Big) = \frac{n^2-3n+2}{2n-3}.
\end{aligned}
\end{eqnarray} 
Therefore, the denominator, the maximum possible sum of the closeness differences in Eq.~\eqref{eq:cgc}, is proved.
\end{comment}

\subsubsection*{\bf Reciprocity} \citet{Newman02-virus} measured a {\em network reciprocity} based on the number of bidirectional edges between two nodes over the total number of possible edges in a network.  In directed networks, for an edge from node $i$ to node $j$, if there is an edge from node $j$ to node $i$, it is said the edge from node $i$ to node $j$ is reciprocated, which is also called {\em co-links} in the World Wide Web context~\cite{Eckmann02}. Formally put, the reciprocity can be denoted by:
\begin{equation} 
\label{eq:reciprocity}
C_{\text{reciprocity}} = \frac{\sum_{ij} A_{ij}A_{ji}}{m} = \frac{\mathrm{Tr} \mathbf{A}^2}{m},
\end{equation} 
where $m$ is the number of edges.

\subsubsection*{\bf $k$-component} This metric refers to a maximal subset of nodes where each node can reach from each of other nodes based on minimum $k$ paths that are vertex-independent. Note that two paths are said to be {\em vertex-independent} if they do not share any of the same vertices~\cite{Newman10}. A variant of the $k$-component can be identified based on edge-independent paths, implying that removing less than $k$ edges cannot make the component disconnected~\cite{Newman10}.

\subsubsection*{\bf $k$-clique} A {\em clique} refers to a maximum subset consisting of vertices in an undirected network where each member of the subset is directly connected to each other~\cite{Seidman78-plex, Tichy73}. If the size of the clique is large, it represents a highly {\em cohesive} network with close connectedness between each other~\cite{Newman10}.

\subsubsection*{\bf $k$-plex} This metric relaxes the condition of the clique as we cannot find a perfect clique in reality.  A $k$-plex refers to the maximum size of the subset of $n$ vertices in a network where each vertex is connected with minimum $n-k$ other vertices~\cite{Seidman78-plex}. 1-plex with $k=1$ is indeed a clique.

\subsubsection*{\bf $k$-core} This metric is a very close concept to the $k$-flex.  It refers to the maximum size of a subset consisting of vertices that have minimum $k$ connections with other vertices in the subset.  In this sense, the $k$-core is a $(n-k)$-flex.  But given a $k$ value, the set of all $k$-cores is not the same as that of all $k$-flexes because $n$ is different for a different $k$-core. Further, different from $k$-flexes, each $k$-core is distinct because when two $k$-cores share one or more vertices, a single, larger-sized $k$-core can be formed~\cite{Newman10, Seidman78-plex}. 

\subsubsection*{\bf Global clustering coefficient} Based on the mean of (local) clustering coefficient for a given graph, \citet{Watts98} also defined the {\em global clustering coefficient} (GCC) as:
\begin{equation} \label{eq:gcc}
GCC (\mathcal{G}) = \frac{\sum_{v \in \mathcal{V}} C_{\text{clustering}}(v)}{n},
\end{equation}
where $C_{\text{clustering}}(v)$ is the local clustering coefficient of node $v$~\cite{Watts98}.  Network transitivity is often defined based on GCC using the concept of transitivity among three nodes in a network~\cite{Holland71, Newman10}. 

\subsubsection*{\bf Degree assortativity} \citet{Newman02-assortativity} first defined the {\em assortativity} of a network as a graph measure to represent to what extent nodes are associated with other nodes in terms of network structural characteristics, such as degree, betweenness, node weight, node coreness as well as node characteristics, such as ethnic, language, and/or culture. In~\cite{Newman02-assortativity}, given a simply undirected, non-weighted network, assortativity is defined as a scalar value $\rho$.  For example, degree assortativity is denoted by $\rho_D$ which can be simply defined based on the linear correlation coefficient between two nodes' excess degrees~\footnote{A node's excess degree is its degree minus 1 (i.e., $d_i-1$), also known as the remaining degree of the node}, which are random variables and given by: 
\begin{equation} \label{eq:assortativity-degree}
\rho_D = \frac{\sum_{jk} jk (e_{jk}-q_j q_k)}{\sigma_q^2},
\end{equation}
where $e_{jk}$ refers to the joint excess degree probability for excess degrees $j$ and $k$.  $q_k$ is a normalized distribution of a randomly selected node and given by $q_k = \frac{(k+1) p_k}{\sum_j j p_j}$, where $\sigma_q$ is the standard deviation of $q_k$ in Eq.~\eqref{eq:assortativity-degree}. \citet{Newman03-assortativity} further defined degree assortativity in non-weighted, directed networks, as $\rho_D = \frac{\sum_{jk} jk (e_{jk}-q_j^{in} q_k^{out})}{\sigma_{in} \sigma_{out}}$, where $e_{jk}$ indicates the probability that a node with out-degree $k$ and a node with in-degree $j$ is connected for $k, j \in \mathcal{N}$, $q_j^{in} = \frac{(j+1)p^{in}_{j+1}}{\sum_j j p^{in}_j} = \frac{(j+1) Pr[D_{in} = j+1]}{E[D_{in}]}$ is the normalized excess in-degree distribution where $D_{in}$ is the in-degree for a randomly selected node, $q_k^{out}$ is defined similarly, and $\sigma_{in}$ and $\sigma_{out}$ are the standard deviations of $q_j^{in}$ and $q_k^{out}$, respectively.  
\citet{Noldus15} discussed multi-layered assortativity to be applied in directed networks, including: (1) {\em in-degree assortivity} measuring the tendency of a particular in-degree node that is connected to the same in-degree or different in-degree nodes; (2) {\em out-degree assortativity} estimating the trend of a particular out-degree node's connectedness with the same out-degree or different out-degree nodes; and (3) {\em overall assortativity} calculated based on both in-degree assortativity and out-degree assortativity.

\subsubsection*{\bf Local Assortativity} \citet{Piraveenan10-local-assortativity} defined {\em local assortativity} to measure an individual node's assortativity based on its degree and its neighbors' degree. The local assortativity is measured by:
\begin{equation} 
\label{eq:local-assortativity}
\rho_i = \frac{(j+1)(j \bar{k}-\mu_q^2)}{2 L \sigma_q^2},
\end{equation} 
where $j$ is the excess degree of node $i$ (i.e., $d_i-1$), $\bar{k}$ is the average excess degree of node $i$'s neighbors (i.e., $[\sum_{j \in N_i} (d_j -1)]/d_i$ where $N_i$ is the set of $i$'s neighbors), $\sigma_q$ is the standard deviation of the distribution of $j$ over all nodes in the network, $\mu_q$ is the average $j$, and $L$ is the number of edges in the network. Note that the sum of all local assortativities is the network assortativity, $\rho = \sum_i \rho_i$.

\subsubsection*{\bf Graph curvature}
One hypothesis to explain the phenomenon observed in many large networks of traffic congestion occurring at a core set of nodes in the network is that the network as a whole is negatively curved. Evidence supporting this hypothesis includes the success in embedding networks in hyperbolic space or deriving various properties using hyperbolic network models~\cite{krioukov2010hyperbolic}.  If the network is negatively curved, then routing paths influenced by shortest path selection are somewhat forced to traverse this core, leading to congestion.  Point centralities are useful in potentially identifying this core set, but they do not measure the network curvature of the graph as a whole.  To address this problem, \citet{narayan2011large} developed a large scale curvature measure by adapting to graphs the ``$\delta$-thin triangle condition''~\cite{gromov1987hyperbolic} that defines negative curvature. For any triple of nodes $i, j, k$, we define the distance function from any other node $m$ to the triangle of nodes by $D(m;i,j,k) = \max \{ d(m;i,j), d(m;i,k), d(m;j,k) \}$ where $d(m;u,v)$ is the minimum distance from the node $m$ to the geodesic between $u$ and $v$. Then, the curvature of a network with respect to the triple can be defined as:
\begin{equation}
\label{eq:graph_curvature}
\delta_{i,j,k} = \min_{m} D(m;i,j,k).
\end{equation}
An infinite network is negatively curved (hyperbolic), if $\delta = \max_{i,j,k} \delta_{i,j,k} < \infty$. Obviously, finite networks would not satisfy this condition, hence comparing $\delta$ to the perimeter length of the triangle formed from the geodesics among the triple $(i,j,k)$. This ratio does not exceed $3/2$ for constant non-positively curved Riemannian manifolds~\cite{jonckheere2007upper}. To relax the constraint that every triple satisfies this condition and for computational reasons, \citet{narayan2011large} considered a random sampling of triples and determine if the ratio $\delta_{\Delta}/\ell$ converges for large $\ell = \min\{ d(i,j), d(i,k), d(j,k) \}$.

\section{Group Selection Metrics} \label{sec:group-selection-metrics}

When a group of nodes is selected for many of the problems in the application space (e.g., influence maximization, network destruction), simply selecting the top-$K$ ranked nodes is a na\"{i}ve approach. Many networks exhibit assortativity, with respect to degree or another centrality, or redundant clustering. A simple example demonstrating the problem with top-$K$ selection strategy is the observation of the importance of the $k$-shell (certainly for influence maximization), as the top-$K$ nodes all may reside in the same $k$-shell and be neighbors. $k$-shell based centrality approaches would only push the selected nodes to the edge of the top $k$-shell, which may be highly localized instead of distributed throughout the network.

One approach to resolving this issue to to iteratively select a single node and recalculate the centrality measure for the remaining network excluding the selected node(s). This strategy has been studied for network robustness \cite{holme2002attack} and the recalculation can be trivial for certain measures (e.g., degree, coreness). For other measures, this recalculation may be expensive. Hence less costly approaches have been developed, seeking to discover a more optimal set of $k$ nodes.

\subsubsection*{\bf DegreeDistance} 
\citet{sheikhahmadi15} introduced a degree-distance metric to ensure the selected nodes are well-dispersed in the network. The strategy first computes the degree of each node and selects the node with highest degree. It then excludes for selection all nodes within a chosen threshold distance $t_{td}$ from any of the previously selected nodes and selects the node with highest degree. Hence, given a current set of selected seed nodes $S$, the next selected node is chosen to be
\begin{equation}
\label{eq:DegreeDistance01}
v = \argmax_{u \in \mathcal{V}|d(u,w)\ge t_{td} \forall w \in S}  C_{\text{deg}}(u) .
\end{equation}
Since this threshold distance can omit from potential selection high degree nodes that are within the threshold distance but have limited common neighbors (or neighbors of neighbors) with the previously selected nodes, the authors introduced two improvements to {\em DegreeDistance}. The first improvement of DegreeDistance (FIDD) does not exclude a node $v$ within the threshold distance provided the number of common neighbors and common neighbors of neighbors with previously selected nodes in $S$ is below a chosen threshold $\theta$. The second improvement of DegreeDistance (SIDD) adds another check to determine an influence score $\mathbb{P}(u,v)+\sum_{w\in CN(u,v)}(\mathbb{P}(u,w)\cdot\mathbb{P}(w,v))$, 
\begin{comment}
\begin{equation} \label{eq:DegreeDistance02}
\mathrm{influence} = \mathbb{P}(u,v)+\sum_{w\in CN(u,v)}(\mathbb{P}(u,w)\cdot\mathbb{P}(w,v)),
\end{equation}
\end{comment}
where $\mathbb{P}(u,v)$ is the activation probability that $u$ will influence $v$, and $CN(u,v)$ is the set of common neighbors of $u$ and $v$. Nodes within the threshold distance with influence above some threshold $\beta$ are excluded from being selected for inclusion in $S$ even when the common neighbors is below the threshold $\theta$. Essentially, sufficient pathways exist for the node to be affected by a seed node indirectly.

\subsubsection*{\bf SingleDiscount} This is essentially the iterative recalculation of degree. \citet{chen09} used this basic heuristic to compare against several greedy approaches to estimate the cascade models of \cite{Kempe03}. The node with maximum degree is selected for the seed set $S$ (ties broken randomly). Each neighbor of a selected node had a unit value reduction in its degree. This selection can be represented by
\begin{equation}
\label{eq:SingleDiscount}
v = \argmax_{u \in \mathcal{V}\backslash S} C_{\text{deg}}(u) - |N(u) \cup S|,
\end{equation}
where $C_{\text{deg}}(u) - |N(u) \cup S| = |N(u)| - |N(u) \cup S|$ is the degree of node $u$ excluding the current links to the seed set $S$.

%\citet{chen09} proposed {\it SingleDiscount}, a degree discount heuristic as an alternative of the original degree-based greedy algorithm~\cite{Kempe03}. The basic rule of this strategy, given that vertex $u$ has been chosen as a seed and vertex $v$ is a neighbor of $u$, is not to count the edge $\overline{vu}$ as a degree of $v$ when considering choosing $v$ as a new seed based on its degree. In this respect, $v$'s degree is discounted as many as the number of neighbors of $v$ which are already in the set of seeds. This can be represented by:
%\begin{equation} \label{eq:SingleDiscount}
%    sd_v = d_v - t_v
%\end{equation}
%where $d_v$ is the original degree of the vertex v and $t_v$ denotes the number of v’s neighbors who have already been selected as spreaders.

\subsubsection*{\bf DegreeDiscount} 
The SingleDiscount approach ignores the probability that a node may be affected by a neighbor in the seed set. \citet{chen09} constructed an alternate heuristic to account for this and better match the {\em independent cascade model} of \cite{Kempe03}. Under the assumption of a small propagation probability of $p$, that $t_v$ neighbors of $v$ are already in the seed set, and that $C_{\text{deg}}(v) = O(1/p)$ and $t_v - o(1/p)$, then the expected number of additional vertices in $N(v)$ that will be influenced by the selection of $v$ can be shown to be $1 + \left(C_{\text{deg}}(v) - 2 t_v - (C_{\text{deg}}(v) - t_v) t_v p + o(t_v)\right) \cdot p$. This is derived via the probability $(1-p)^{t_v}$ that $v$ would not be influenced by nodes already in the seed set and the expected number of vertices $1 + (C_{\text{deg}}(v)-t_v) \cdot p$ that $v$ influences its neighbors that are not in the seed set. This ignores indirect influences, which would be expected to be minimal for small $p$. Hence, the selection criteria, using an appropriate {\em DegreeDiscount} is
\begin{equation}
\label{eq:DegreeDiscount02}
v = \argmax_{v \in \mathcal{V}\backslash S} C_{\text{deg}}(v) - 2 t_v - (C_{\text{deg}}(v) - t_v) t_v \cdot p ,
\end{equation}
where $S$ is the current seed set.

%\citet{chen09} derived an advanced degree discount heuristic from {\it SingleDiscount} for the {\it independent cascade model} \cite{Kempe03} with a small propagation probability $p$. Vertex $v$ will not be selected as a seed in case $v$ is influenced by $u$ with probability of $p$ given that $v$ is a neighbor of a seed vertex $u$. Based on this idea, this strategy applies additional degree discount. Let $N(v)$ be the subgraph, including $v$ and all vertices connected with $v$ via a single edge.  The expected number of additional vertices in $N(v)$ influenced by selecting $v$ as a seed is:
%\begin{equation} \label{eq:DegreeDiscount01}
%1+(d_v-2t_v-(d_v-t_v)t_v p+o(t_v))\cdot p,
%\end{equation}
%where $t_v$ is the number of neighbors of $v$ which are already included in the set of seeds. Assuming that $d_v = O(1/p)$ and $t_v = o(1/p)$ which are applicable in real-life networks, vertex $v$ will be influenced by its adjacent vertices at probability of $1-(1-p)^{t_v}$. In case $v$ is not influenced with probability of $(1-p)^{t_v}$, the additional vertices influenced by choosing $v$ as a seed is $1+(d_v-t_v)p$. Thus, the above equation is derived by:
%\begin{equation} \label{eq:DegreeDiscount02}
%(1-p)^{t_v}\cdot(1+(d_v-t_v)\cdot p).
%\end{equation}
%If $t_v = 0$, the expected number of influenced vertices is $1+d_v\cdot p$ which implies that the strategy discounts $v$'s degree by $2t_v+(d_v-t_v)t_v p$ when there are $t_v$ neighbors already in the seeds set. This strategy ignores the indirect influences of $v$ based on the assumption that the influences are negligible for small $p$.

\subsubsection*{\bf DegreePunishment} 
To account for indirect influence from nodes in the seed set, \citet{wang2016effective} introduced a strategy that punishes nodes near the seed set. The punishment is determined by how many short paths the node is on, the penalty more severe if the node is closer to a seed and, consequently, closer to the seed on the paths. This punishment is $p_{u \to v} = C_{\text{deg}}(u) \sum_{h=1}^{r-1} (\mathbf{A}^h)_{uv} \omega^h$, where $\mathbf{A}$ is the adjacency matrix, $\omega$ is a weaken factor (typically assigned to be the propagation probability), and $r$ is the radius of influence or length of the considered paths. Then given the current seed set $S$, the {\em DegreePunishment} selection of the next node is given by
\begin{equation}
\label{eq:punisheddegree}
v = \argmax_{v \in \mathcal{V} \backslash S} C_{\text{deg}}(v) - \sum_{u \in S} p_{u \to v} .
\end{equation}
The complexity of this process grows with the radius $r$ of the paths from the seed set, so \citeauthor{wang2016effective} limited the radius to $r=2$ in their simulations.

%\citet{wang2016effective} introduced a punishment strategy for the degrees of nodes nearby each successively selected node. At each step (or node selection) of the process, the nodes have a {\em punished degree}, defined as:
%\begin{equation}
%\label{eq:punisheddegree}
%dp_v = d_v - \sum_{u \neq v} p_{u \to v}\sigma_u,
%\end{equation}
%where $\sigma_u$ is an indicator function on the previously selected nodes (i.e., $\sigma_u=1$ if $u$ has been selected and $\sigma_u=0$, otherwise) and the punishment is given by:
%\begin{equation}
%\label{eq:punishment}
%p_{u \to v} = d_u \sum_{h=1}^{n-1} (A^h)_{uv} \omega^h.
%\end{equation}
%The punishment from a selection extends to radius $n-1$ from each selected node with diminishing effect dependent on $\omega \in (0,1]$.  \citet{wang2016effective} set $\omega$ to be the propagation probability and set $n-1=2$ so the process is semi-local and remains efficient.

\subsubsection*{\bf Collective influence}
\citet{morone2015influence} introduced a scheme to capture the {\em collective influence} (CI) of a set of nodes using the concept of optimal percolation. The influence of a single node is determined by its corona, defined in a similar manner as volume centrality (see Eq. \eqref{eq:volume-centrality}). This influence of a node $v$ is $C_{\text{collective-inf}}(v,\ell) = (C_{\text{deg}}(v)-1) \sum_{u\in\partial B(v,\ell)}(C_{\text{deg}}(u)-1)$, where $\partial B(v,\ell)$ is the set of nodes within the distance of $\ell$ from node $v$. Hence, given the current seed set $S$, the next node selected is
\begin{equation}
\label{eq:collectiveinfluenceselection}
v = \argmax_{v\in \mathcal{G'} = \mathcal{G}\backslash S} C_{\text{collective-inf}}(v,\ell) ,
\end{equation}
where the collective influence is in the remaining graph with the nodes in S removed. \citet{morone16} also provided a stopping criteria for their approach by updating an estimate of a lower bound on the minimum eigenvalue of the non-backtracking matrix when a fraction of $q$ nodes are removed. This estimate is given by $\lambda(\ell;q) = \left(\frac{\sum_v C_{\text{collective-inf}}(v,\ell)}{n \langle k\rangle}\right)^{1/(\ell+1)}$, where $\langle k\rangle$ is the mean degree of original network. When $\lambda(\ell;q)=1$, the selection process is finished.

\begin{figure}[!t]
    \centering
\begin{tikzpicture}[font=\footnotesize]
\begin{axis}[
    ybar,
    legend style={
        cells={anchor=west},
        legend pos=north west},
    bar width=.15cm,
    width=.70\textwidth,
    height=.40\textwidth,
    ymin=0,
    ymax=32,
    enlarge x limits=0.05,
    ylabel={\# centrality metrics},
    ylabel near ticks,
    symbolic x coords={~1960s or earlier, 1970s, 1980s, 1990s, 2000s, 2010s},
    xtick=data,
    nodes near coords,
    nodes near coords align={vertical},
    ]
\addplot coordinates {(~1960s or earlier, 4) (1970s, 2) (1980s, 2) (1990s, 7) (2000s, 14) (2010s, 25)};
\addplot coordinates {(~1960s or earlier, 1) (1970s, 8) (1980s, 0) (1990s, 2) (2000s, 2) (2010s, 4)};
\addplot coordinates {(~1960s or earlier, 0) (1970s, 0) (1980s, 0) (1990s, 0) (2000s, 2) (2010s, 3)};
\legend{Point centrality, Graph centrality, Group selection centrality}
\end{axis}
\end{tikzpicture}
\caption{Centrality metrics developed under each category from the 1960s or earlier to the 2010s. }
\label{fig:evolution-centrality-metrics}
\vspace{-5mm}
\end{figure}
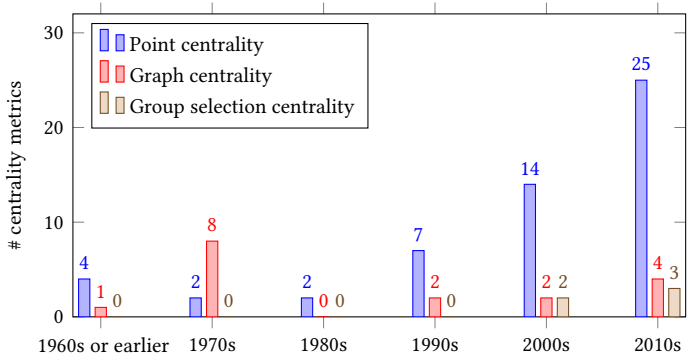

Based on our comprehensive survey on centrality metrics conducted in Sections~III-V, we summarized them based on their published years in order to capture the overall evolution of centrality metrics in Table~1 of the supplement document due to the space constraint. Instead, we summarized how many metrics are studied over time from the 1960s or earlier until the 2010s in Fig.~\ref{fig:evolution-centrality-metrics}.  From Table~1 of the supplement document, we observed that the centrality metrics developed in the 1960s or earlier until the 1980s (e.g., degree, betweenness, closeness, eigenvector centrality) have been still commonly used in the research under various network domains. But we can also clearly notice from Fig.~\ref{fig:evolution-centrality-metrics}, various types of centrality metrics have been significantly studied since the 2000s and more actively in the 2010s.

\section{Applications of Centrality Metrics in Various Network Types} \label{sec:applications}

In this section, we give an overview of how centrality metrics have been applied in various types of networks, including social networks, contact networks, computer communication networks, and biological networks.

\subsection{Social Networks} \label{subsec:social-networks}

{\bf Information Diffusion.} This problem involves determining the initial set of nodes that efficiently propagates information throughout the network. \citet{Kim12-volume} and \citet{Kim15-neighbor-influence} investigated this selection process under different information diffusion strategies. They found that when the initial set of seed propagators are high-degree nodes, then the choice of which neighboring nodes to spread the information does not affect the long-term propagation significantly. %This implies that the initial selection seed nodes to start the information spread is the more critical component.

Network structure features, such as network topology, node in-degree, out-degree, edge weight, and clustering coefficient have also been considered in studies of false information propagation~\cite{Cho19-tsc, kumar2016disinformation, ratkiewicz2011truthy, Wu16-bc}.  \citet{Cho19-tsc} built a uncertainty-based subjective opinion model using a belief model, called Subjective Logic.  They developed different types of agents that can propagate false information intentionally (i.e., disinformers) and mistakenly (i.e., misinformers), where true information is also propagated to counter the false information.  %The authors investigated the effect of different types of centrality metrics used in the selection of sources propagating the false or true information.  
\citet{kumar2016disinformation} developed four feature sets including network features to identify hoaxes in Wikipedia. The network features measure the relation between the references of the article in the Wikipedia hyperlink network.  \citet{ratkiewicz2011truthy} built a `Truthy' system to enable the detection of `astroturfing' (fake grass root campaigning with hidden sponsors) on Twitter.  \citet{Wu16-bc} summarized false information spreader detection based on network structures.%  The authors examined how the so-called `forceful' individuals (not changing their opinions) can affect information diffusion depending on how they are connected with other individuals or when they bridge multiple communities. 

\citet{Kimura07-info-diffusion, Kimura10} considered the problem of identifying the most influential nodes in a large-scale social network as a combinatorial optimization problem.  %The authors proposed an efficient greedy algorithm based on bond percolation and graph theory and demonstrated its  superior performance over conventional methods in terms of computational cost.  
\citet{tang2010information} investigated an email dataset as a dynamic, social network in order to study dynamic interactions using a proposed `temporal centrality metric.'  %Particularly, the authors measured information dissemination using the centrality metric and examined the role of `information mediators' to better understand dynamics of the social network and accurately identify central people compared to only using conventional, static centrality metrics.  
\citet{kandhway2016} studied information diffusion using an epidemic model to maximize information diffusion for a certain period of campaign running in a social network.  %The authors proposed an optimal control framework that can maximize the information diffusion using controls (i.e., advertisement) over the campaign period. They examined the effect of various types of centrality metrics for initial spreaders on the information diffusion.

{\bf Influence Maximization.}
\citet{bae2014identifying} focused on classifying the ability of influential nodes in order, avoiding the assignment of multiple nodes to the same order, using neighborhood coreness centrality. %This measure is closely related with epidemic models in that higher influence implies a broader scope of epidemic spreading.  
\citet{bian2017identifying} adopted the SI (Susceptible-Infected) model to identify influential nodes spreading a disease in complex networks by using the AHP decision making strategy that combines different centrality metrics which typically include degree, closeness and betweenness.  \citet{chen2012identifying} introduced semi-local centrality metric and used a modified version of the SIR model to verify its correctness. %The difference between the original SIR model is that not every neighbor of an infected node will be infected with a particular propagation probability but rather one neighbor is chosen randomly and infected with certainty.
\citet{bavelas1950communication} indicated that centrality position in small groups influences the perceptions of leadership (as well as morale).  \citet{newman2005measure} demonstrated how random-walk betweenness is a better measure than degree in the Florentine families intermarriage network~\cite{padgett1993robust}.  \citet{Mochalova2013} examined the relationships between the influence of key members and the attitude the remaining members have towards information and how the relationship impacts information diffusion and its outcome.  %This work is unique in that influence is captured by network centrality metrics without consideration other data features describing an individual.

A key goal in marketing or information diffusion research is to identify influentials, a small set of nodes that can significantly affect a large portion of their network.  \citet{Watts07-influentials} questioned this hypothesis and studied if the size of influence cascades is truly caused by the information propagated from the influentials.  %In their extensive computation simulations, considering the interplay between influencers and influencees, they found that a large component of the influence is introduced by those who tend to be easily influenced, or `easily influenced individuals,' rather than from the influentials themselves. 
\citet{Saito13} studied the identification of super-mediators, nodes playing a significant role in receiving or passing information between other nodes in social networks.  %The authors formulated the maximization problem in terms of the maximum difference between when a node is removed and when the node remains present in the network.  Via extensive experiments based on three real-world datasets, they observed in-degree as a good indicator of information diffusion under a small diffusion probability while betweenness was better under a large diffusion probability.  However, no single centrality metric was as informative when the diffusion probability is widely varied.
\citet{Goyal10} studied a fundamental problem in terms of where or how the input parameters to study an influence model in social networks can be obtained. % Since the expected finding would be vastly different depending on those parameter values characterizing the social network and interactions between users, this work proposed learning algorithms to obtain those input values and validated the decisions using a large real social network topology. 

{\bf Influence Minimization.} 
\citet{Kimura09} solved an influence minimization problem by blocking a limited number of links that spread false information or rumors, where betweenness and out-degrees are used to identify links or nodes to remove.  This study found that removing high out-degree nodes is not necessarily effective compared to blocking a limited number of links to maximize the containment.  \citet{Dey17-inf-min} also studied what nodes to block in order to minimize information propagation. This work used betweenness, edge betweenness, degree, and closeness to block influential nodes. Similarly, \citet{Qipeng15-www} solved the same problem but by blocking a limited number of nodes where the centrality metrics considered are out-degree and betweenness.  \citet{Luo14} proposed an algorithm that identifies a set of critical nodes to minimize disinformation in time-varying online social networks.  The authors conducted a comparative performance analysis and demonstrated that their proposed algorithm outperforms a centrality-based heuristic counterpart, particularly using degree and closeness. 

\subsubsection*{\bf Behavior Adoption for Marketing}  Centrality metrics have been also studied as a way to identify initial target populations as a marketing strategy.  In adopting technological innovations or purchasing some products, word-of-mouth processes are also modeled using information diffusion models~\cite{Czepiel74}. In particular, as marketing tools, what population to focus advertising is a major concern, wherein centrality metrics are adopted to identify the target populations~\cite{Kempe03}. Many marketing applications aimed to leverage social networks or media by targeting populations using simple centrality metrics, such as degrees~\cite{Dinh14-marketing, Yan14}, betweenness~\cite{Shao13-marketing, Yan14}, closeness~\cite{Shao13-marketing, Salavati19-marketing}.

To study the spreading process of technology adoption, various information maximization algorithms have been proposed and applied to investigate the effect of word of mouth in markets, or game theoretic strategies~\cite{Kempe03}. \citet{Kempe03} showed that the influence maximization problem is NP-hard and many heuristic or greedy algorithms to solve this problem can provably guarantee a solution to within 63\% of the optimal solution, with performance guarantees close to $1-1/e$.  %The authors showed that node selection based on structural centrality metrics outperforms the-state-of-the-art heuristic approaches in selecting influential nodes.

{\bf Community Detection.} 
\citet{nikolaev15detection} developed a variant of entropy centrality to understand `the entropy of flow destination' in networks and showcased how the new entropy centrality is more useful over the original entropy centrality in community detection applications.  \citet{jiang2013detection} proposed an efficient centrality measure, called $K$-rank, designed for selecting the top-$K$ nodes with the highest centrality. The top $K$ nodes are used as the initial seeding nodes and updated based on $K$-means iterations.  %The authors applied the $K$-rank to derive a directed, weighted network for detecting overlapping communities.

\subsection{Contact Networks}
\citet{christley2005infection} attempted to identify the risk of disease infection of nodes using centrality metrics, such as degree, random-walk betweenness, shortest-path betweenness, and farness.  \citet{Dekker13} also used six different centrality metrics, including degree, betweenness, two types of closeness, distance-based centrality, and eigenvector centrality in order to identify the super spreaders of infectious diseases.  \citet{bell1999} investigated the co-relationships between various types of centrality metrics and their variants such as degree, betweenness, closeness, eigenvector centrality, information centrality, and power prestige.  \citet{gomez2013diseases} studied high-risk hosts for emerging infectious diseases based on various centrality metrics (e.g., strength, degree, betweenness, closeness, eigenvector centrality) for their control and surveillance.  The authors used network tools to predict parasitism and the host spreading future infectious diseases. 

\subsection{Communication Networks}

Centrality metrics have also been used to make decisions to solve various problems in communication networks. Centrality metrics have been used to select critical nodes to prevent or mitigate computer virus or malware spreads.  \citet{Newman02-virus} conducted an empirical study of investigating the email network structure to examine what nodes can significantly contribute to spreading computer viruses. \citet{Kim19} measured the risk of websites exposing security vulnerability (e.g., malware, fake infectious sites) based on degree, betweenness, eigenvector, and closeness. 

%Targeted attacks based on centrality metrics have been popularly considered to investigate the impact of the targeted attacks on network resilience using the size of the giant component.  

\citet{albert2000error} showed scale-free networks, following a power-law degree distribution, are highly robust to random attacks while highly vulnerable to targeted attacks on high degree nodes.  \citet{holme2002attack} also investigated the network resilience in complex networks when targeted attacks are applied based on degree or betweenness.  \citet{Yoon17} developed a scalable centrality-based traffic measurement based on software defined networking functionalities. %To ensure secure traffics, the authors proposed a packet sampling technique that determines the traffic sampling points among the switches. This work proposed a degree-based algorithm to detect malicious traffics and demonstrated its performance in terms of an intrusion detection accuracy.

\subsection{Geographic Networks}

\citet{crucitti2006centrality} analyzed spatial networks based on different centrality metrics to characterize the geographic properties of cities as networks. %Substituting an undirected graph for urban streets of a city and measuring the different centralities, they presented a spatial distribution of centralities that show the main structures of the city: centric areas and major routes, depending on the type of measures.  
\citet{Gao13} used the betweenness centrality to measure urban traffic flow with GPS-enabled taxi trajectory information in Qingdao, China.  This study demonstrated that betweenness is not necessarily a good metric to measure the traffic flow distributions. %The authors suggested combining a network structure with other information such as different patterns of human activities depending on location, power law distance-decay, and human mobility patterns.
\citet{Porta08} developed a `Multiple Centrality Assessment (MCA)' framework that uses centrality metrics to understand why the current design features of a city do not attract more people or increase social life.  %This work used closeness, betweenness, straightness (or degree), and information centrality to understand the current attractiveness of the city. 
\citet{Guimera05} examined the impact of a city's global role based on degree and betweenness. %They found that a city's betweenness is more closely related to the city's global role with intercommunity and intracommunity connections than the city's degree centrality.  
\citet{Li15} examined how centrality of each shipping area, with 25 geographical areas, plays a key role in changing the centrality of the global shipping networks (GSNs) during the years 2011-2012.  %This study used degree, betweenness, and closeness as centrality metrics to analyze the dynamics of the GSNs. 

\subsection{Biological Networks}
\citet{Estrada05} used centrality to study the removal of proteins from the yeast {\em S. cereviciae}. The lethality of protein removal has been shown to correlate with the degree of the protein.  \citet{jeong2001lethality} conducted an experiment of arranging proteins in order of the degree they have and testing the consequences after each protein has been removed. %From this experiment, they found that the removal of proteins with high degrees tends to show a greater effect on the topology than the proteins with relatively low degree. This study demonstrated that the robustness of networks depends on not only the functionality of individuals, but also the inter-node interactions within the network itself and its topological properties.  
\citet{Koschutzki08} analyzed the structure of gene regulatory networks based on the ranks of nodes, which are measured by centrality metrics. %They used degree, betweenness, integration, radiality, Katz status index, PageRank and various types of motif-based centralities.  
\citet{Karabekmez16} proposed a new centrality metric called {\em weighted sum of loads eigenvector centrality} (WSL-EC) in order to identify critical nodes in biological networks. %The examples are to identify central nodes such as pathogen-interacting, cancer, aging, HIV-1 or disease related proteins, proteins involved in immune system processes, and auto-immune diseases in the human interactome.  The authors showed WSL-EC outperforms other centrality metrics such as degree, betweenness, subgraph centrality, and eigenvector centrality in this identification problem.  They observed that other centrality metrics are biased towards super-hubs while the WSL-EC is not.  
\citet{Mistry17} developed a new centrality metric to predict central and critical genes and proteins based on a protein-protein interaction network.  %The proposed centrality metric considers both the amount of a protein's interaction and the gene coexpression values of genes.

We summarized what centrality metrics have been used in various network types based on our discussions in this work in Table~2 of the supplement document.  Although our discussions on the applicability of centrality metrics are limited, this table shows a trend of what centrality metrics have been substantially utilized in contact and biological networks compared to other network domains. %Obviously the use of centrality in communication networks is in its infancy and their remains potential for leveraging their merits. 
Despite a large volume of centrality metrics studied in the literature (see Sections~\ref{sec:point-centrality}, \ref{sec:graph-centrality-metrics}, and \ref{sec:group-selection-metrics}, we clearly observe that the uses of centrality metrics have been mostly limited to several common centrality metrics, such as degree (including in/out-degree), betweenness, closeness, and eigenvector centrality.

\section{Concluding Remarks} \label{sec:conclusion}

In this section, we discuss what we learned from this present study and how to improve the limitations of the existing centrality metrics by suggesting future research directions. 

In particular, we implemented over 60 centrality metrics surveyed in this work under the three centrality metrics categories (i.e., point, graph, and group selection centrality metrics). We tested their effect on network resilience based on a size of the giant component when each centrality metric is used to model targeted attacks. We evaluated the performance of each metric under two undirected real network datasets and two directed real network datasets.  Due to the space constraint, the details and experimental results along with the explanations of observed trends are addressed in Section 4 of the supplement document. In this section, we also discuss some insights learned from the findings obtained from the extensive simulation results.

\subsection{Limitations, Insights, and Lessons Learned} \label{subsec:limitations-insights}
We have found limitations of the existing centrality metrics surveyed in this work, learned lessons and obtained the insights from them as follows:
\begin{itemize}
\item The meaning of centrality is not only limited to how a node is connected to other nodes, but also implies how actively the node communicates to each other and how it can control or influence other nodes in their centrality or vulnerability.  In brief, node centrality determines influence in terms of {\em connectivity}, {\em communicability}, and {\em controllability} in a given network.  However, node connectivity is not commonly aligned with the capacity to deal with traffic (e.g., communicability) because nodes with high connectivity are often congested.

\item Centrality metrics can be applicable in various disciplines with different purposes. In addition, there is a rich volume of centrality metrics available that can be used for various design goals.  For example, we may want to investigate how to balance traffic loads, how to set edges between nodes to make a network robust against faults or attacks, what types of targeted attacks to develop, how to identify vital nodes based on various criteria, or what is the most (least) influential or vulnerable node in a given network.

\item We investigated the effect of each centrality metric on network resilience in terms of a size of the giant component.  We found that if a centrality metric measures how well a node is connected with its close neighborhood (i.e., locally well connected), its impact upon removing the node with high centrality tends to be limited. For example, removing nodes with high clustering coefficient or volume centrality is not as severe as the random removal of nodes in network resilience (i.e., the size of the giant component).  However, if the centrality metric refers to how well the node is globally linked with other nodes which may belong to another cluster of the network (e.g., another community), when the node fails, the network is highly impacted by the node's failure.

\item We found that when an attack using a given centrality metric is non-infectious, what metric to choose is highly critical because the effect of a different centrality metric can be vastly different.  However, when the attack is infectious, using different centrality metrics doesn't introduce a significantly different impact on network resilience as the infectious attack itself may be powerful. In addition, we found how a node is connected in a given network (i.e., network topology characteristics such as network density) is a more important factor that influences the network resilience (i.e., a smaller size of the giant component).

\item Although a large volume of centrality metrics has been developed so far, only common centrality metrics have been used, such as degree, betweenness, closeness, clustering coefficient, or pagerank, which has been developed for several decades ago.  Although degree is a simple metric, other metrics, such as betweenness or clustering coefficient, require high complexity with high running time.  It was interesting to observe that even if there have been many centrality metrics developed in the 2010s, not many of them have been used in the existing network applications while the metrics developed from the 1970s to the 1990s have commonly been used in the literature.  

\item Unlike centrality metrics that are applicable in undirected networks, centrality metrics in directed networks may not be appropriate to study their effect on network resilience.  This is because even a node's failure with high centrality (e.g., hub, authority, or leaderrank) in sparse networks may not introduce any significant impact where centrality is mainly measured based on in-degree, not out-degree.

%\item Since each centrality metric measures centrality differently and the range of the centrality value is vastly different depending on the centrality metric, we cannot use the largeness of the centrality value as an absolute standard to determine the resilience of a given network (i.e., a size of the giant component). In addition, we observed that all centrality values do not necessarily decrease as a size of the giant component decreases. Some centrality metrics even show smaller centrality values when a size of the giant component becomes smaller by removing more nodes.  In addition, the characteristics of a given network, such as node density (i.e., a number of nodes), network density (i.e., a number of edges), or how nodes are being connected to each other (e.g., preferential attachment in scale-free networks or mean degree with less variance in random graphs) significantly affects the largeness of a centrality value under some centrality metrics.  If we want to use the largeness of a centrality value as an indicator to represent a size of the giant component in a given network, we should ensure if the centrality metric shows a consistent trend in its largeness under various sizes of the giant component.

\item We used the size of the giant component as an indicator to represent network resilience. A size of the giant component is a conventional network resilience metric in the Network Science domain.  However, it does not necessarily indicate how many nodes are compromised as a metric to measure system vulnerability in terms of a cybersecurity perspective.  Even if the size of the giant component is small, it does not necessarily imply that the network has more compromised nodes because there could be healthy nodes in smaller components in the network.  
%The reason of having a smaller size of the giant component can be just because of a lack of connectivities between nodes, but because of the nodes being compromised in each piece of the components.

\item We investigated the running time of all centrality metrics surveyed in this work (see Section 5 in the supplement document). The overall trend is that centrality metrics tested under directed networks (e.g., SALSA authorities, SALSA hubs, leaderrank, clusterrank) tend to show higher running time than centrality metrics tested under undirected networks.   This may be because undirected networks innately have higher connectivity than directed networks. Recall that many centrality metrics rely on the (shortest) path distances between two nodes as part of the metric calculation.

%\item Based on our observation of the running times of all centrality metrics surveyed in this work, it seems there is no clear relationship between global centrality metrics (i.e., measuring the connectivity between a node and other nodes in another cluster, rather than in a local neighborhood) and the high running time.  This implies that even if we aim to measure a global centrality metric, we can still develop an efficient algorithm to measure the global centrality of a node or graph. \tjm{This is not a fair point. The times are based on networks of approximately 1000 nodes. But people or companies want to run some operations on networks with millions of nodes. At that scale, certain approaches even in sparse networks become untenable, e.g., betweenness. In order to judge the practical complexity, we really need to compare performance on different sized networks. (From Terry)}
%\textcolor{red}{
\item The running time of each metric (see Figs. 10-14 of the supplement document) is mainly influenced by network size, network or node density. In addition, in some metrics, we optimized the code to expedite the running time while others may not.  Therefore, there may be an inaccuracy introduced in the running times of centrality metrics demonstrated in this work.  However, we believe that this imperfect code optimization won't significantly affect the order of running time performance of centrality metrics compared in this work. %\jhc{In addition, the running time may be impacted by a subjective factor, which is that we have no guarantee that all metrics are emulated in optimum. In other words, even we provide the same computational resources for all metrics, we cannot guarantee that those metrics are maximizing the use of these resources. -- cannot understand; Zelin can you clarify this?}%(From Zelin)}

%\textcolor{brown}{\item Aligning with the first bullet point, the meaning of centrality varies depending on the network and its context. A node connected to many powerful nodes could be a powerful node regarding connectivity, on the other hand it could also be interpreted as having insufficient resources in other network such as exchange network regarding its bargaining power. (From BW)}

%\textcolor{brown}{\item Considering the size of giant component, although I am not really sure if it really does, I think the meaning of `giant' could vary as well in a specific case. The giant component is normally defined as the component that has the largest number of nodes in a network. Depending on the centrality metric being applied to the network, maybe I could suggest that the giant component of the network is the component that has the biggest total value of centrality of all nodes in the component. But still, the component that has the biggest number of nodes is highly likely to be `giant' as well regarding most centrality metrics. (From BW)} -- already discussed this on a size of the giant component.

\item Most point centrality metrics are extensions from notions of degree of the node or its neighbors (e.g. semi-local, $k$-shell, $h$-index), connections between neighbors of the node (e.g., Burt's redundancy, clustering coefficient), path finding processes involving the node (e.g., betweenness, closeness), or iterative processes between the node and its neighbors (e.g., eigenvector, pagerank).  The extensions attempt to capture something missing or ignored in a fundamental metric, e.g. the degree of the node by itself ignores the degree of its neighbors whereas semi-local centrality aggregates that information and both $k$-shell and $h$-index consider threshold effects on that information. New centrality metrics can be considered by supplementing an existing approach with missing information that may be relevant to the particular problem criteria. % (e.g., spreading, robustness, etc.). %(From Terry)}

\item For an insightful comparison of network resilience under infectious attack using different centrality metrics, the infection rate variability is highly dependent on the characteristic of the network (e.g., network or node density or network topology).  Infection is spread more easily in a dense network wherein all the nodes are more easily accessible.  On the other hand, a sparse network has a structural insulation protecting itself from an infectious attack.  %(From Yash)}
\end{itemize}

\subsection{Future Research Directions} \label{subsec:future-research}

\begin{itemize}
\item {\bf More efficient centrality metrics are needed}: Since there are many centrality metrics that suffice to meet certain tasks but require less complexity (i.e., low running time), we can leverage these or perhaps modify to enhance their effectiveness for the task (e.g., increasing the effect of removing a node with high centrality) or efficiency (e.g., running time).  Some metrics are representative of a broader meaning of centrality, such as communicability or controllability (e.g., load centrality in Eq.~\eqref{eq:load}), in addition to a simple connectivity.  However, their high complexity hinders applicability in various domains. 

\item {\bf More meaningful metrics are needed to measure network resilience}: The size of the giant component, as a common metric to measure network resilience, does not reflect a broader concept of network resilience.  Network resilience can be defined in terms of how {\em adaptable} a network is to deal with sudden changes or attacks/failures (i.e., adaptability), how {\em tolerant} the network is to prevent its failure against attacks or failures (i.e., fault tolerance), and how easily {\em recoverable} the network is from attacks or failures (i.e., recoverability)~\cite{cho19-stram}. %However, a size of the giant component as a network resilience metric only captures the connectivity of a node where the connectivity is assumed the only key factor of network resilience. In addition, when multiple times of attacks are applied, a size of the giant component cannot be a good indicator for network resilience because the connectivity is only ensured even with a single path between two nodes and does not capture how strong the two nodes are being connected (e.g., a number of paths between two nodes). However, some metrics may represent more than a size of the giant component, such as redundancy we studied in Eq.~\ref{eq:redundancy}. 
As a future work direction, we need to develop metrics that can measure network resilience embracing adaptability, fault tolerance, and recoverability, or other properties based on system requirements.% of various domains.

\item {\bf Graph centrality metrics can be enhanced as a novel measure of network resilience}: Graph centrality metrics measure certain characteristics of a given network, such as the distances between nodes, connections between neighbors, or redundant paths between nodes.  However, as we observed in Tables~4 and 5 of the supplement document, it is not necessarily correlated to the size of the giant component, which is a conventional metric measuring network resilience in some graph centrality metrics.  We can improve the existing graph centrality metrics or invent ones that can be used as indicators related to the key properties to network resilience. For example, when a certain graph centrality value is high, it may indicate the network has the ability to easily recover from attacks or failures. 
 
\item {\bf Centrality metrics embracing a broader concept of influence need to be developed}: Although a rich volume of centrality metrics has been explored in the literature, most of them rely on the concept of centrality based on connectivity.  However, in reality, being connected with less critical nodes does not introduce high impact on network resilience, as long as a small set of critical nodes are still kept safe and operating in a reliable manner.  In addition, although controllability is one of the key centrality concepts as discussed in Section~\ref{subsec:multidimensional-concept-centrality}, not many centrality metrics are developed without explicitly considering a node's controllability over a given network. There should be more efforts to develop centrality metrics that can fully consider its ability to control the network.

\item {\bf Enhancement of the infection process for modeling infectious attacks:} In the infection process considered in this work, a node is infected with a given probability. If the node is not infected with the probability, we simply assumed that it is immune to the attack and are not infected again.  However, in real world scenarios, various types of attacks are spread out in a network and there is the possibility that a node can be attacked by multiple or different types of attackers, which allows the same node to be infected multiple times easily.  Hence, as a future research direction, a more realistic infection process can be considered where an infected node can recover and be reinfected.

\item {\bf In-depth analysis of network resilience under various network conditions is important}:  Due to the space constraint, we have not demonstrated more sensitivity analyses to investigate the effect of using a different centrality metric under various network conditions in terms of network density (i.e., the number of edges), node density (i.e., the number of nodes in a given area), or the variance in the number of degrees (e.g., for a scale-free network or a random graph).  We can take another in-depth analysis of network resilience by using a different centrality metric in order to identify what metric would be more powerful under what network conditions.  In addition, more comprehensive, diverse, larger, and real network topologies can be considered to obtain more meaningful findings to provide generalizable guidelines for selecting useful centrality metrics in a given application.% domain.
%\textcolor{brown}{\item {\bf Context-aware analysis and metrics adoption}: As mentioned in the earlier part, the effect of centrality metrics depends on the network topology or the attack scenario: whether the type of attack is infectious or not, whether the metric is designed for directed networks or undirected networks. Considering this, we may further investigate various types of network topology and its real example. We could even suggest different metrics based on the topological characteristics of given cases, with verification of impacts of suggested metrics. (from BW) }

%\textcolor{red}{\item {\bf Improvement to comprehensive analysis:} One improvement to our study is that instead of analyzing metrics in four networks which are approximately 1000 nodes,various real networks could be utilized to analyze metrics and summarize their performance on multiple dimensions, such as time, SGC, or other metrics for network resilience. This improvement could give us a comprehensive view of metrics' performance without worrying about the impact of the features from a single network (like Terry's worry in subsection A, bullet point \# 10). (from Zelin. I feel like this point is not as strong as the first three points in this subsection.)}

\end{itemize}

\bibliographystyle{ACM-Reference-Format}
\bibliography{ref}

\newpage
\appendix
\begin{appendices}

\section{Centrality Metrics Research in Multidisciplinary Domains}

As the extensions of Section II.B of the main paper, we discuss how different disciplines have studied centrality metrics and applied them to solve critical problems in their domains.

{\bf Mathematics.}  The study of networks has its origins in the analysis of data with certain relations in various disciplines. For mathematics, this dates back to the 1730s with Leonhard Euler's solution to the Seven Bridges of K\"{o}nigsberg problem, which is the foundation of graph theory. Centrality metrics are explored based on graph theory, which has been described as the study of networks~\cite{Koenig90}.  

{\bf Chemistry.} Graph theory has been applied in Chemistry since the 1870s~\cite{Sylvester1878}.  Chemical process plants can be represented by networks in which centrality metrics are used to identify more important units and controllers~\cite{Klein10, Restrepo18}. 

{\bf Anthropology.} Network centrality was first investigated in Anthropology by studying human behaviors in groups~\cite{Bavelas48}.  Many human organization or group-based decision making research communities have studied centrality metrics to measure influence and/or power of a group or organization~\cite{Cohn58}.  In the recent Anthropology research, \citet{collins2014networked} discussed `networked anthropology' by using diverse multimedia and OSN platforms.  In addition, how community centrality affects scholarly activities in social science has been studied in Anthropology~\cite{Allman15}. 

{\bf Physics.} Network centrality metrics have been heavily studied in the area of complex networks/systems by physicists~\cite{Newman10}.  In particular, physicists have been major players in the area of Network Science, which has been studied in multiple disciplines, including all these disciplines discussed above.  {\em Network science} is defined as ``the study of network representations of physical, biological, and social phenomena leading to predictive models of these phenomena''~\cite{NAP11516}. 

{\bf Geography.} Historical geographers were interested in how the centrality of a region (e.g., Moscow) can affect dominance and evolution of the region in which the area can be described based on graph theory~\cite{Pitts65}.  \citet{Agryzkov19} studied urban street networks based on graph theory in order to identify important areas in terms of the influence of topology and geo-referenced data extracted from the network.

{\bf Economics.} \citet{souma03economics} studied business networks to investigate the probability of business networks becoming scale-free and the effect of the merger among banks on the cliquishness of companies or the separation between two companies.  \citet{Mayer2009economics} also investigated how social and economic factors (e.g., economic incentives or socioeconomic background) can introduce the changes in social network structure and its composition which were measured by centrality metrics (e.g., Bonacich centrality).

{\bf Psychology.} Centrality metrics have been used to measure socio-cognitive aspects of human behavior in various contexts.  \citet{kameda97psychology} defined a person's power in a group based on his/her centrality measured by the degree of information the person shares with others. The person's influence based on network centrality has been shown to be critical to forming consensus in the decision making process.  \citet{Lee10-psychology} looked at how a person's centrality in a network position affects consumer influence as well as susceptibility to the influence of others. \citet{Epskamp18} also provided how to measure centrality in psychological networks. 

{\bf Sociology.}  Centrality metrics have been used in Sociology for a long time in order to examine various types of social networks. The Bonacich centrality metric has been studied in~\cite{Bonacich72, Bonacich87} in order to measure status and power in society. Borgatti's centrality metrics have been used to investigate the relationships between a person's centrality and other significant factors~\cite{Borgatti06, Borgatti97, Borgatti95centralityand}.  Metrics measuring social relationships are also developed such as social proximity~\cite{Freeman91} based on betweenness measure and faster betweenness algorithm~\cite{brandes2001faster} in the mathematical sociology domain.

{\bf Biology.} Centrality metrics have been used in Biology in selecting central nodes, such as pathogen-interacting, cancer, ageing, HIV-1 or disease-related or immune-related proteins~\cite{Karabekmez16, Koschutzki08, Mistry17} in gene regulatory networks, protein-interaction networks, and metabolic networks.

{\bf Management.} Centrality metrics have been investigated to identify the key factors to be successful in business management.  The management research has investigated how a founder's centrality affects top management group, the group's culture and vision~\citet{Athanassiou02, Louise08} and how network centrality is critical to increasing financial performance~\cite{Ho14}. 

{\bf Computer Science.}  Centrality metrics have been highly leveraged and investigated for diverse applications in the computer science domain.  For example, centrality metrics are used in mobile social network applications~\cite{Zhou18-msn}, visual reasoning in online social networks~\cite{Correa12}, water network distribution~\cite{Narayanan14-water}, or traffic management for space satellite network~\cite{Zhang18-satellite}.

{\bf Political Science.} Graph centrality measures have been considered in identifying power and/or influence of individuals and/or attracting resources in political networks since the 2010s~\cite{Hafner-Burton10}.  As social media and social network services (SNSs) become more and more popular, the availability of social network data allowed the analysis of political views and/or attitudes with respect to various centrality measures~\cite{Lazer10, Miller15}.  

{\bf Psychiatry.} Network science has been applied in Psychiatry under the name of {\em Network Psychiatry}~\cite{Saxe17} based on computational models to investigate the structure of psychiatric disorders which are treated as complex systems.  \citet{zuo11connectome} considered centrality metrics to measure `functional connectivity' in a brain connectome.  They investigated the relationship between the extent of centrality and certain disease or body conditions/characteristics (e.g., age and sex).  Their findings backed up how the centrality in the brain connectome can be used as the underlying physiological mechanisms to study `neurodegenerative and psychiatric disorders.'  \citet{fried2016} also used centrality metrics to determine the centrality of the Diagnostic and Statistical Manual of Mental Disorders (DSM) symptoms and non-DSM symptoms where a network consists of 28 depression symptoms.  In this work, centrality is used as an indicator of the relationships between different depression symptoms.  

\section{Evolution of Centrality Metrics}

Based on our comprehensive survey on centrality metrics conducted in Sections~III-V of the main paper, we summarized them based on their published years in order to capture the overall evolution of centrality metrics in Table~\ref{tab:chronological-centrality-evolution}. As discussed in the main paper, we observed that the centrality metrics developed in the 1960s or earlier until the 1980s are still commonly used in the research literature under various network domains. However, we can also clearly notice that various types of centrality metrics have been developed since the 2000s and more in the 2010s.

\begin{footnotesize}
\begin{table*}[htbp]
\centering
\caption{Evolution of centrality metrics from the 1960s or earlier to the 2010s.}
\label{tab:chronological-centrality-evolution}
\vspace{-2mm}
\begin{tabular}{|P{1.2cm}|P{1.5cm}|P{1.5cm}|P{1.5cm}|P{1.4cm}|P{2.1cm}|P{2.9cm}|}
\hline
{\bf Centrality metrics} &  {\bf 1960s or earlier} & {\bf 1970s} & {\bf 1980s} & {\bf 1990s} & {\bf 2000s} & {\bf 2010s} \\
\hline
\hline
{\bf Point centrality} 
     & Katz centrality~\cite{Katz53}; 
Farness~\cite{bavelas1950communication}; Betweenness~\cite{Bavelas48,Leavitt51}; Closeness~\cite{Sabidussi66}

  & Degree~\cite{Garfield72, Freeman78}; Betweenness~\cite{Freeman78};  
  
  & Eigenvector centrality~\cite{Bonacich87}; Information centrality~\cite{Stephenson89}
 
 & Flow betweenness~\cite{Freeman91}
; Degree~\cite{wasserman1994social}; Eccentricity~\cite{hage1995eccentricity};  Redundancy~\cite{Burt95, Borgatti97}; Clustering coefficient~\cite{Watts98}; PageRank~\cite{Brin98theanatomy}; Authority and Hub centralities~\cite{Kleinberg99}; 
 
 & Cumulative nomination~\cite{poulin2000nomination}; SALSA~\cite{lempel2000salsa}; Gaussian curvature on planar graphs~\cite{higuchi2001combinatorial}; Load centrality~\cite{Goh01};  Curvature~\cite{Eckmann02}; Degree~\cite{Brandes03, Hanneman05}; H-index~\cite{Hirsch05hindex}; Eigenvector centrality~\cite{Hanneman05}; Subgraph centrality~\cite{Estrada05}; Communicability~\cite{Estrada06, Estrada08}; Random-walk betweenness~\cite{newman2005measure}; Current-flow betweenness and closeness~\cite{brandes2005centrality}; Residual closeness~\cite{dangalchev2006residual}; Spatial centrality~\cite{crucitti2006centrality}; Ricci culvature~\cite{ollivier2007ricci}; 
 
 & Generalized degree and shortest paths~\cite{opsahl2010node}; Decay centrality~\cite{jackson2010social}; $L$-betweenness~\cite{ercsey2010centrality}; Degree~\cite{Newman10}; Routing betweenness~\cite{Dolev10}; $k$-shell index or coreness~\cite{kitsak2010identification}; Leader Rank~\cite{Lu11}; Semi-local centrality~\cite{chen2012identifying}; Dynamical influence~\cite{Klemm12};  Volume~\cite{Kim12-volume, wehmuth2012distributed}; Gaussian curvature on planar graphs~\cite{keller2011curvature, wu2015emergent}; Cluster Rank~\cite{chen13cluster}; Diffusion centrality~\cite{Banerjee13}; Mixed degree decomposition~\cite{zeng2013ranking}; Percolation centrality~\cite{Piraveenan13}; Improved method~\cite{liu2013ranking}; Ricci culvature~\cite{jost2014ollivier}; Neighborhood coreness~\cite{bae2014identifying, kitsak2010identification}; Contribution centrality~\cite{Alvarez-Socorro15}; Mapping entropy~\cite{nie2016using}; Hybrid degree~\cite{ma2017identifying}; Weight neighborhood centrality~\cite{wang2017novel}; AHP-based centrality~\cite{bian2017identifying}
 \\
\hline
{\bf Graph centrality} & Distance-based GC (e.g., dispersion~\cite{Shimbel53}) & $k$-clique~\cite{Tichy73}; Degree-based GC~\cite{Nieminen74}; Betweenness-based GC~\cite{Freeman78}; Closeness-based GC~\cite{Freeman78}; $k$-clique~\cite{Seidman78-plex}; $k$-plex~\cite{Seidman78-plex}; $k$-core~\cite{Seidman78-plex}
; Distance-based GC (e.g., compactness~\cite{Freeman78}) & & Flow betweenness-based GC~\cite{Freeman91}; Global clustering coefficient~\cite{Watts98} & Reciprocity~\cite{Newman02-virus}; Degree assortativity~\cite{Newman02-assortativity}  & Local Assortativity~\cite{Piraveenan10-local-assortativity}; $k$-component~\cite{Newman10}; Graph curvature~\cite{narayan2011large};
 \\
\hline
{\bf Group Selection centrality} 
       & & & & & SingleDistance~\cite{chen09}; DegreeDiscount~\cite{chen09} & DegreeDistance~\cite{sheikhahmadi15}; collective influence~\cite{morone2015influence} DegreePunishment~\cite{wang2016effective} \\
\hline
\end{tabular}
\end{table*}
\end{footnotesize}

\begin{footnotesize}
\begin{table*}[t]
\centering
\caption{Applications of centrality metrics}\label{tab:app-metrics}
\vspace{-3mm}
\begin{tabular}{|P{1.8cm}|P{3.2cm}|P{5cm}|P{2cm}|}
\hline
{\bf Network Type} & {\bf Research Problem} & {\bf Centrality metrics used} & {\bf Ref. No.} \\
\hline
\multirow{5}{*}{Social Networks} & Information diffusion & In-degree; out-degree; clustering-coefficient; temporal centrality; betweenness; closeness; proximity & \cite{Cho19-tsc, Kim12-volume, Kim15-neighbor-influence, Kimura07-info-diffusion, kandhway2016, Kimura10, kumar2016disinformation, ratkiewicz2011truthy, Wu16-bc, tang2010information} \\
\cline{2-4}
 & Influence maximization & Coreness; random-walk betweenness; in-degree & \cite{bae2014identifying, bavelas1950communication, bian2017identifying, chen2012identifying, Goyal10, Mochalova2013, newman2005measure, padgett1993robust, Watts07-influentials, Saito13}  \\
\cline{2-4}
 & Influence minimization & Betweenness; out-degree; degree; closeness  &  \cite{Dey17-inf-min, Kimura09, Luo14, Qipeng15-www} \\
\cline{2-4}
 & Behavior adoption for marketing & Degree; betweenness; closeness &  \cite{Czepiel74, Dinh14-marketing, Kempe03, Salavati19-marketing, Shao13-marketing, Yan14}  \\
\cline{2-4}
 & Community detection & Entropy centrality; $K$-rank & \cite{jiang2013detection, nikolaev15detection} \\
\hline
Contact Networks & Identification of high-risk hosts or super spreaders & Degree; random-walk betweenness; betweenness; shortest-path betweenness; farness; closeness, distance-based centrality; eigenvector centrality; information centrality; power prestige; strength & \cite{bell1999, christley2005infection, Dekker13, gomez2013diseases} \\
\hline
Communication Networks & Selecting critical nodes to prevent or mitigate computer virus or malware spreads; modeling targeted attackers & In-degree; out-degree; degree; betweenness; eigenvector centrality; closeness centrality &  \cite{albert2000error, holme2002attack, Kim19, Newman02-virus, Yoon17} \\
\hline
Geographic Networks & Characterizing the geographic properties of cities as networks & Betweenness; closeness; degree; information centrality & \cite{crucitti2006centrality, Gao13, Guimera05, Li15, Porta08} \\
\hline
Biological Networks & Removing critical proteins; identifying central nodes such as pathogen-interacting, cancer, aging, HIV-1 or disease related protein & Degree; betweenness; integration; radiality; Katz status index; PageRank; motif-based centralities; weighted sum of loads eigenvector centrality; subgraph centrality; eigenvector centrality & \cite{Estrada05, jeong2001lethality, Koschutzki08, Mistry17} \\
\hline
\end{tabular}
\end{table*}
\end{footnotesize}

\section{Applications of Centrality Metrics}

In~Table~\ref{tab:app-metrics}, we summarize what centrality metrics have been used in various network types based on what we discussed in this work. The details of each work summarized in this table were discussed in Section~VI of the main paper.

\section{Network Resilience Analysis of the Surveyed Centrality Metrics} \label{sec:analysis-centrality}
%\begin{comment} 

\begin{footnotesize}
\begin{table}[th!]
\centering
\caption{Characteristics of the used datasets}
\label{tab:datasets}
\vspace{-2mm}
\begin{tabular}{|P{2cm}|P{2.5cm}|P{2cm}|P{2cm}|P{2cm}|}
\hline
Network characteristics &  UCI Social Network~\cite{panzarasa2009patterns-dataset} & Rocketfuel Network~\cite{router-dataset} & URV Email Network~\cite{Rossi15-dataset}  & EU Email Network~\cite{Paranjape2017-dataset}   \\
\hline
\hline
Network type & Directed & Directed & Undirected & Undirected \\
\hline
\# of nodes & 1893 & 2113 & 1133 & 930\\
\hline
\# of edges & 59835 & 6632 & 5451 & 24929 \\
\hline
Average degree & $\sim$ 63 (in+out) & $\sim$ 6 (in+out) & $\sim$ 10 & $\sim$ 27 \\
\hline
Max degree & 558 (in), 1091 (out) & 79 (in), 85 (out) & 71 & 319 \\
\hline
\end{tabular}
\end{table}
\end{footnotesize}

\subsection{Experimental Setup}
This section explains the experimental setup used for evaluating the performance of each centrality metric surveyed in this work in terms of the size of the giant component as the indicator of network resilience.  To be specific, we provide datasets, metrics, and attack scenarios used for evaluating the surveyed centrality metrics in this work.

\subsubsection{\bf Datasets} \label{subsec:datasets}

We selected the following real datasets for network topologies used in the performance demonstration of the surveyed centrality metrics:
\begin{itemize}
\item {\em Directed Network Topologies}: (1) The UCI Social Network~\cite{panzarasa2009patterns-dataset} is a collection of interactions from private messages sent over an online social network at The University of California, Irvine. (2) The Rocketfuel Network~\cite{router-dataset} is a snapshot of router connections on an Internet Service Provider (ISP) topology from measurements.

%This dataset~\cite{panzarasa2009patterns-dataset} refers to a network dataset where the private messages are sent on an online social network at The University of California, Irvine. Edge edge $(u, v, t)$ means user $u$ send $v$ a message at time $t$; and (2) This dataset~\cite{router-dataset} dataset is a publicly released dataset of ISP Network Maps inferred from the measurements by \citet{router-dataset}. The dataset maps the connectivity of the routers in the ISP network. 

\item {\em Undirected Network Topologies}: (1) The URV Email Network~\cite{Rossi15-dataset} captures the email communication for the Universitat %NOTE UNIVERSITAT IS NOT A TYPO!!!!!!!!!!!!!!!!!!!
Rovira i Virgili in Spain.
(2) The EU Email Network~\cite{Paranjape2017-dataset} captures the internal (or core) email communication for a large European research institution.

%This dataset~\cite{Rossi15-dataset} is an university email communication network for the Rovira i Virgili University in Spain; and (2) The email dataset~\cite{Paranjape2017-dataset} includes the email network from a large European research institution where an edge represents a communication between institution members (the core). The dataset does not contain incoming messages from or outgoing messages to the rest of the world, which represents a directed network.
\end{itemize}

\begin{figure*}[th]
    \centering
\subfigure[UCI Social Network with 1,893 nodes and 59,835 directed edges]{
\includegraphics[width=0.46\textwidth, height=0.3\textwidth]{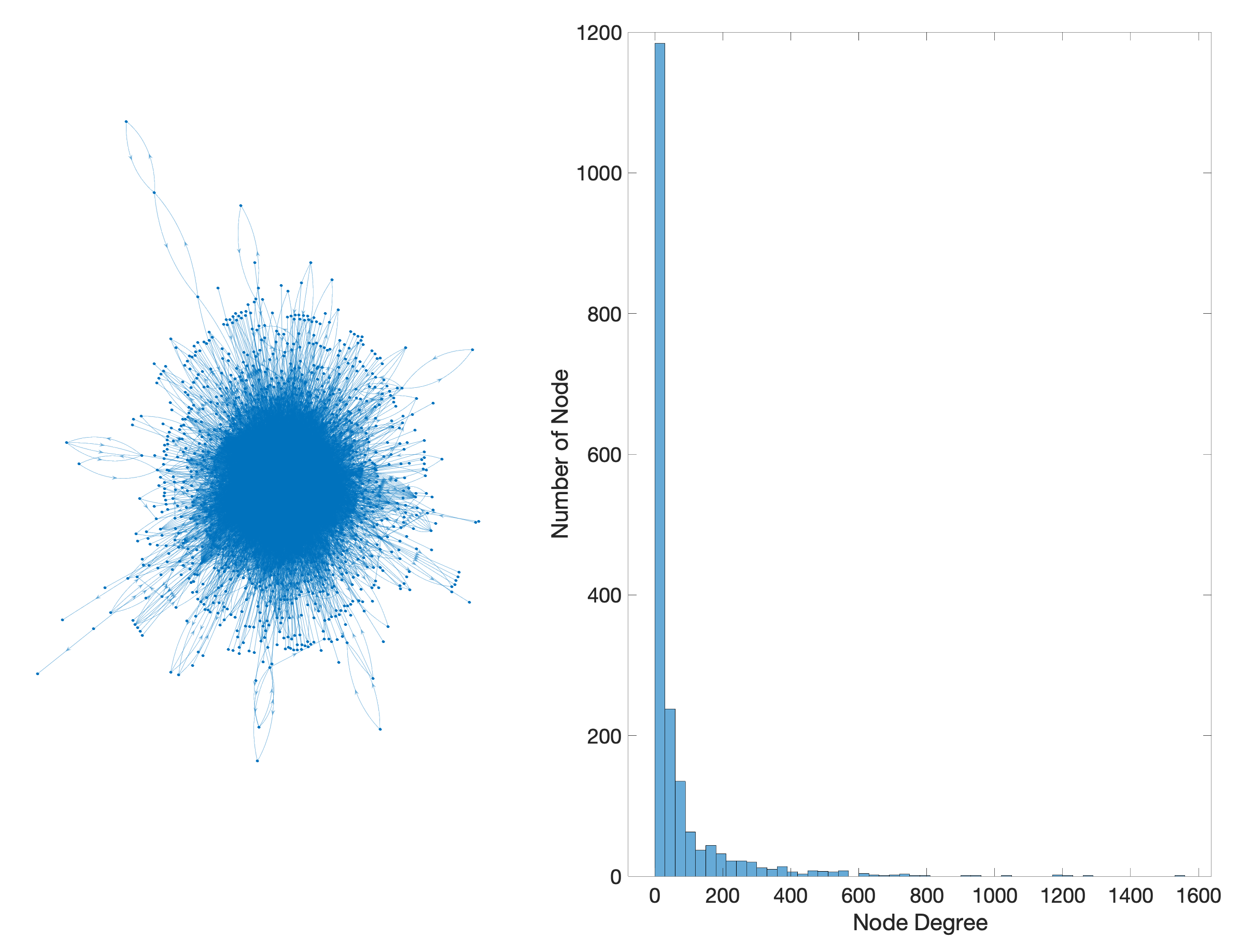}}
\hspace{5mm}
\subfigure[Rocketfuel Network with 2,113 nodes and 6,632 directed edges]{
\includegraphics[width=0.46\textwidth, height=0.3\textwidth]{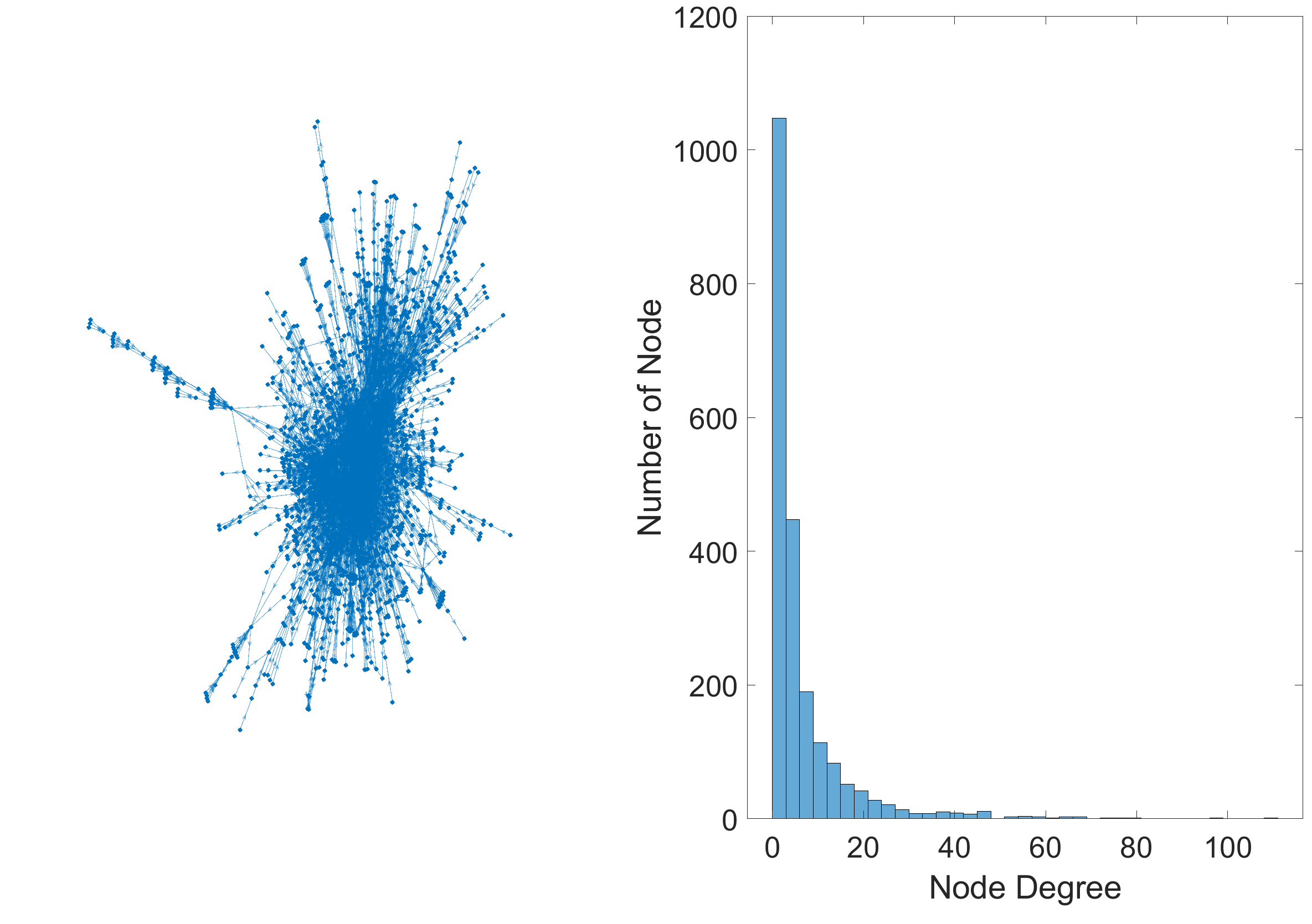}}
\subfigure[URV Email Network with 1133 nodes and 5451 undirected edges]{
\includegraphics[width=0.46\textwidth, height=0.3\textwidth]{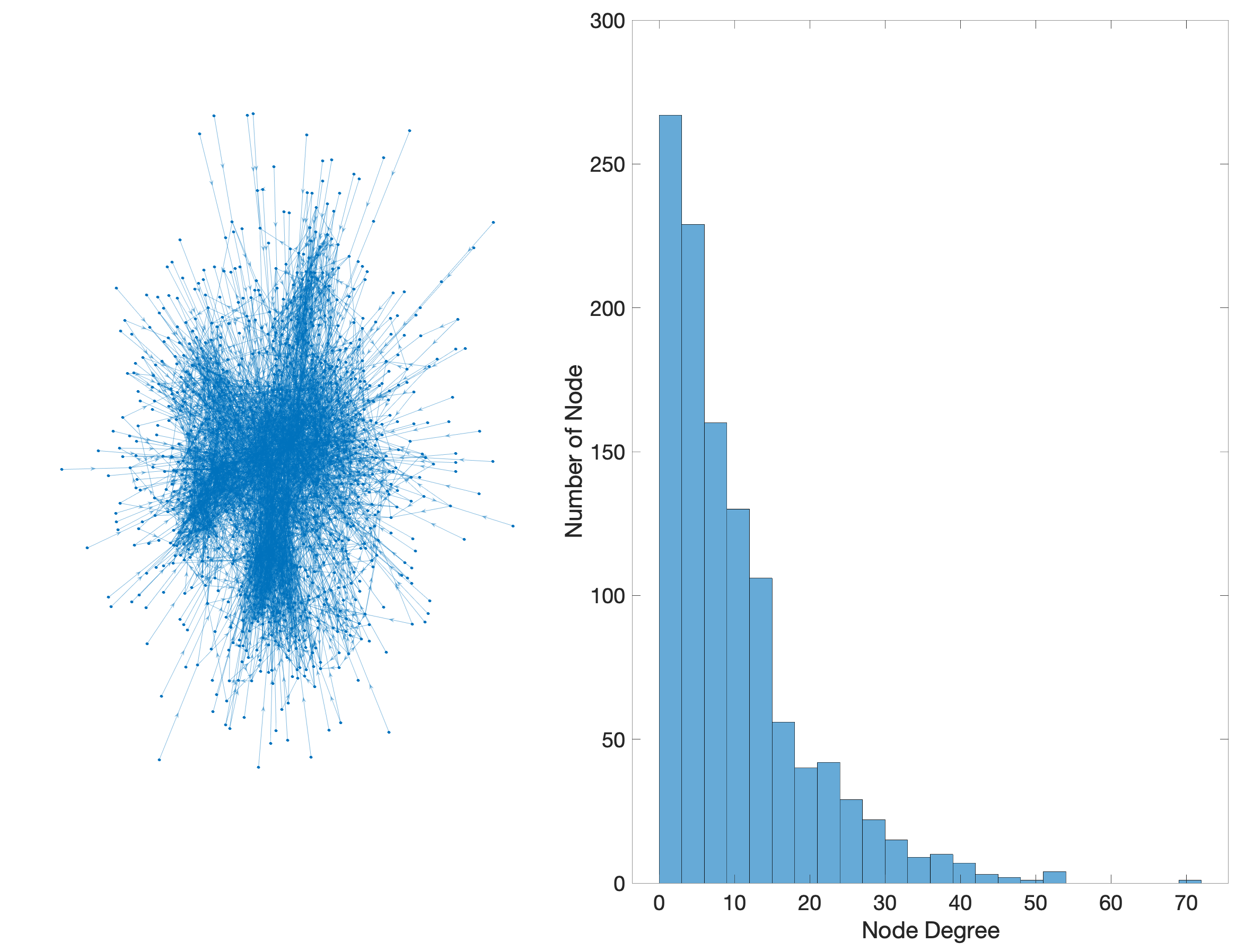}}
\hspace{5mm}
\subfigure[EU Email Network with 930 nodes and 24,929 undirected edges]{
\includegraphics[width=0.46\textwidth, height=0.3\textwidth]{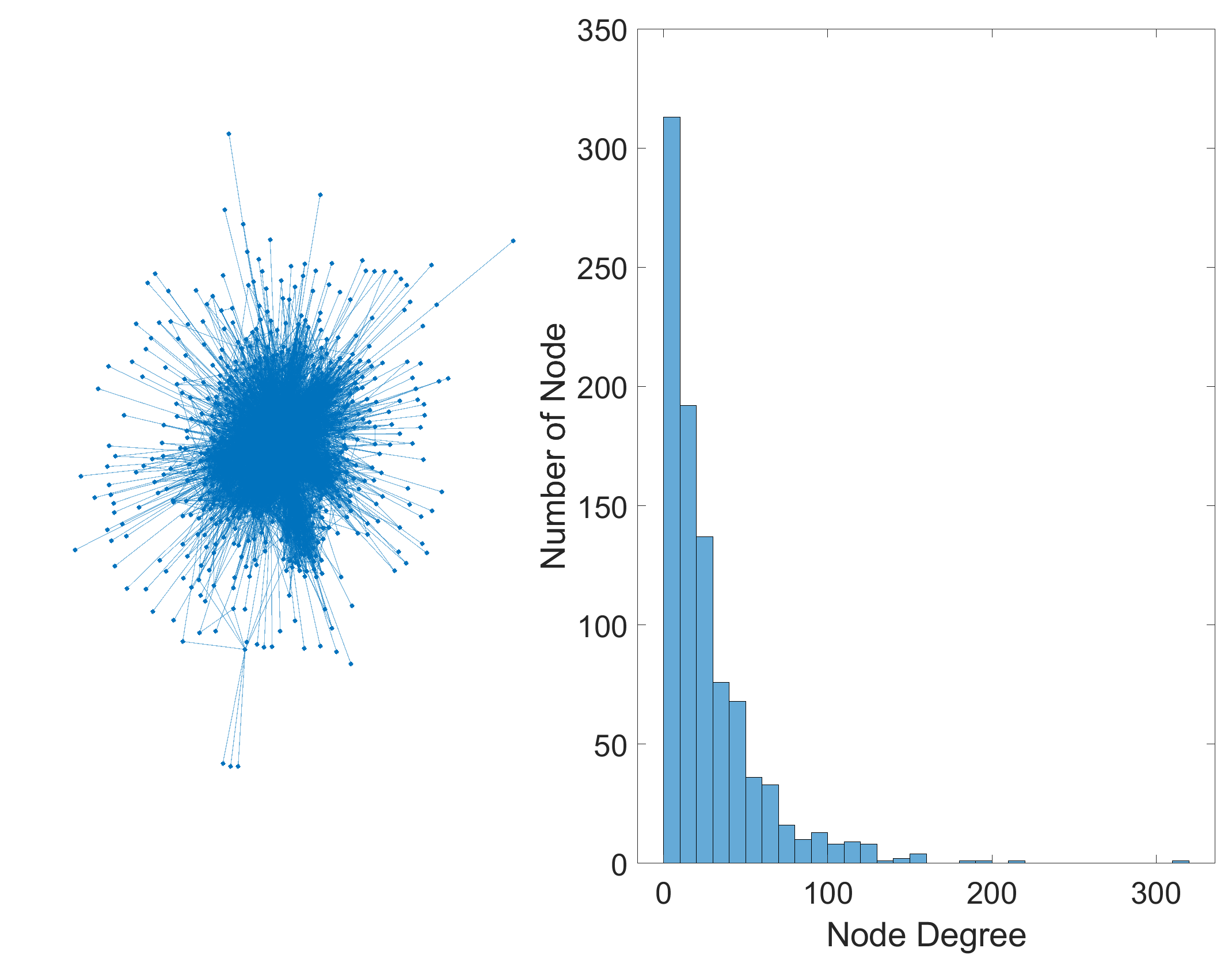}}
    \caption{Network Topologies and Degree Distributions for the Datasets Used.}
    \label{fig:datasets}
\end{figure*}

In Fig.~\ref{fig:datasets}, we described the topologies and degree distributions of all four datasets used in this work.

\subsubsection{\bf Metrics} \label{subsec:metrics}
We use the following metrics to evaluate centrality metrics discussed in this work:
\begin{itemize}
\item {\em Size of the giant component}:  This metric measures the fraction of nodes in the giant component.  This metric is commonly used as an indicator of network resilience in the Network Science~\cite{barabasi16-book}.
\item {\em Mean fraction of infected nodes}: This metric measures the mean number of infected nodes by an initial attacker.
\item {\em Running time}: This measures the simulation time in seconds to calculate the centrality metrics in the given datasets.
\end{itemize}

\subsubsection{\bf Attack Scenarios} \label{subsec:attack-scenarios}
We consider the two attack types as:
\begin{itemize}
\item {\em Non-infectious attacks}:  This attack type reflects node failures without infecting the node's neighbors. The practical examples include partial physical destruction of a system~\cite{Alam14}, non-critical nodes that are not functioning due to denial-of-service (DoS) attacks~\cite{Mavoungou16}, or a node accessed by a unauthorized party aiming to illegally obtain credentials~\cite{Mavoungou16}. The fraction of removed nodes, $\phi$, is the same as the number of attackers without propagating infections.

\item {\em Infectious attacks}: Unlike the above non-infectious attack, this attack propagates infections towards other nodes. The common examples are malware or virus spreads.  Botnets can propagate malwares or viruses through mobile devices, which can use mobile malware such as a Trojan horse, which acts as a botclient to obtain a command and control from a remote server~\cite{Mavoungou16}.  We model this infectious attacks by selecting the initial attackers with $\phi$, a fraction of nodes being selected as initial seeding attackers.  We assume that the infectious attackers follow the Susceptible-Infected-Removed (SIR) epidemic model~\cite{Newman10}.  Nodes in the susceptible state (S) refer to healthy nodes, not being infected by the attackers yet. Nodes in the infected state (I) are the compromised nodes, becoming an inside attacker, which can also replicate infections to their neighboring nodes.  Nodes in the removed state (R) are the nodes detected and isolated from the network by cutting all edges of the detected node. The compromised and detected nodes are treated as failed nodes.  A susceptible node (S) can become infected (I) and later recover or be removed (R). When the size of the giant component is captured, we only consider healthy nodes, which are still in the $S$ state.  We consider the probability that a node is infected as the infection rate, $\beta$.

\end{itemize}

\subsubsection{\bf Centrality Metrics Tested and Parameter Settings}  
For the volume and flow betweenness centrality metrics, we used the number of hops ($h$) set to 2.  In the group selection metrics, we used $d_{td}=4$ in the degree distance metric and each group is defined with 10 nodes.  Due to the high complexity of some metric computations (i.e., too slow even for one simulation run), we excluded the following point centrality metrics: random-walk betweenness, routing betweenness, dynamical influence, load centrality, and curvature.  In the point centrality metrics, we didn't show communicability centrality as it is the same as subgraph centrality when it is used to measure node centrality.  In the graph centrality metrics, since reciprocity was the only metric that can be measured in a directed network, we excluded it. 

%\jhc{list what metrics are not implemented}
%\begin{table}[h]
 %   \centering
  %  \caption{Metrics Unimplemented}
   % \label{tab:metrics-implemented}
    %\begin{tabular}{|P{2cm}|P{4cm}|}
    %\hline
     %    &  Unimplemented Metrics\\
    %\hline
    %Point centrality metrics & random-walk betweenness, Routing betweenness, Dynamical influence, Load centrality, Curvature \\
    %\hline
    %Graph centrality metrics &   Reciprocity \\
    %\hline
    %Group selection metrics & \textcolor{red}{None}  \\
    %\hline
    %\end{tabular}
%\end{table} 

\subsection{Network Resilience Analysis of Point Centrality Metrics} \label{subsec:exp-point-centrality}

\subsubsection{\bf Under Non-Infectious Attacks}

%\begin{comment}
\begin{figure*}
\centering
\subfigure[Noninfectious attacks with degree, closeness, betweenness, pagerank, eigenvector, local entropy and mapping entropy]{
\includegraphics[width=0.4\textwidth]{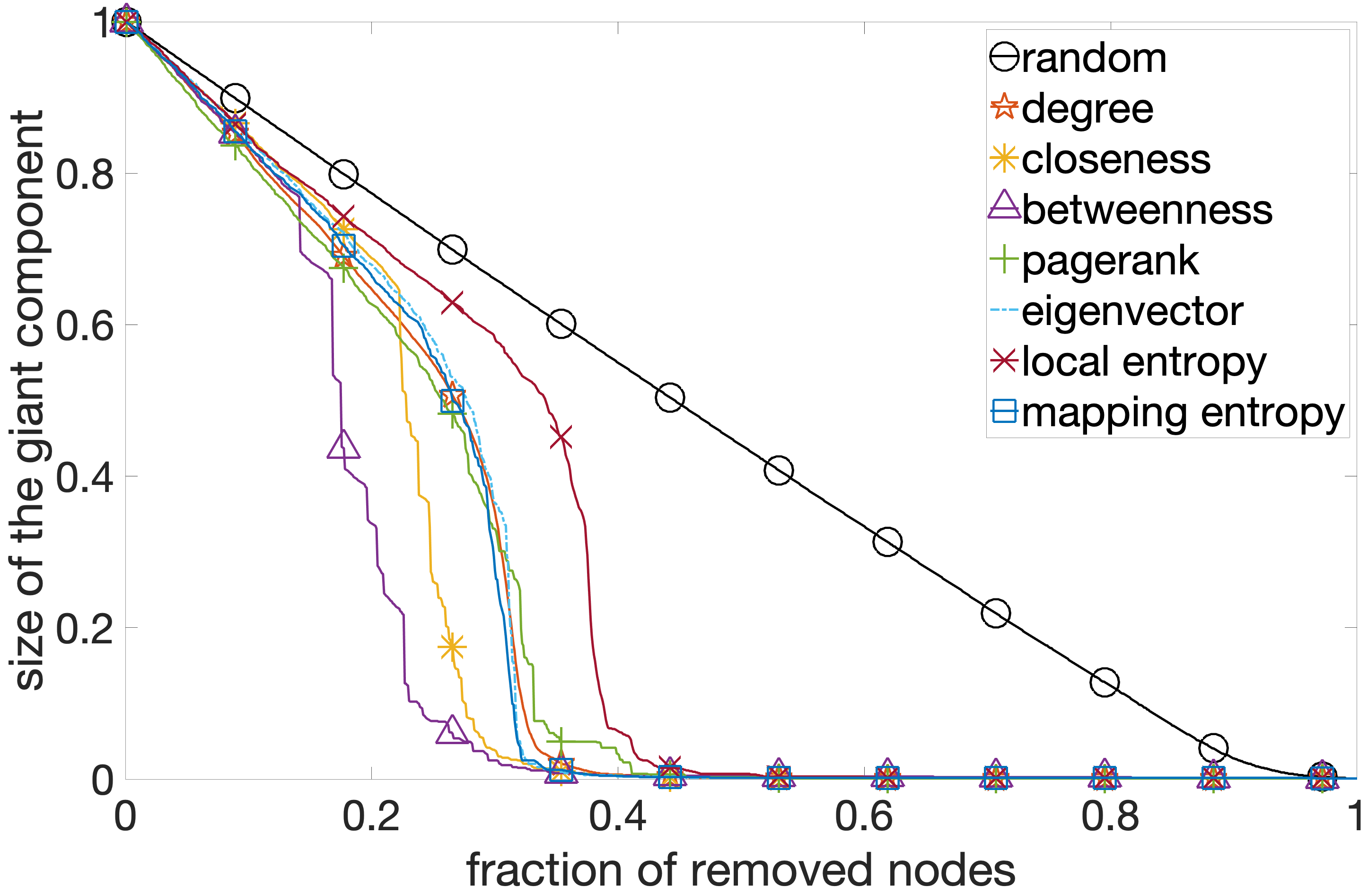}} \hspace{5mm}
\subfigure[Noninfectious attacks with local betweenness, volume, redundancy, kshell, improved kshell, percolation and hybrid degree]{
\includegraphics[width=0.4\textwidth]{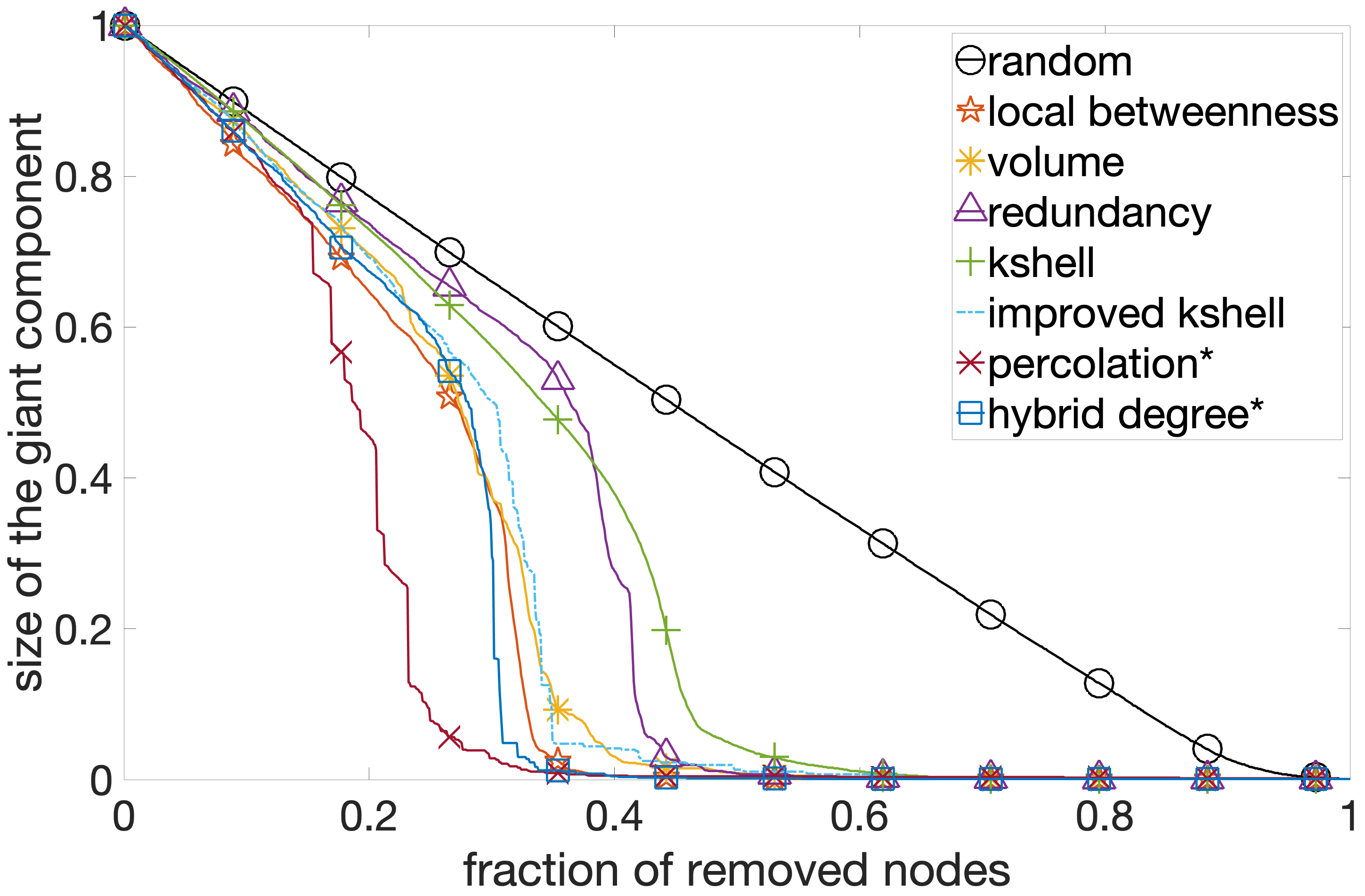}} \hspace{5mm}
\subfigure[Noninfectious attacks with neighborhood coreness, flow betweenness, katz, diffusion centrality, subgraph and clustering coefficient]{
\includegraphics[width=0.4\textwidth]{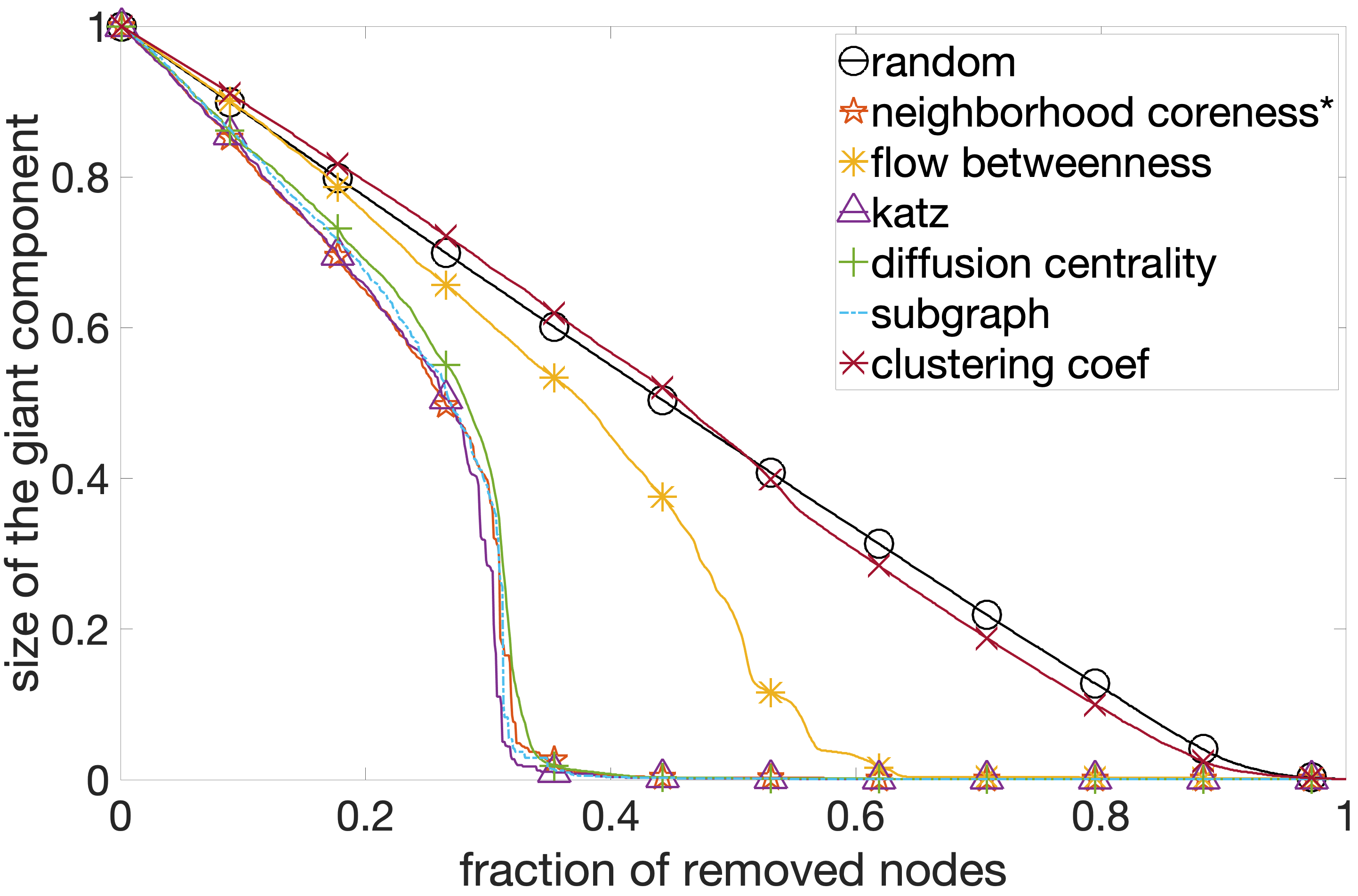}} \hspace{5mm}
\subfigure[Noninfectious attacks with information centrality, residual closeness, semi local, mixed degree decomposition, dynamic influence and weight neighborhood]{
\includegraphics[width=0.4\textwidth]{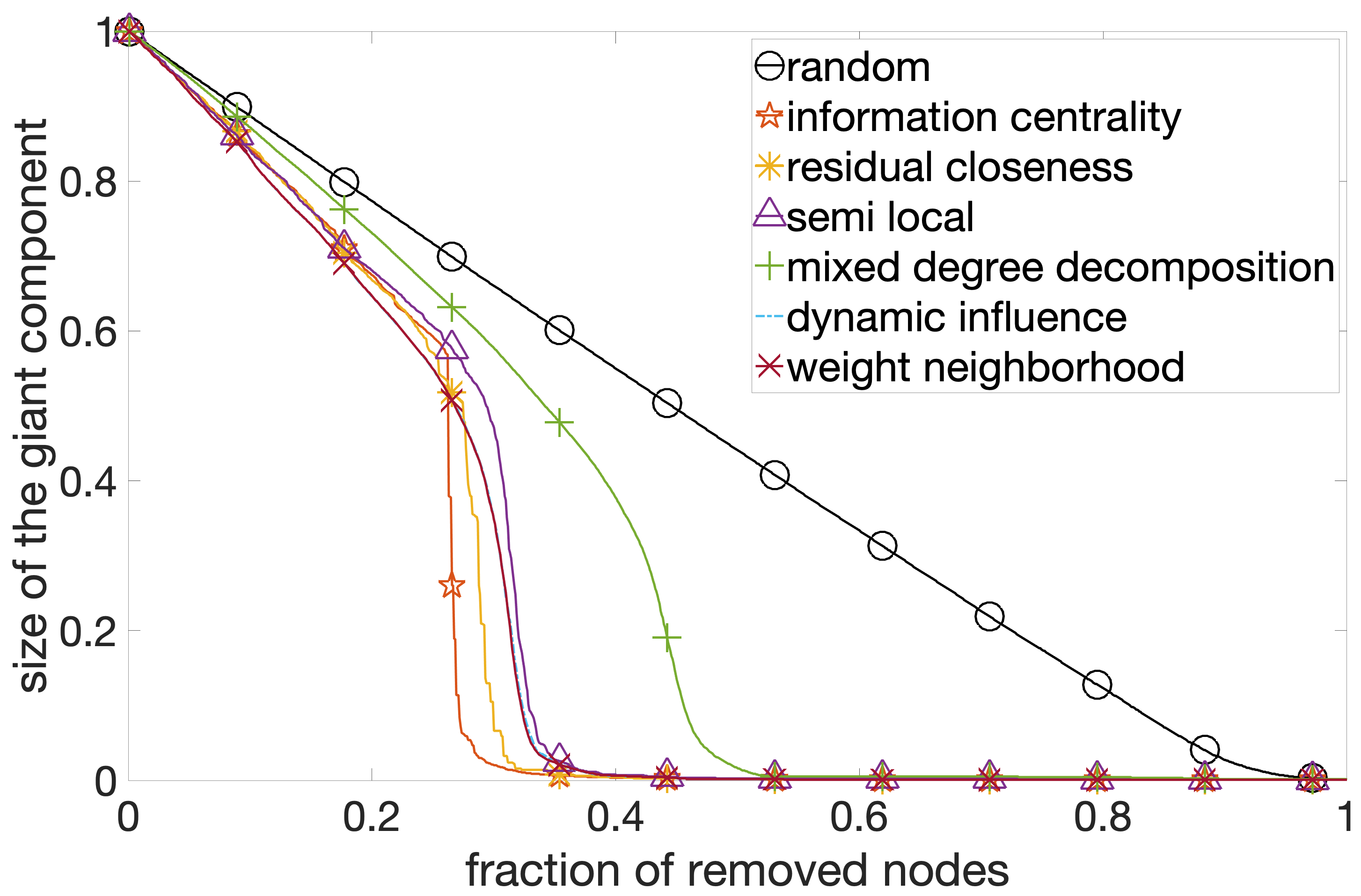}} \hspace{5mm}
\subfigure[Noninfectious attacks with GDSP degree, GDSP closeness, GDSP betweenness, eccentricity, cummulative nomination, h index and contribution]{
\includegraphics[width=0.4\textwidth]{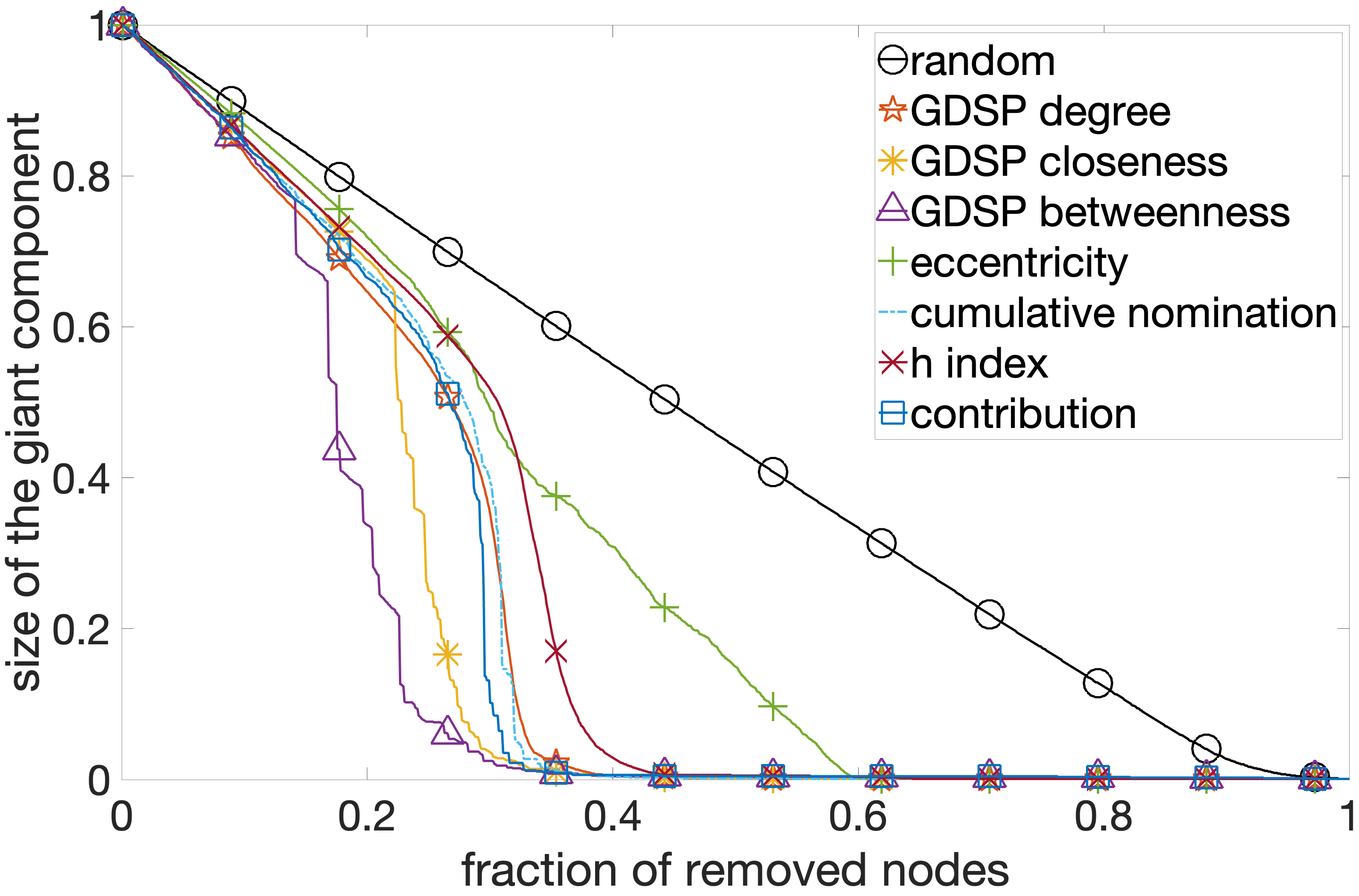}} \hspace{5mm}
\subfigure[Noninfectious attacks with hubs, authorities, clusterrank, SALSA authorities, SALSA hubs and leaderrank in the directed UCI Social Network]{
\includegraphics[width = 0.4 \textwidth]{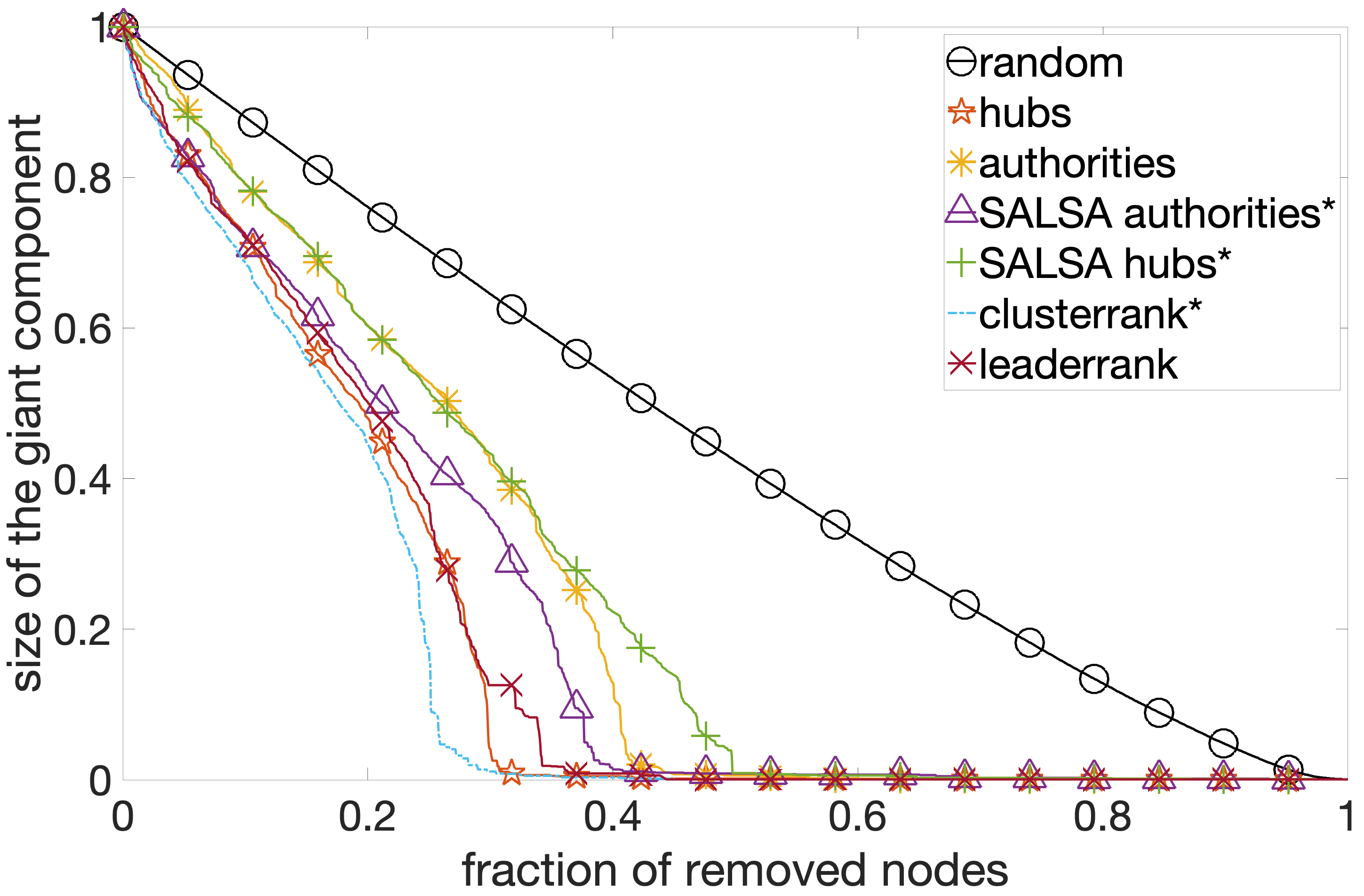}}

\caption{The size of the giant component after removing the initial non-infectious attacker nodes based on the surveyed centrality metrics (39 point centrality metrics tested) in the undirected URV Email Network for (a)-(e) and the (directed) UCI Social Network for (f) where the random node removal is included as a baseline model. The star notation(*) in the legend indicates the result was obtained with only a single simulation run due to too high running time. Otherwise, 100 simulation runs are used to obtain the mean size of the giant component.}
\label{fig:ia-email-point}
\end{figure*}
%\end{comment}

\begin{figure*}
\centering
\subfigure[Noninfectious attacks with degree, closeness, betweenness, pagerank, eigenvector, local entropy and mapping entropy]{
\includegraphics[width=0.4\textwidth]{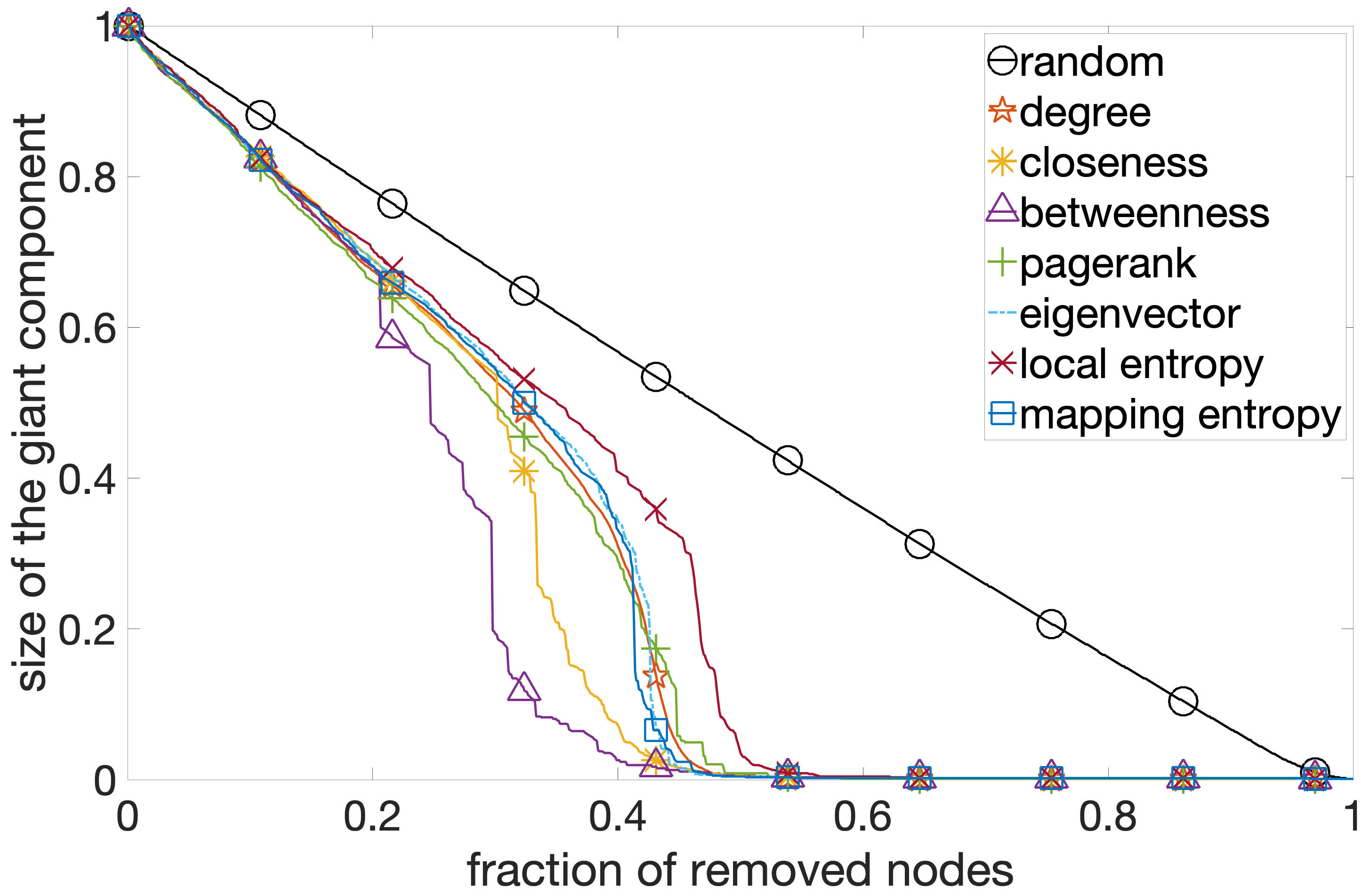}} \hspace{5mm}
\subfigure[Noninfectious attacks with local betweenness, volume, redundancy, kshell, improved kshell, percolation and hybrid degree]{
\includegraphics[width=0.4\textwidth]{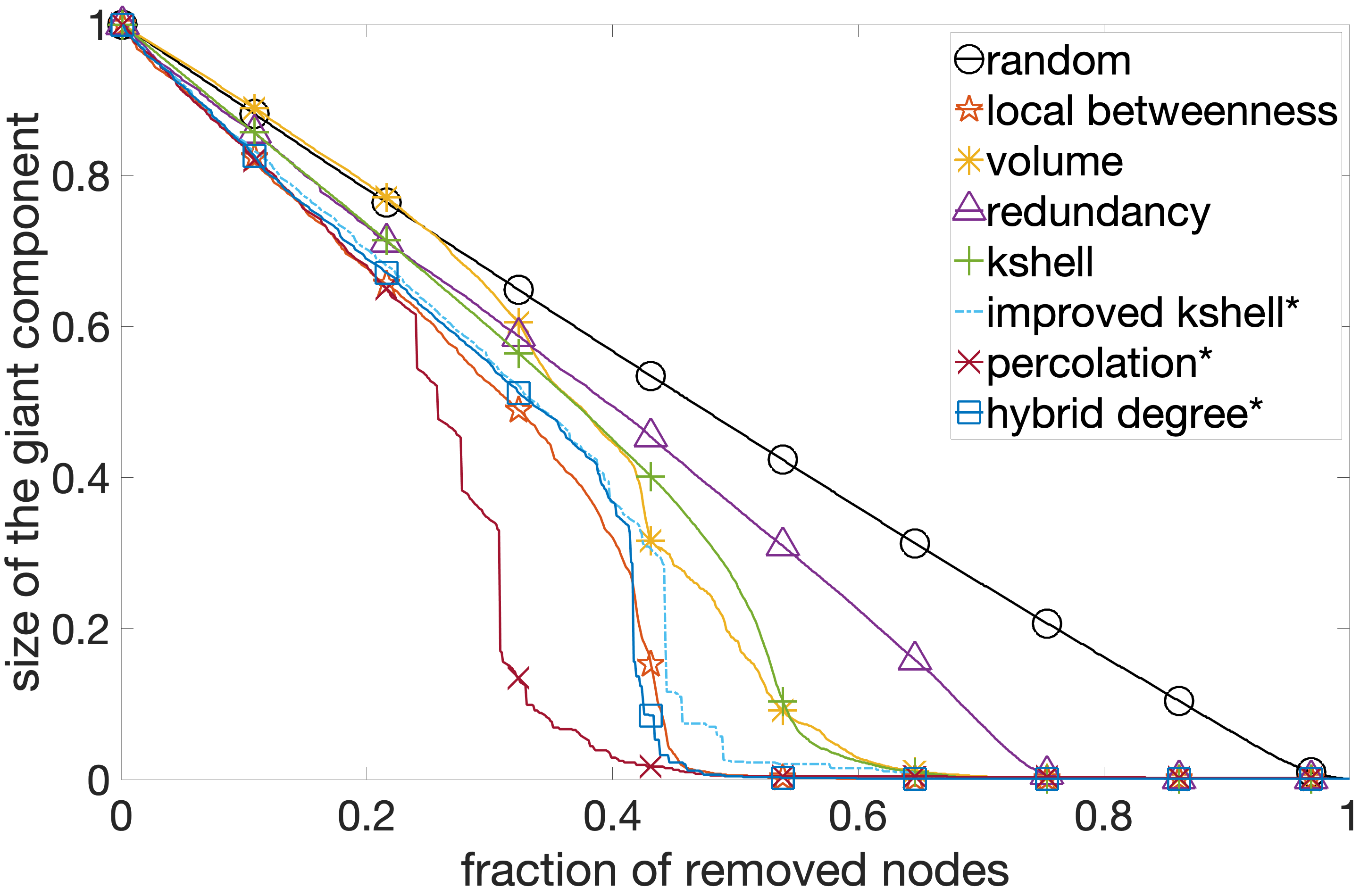}} \hspace{5mm}
\subfigure[Noninfectious attacks with neighborhood coreness, flow betweenness, katz, diffusion centrality subgraph and clustering coefficient]{
\includegraphics[width=0.4\textwidth]{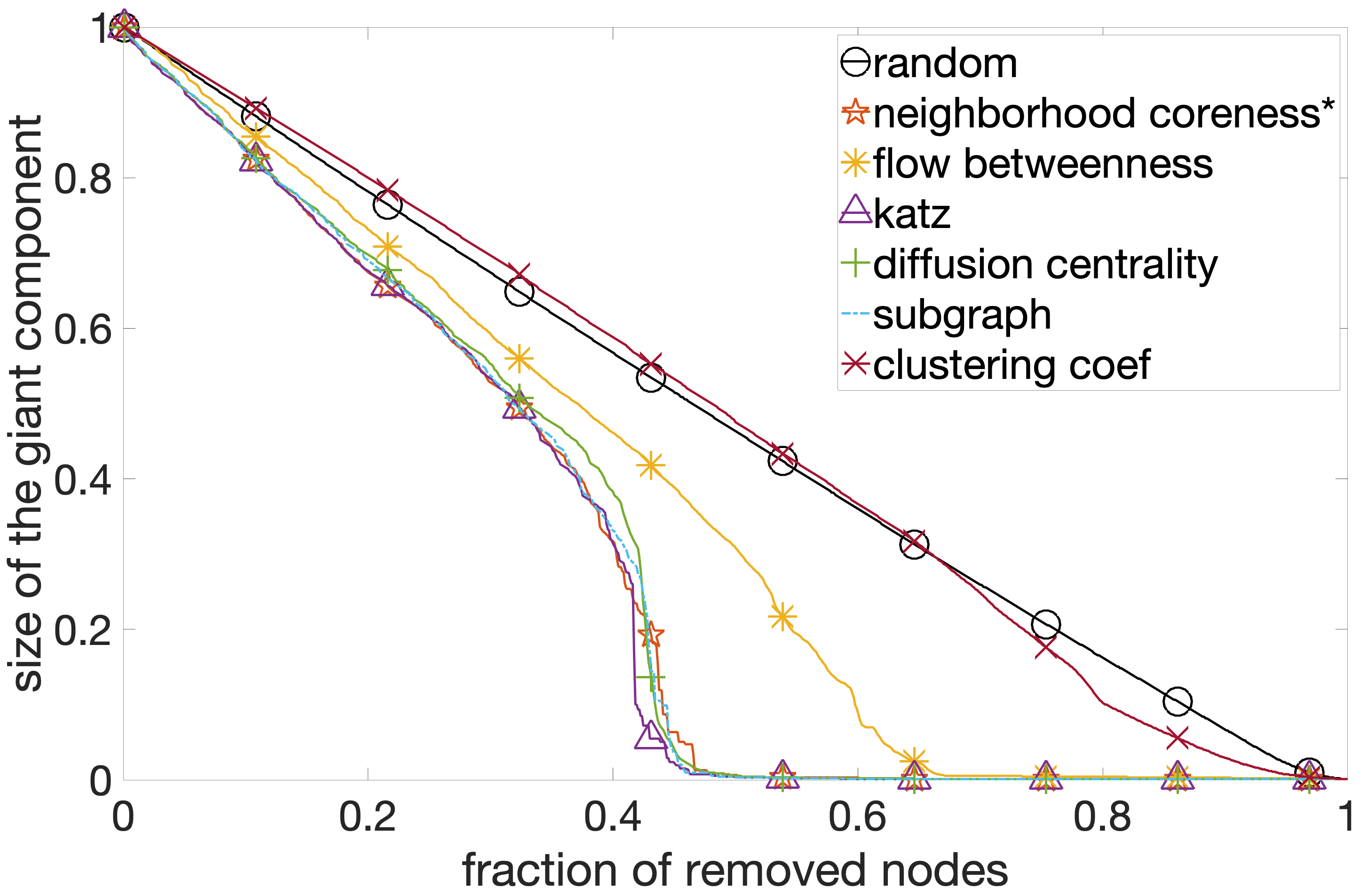}} \hspace{5mm}
\subfigure[Noninfectious attacks with information centrality, residual closeness, semi local, mixed degree decomposition, dynamic influence and weight neighborhood]{
\includegraphics[width=0.4\textwidth]{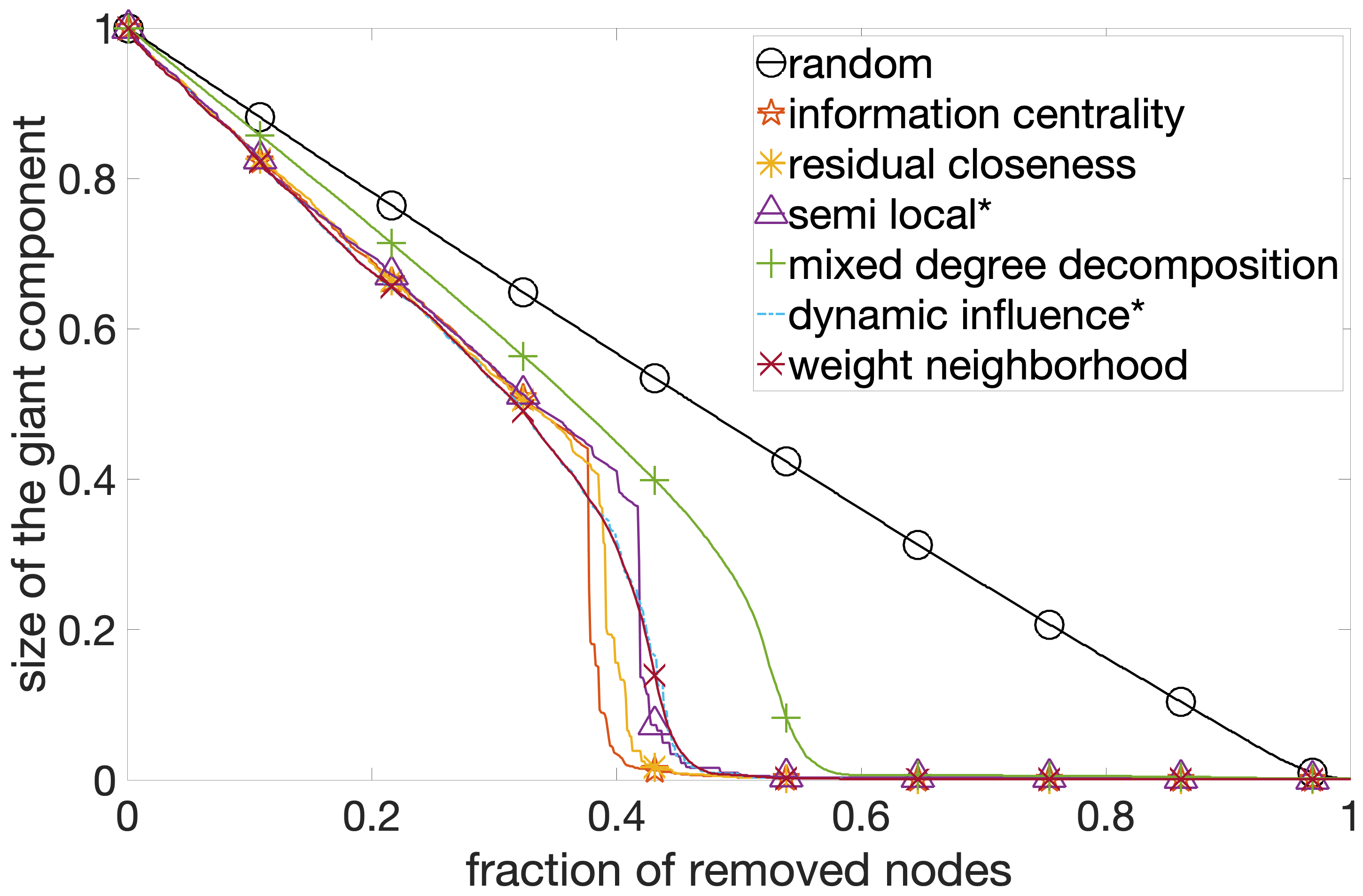}} \hspace{5mm}
\subfigure[Noninfectious attacks with GDSP degree, GDSP closeness, GDSP betweenness, eccentricity, cumulative nomination, h index and contribution]{
\includegraphics[width=0.4\textwidth]{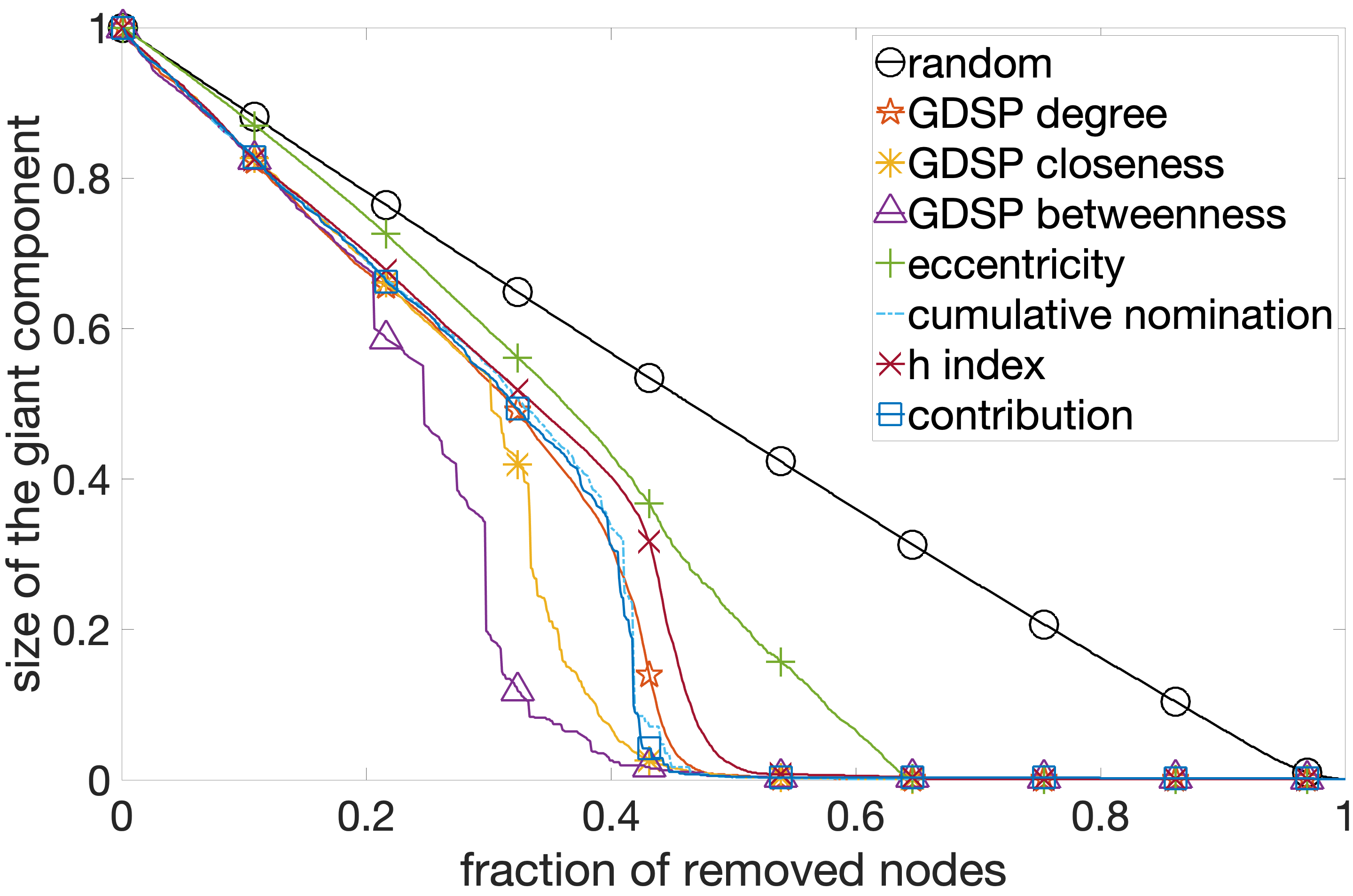}} \hspace{5mm}
\subfigure[Noninfectious attacks with hubs, authorities, clusterrank, SALSA authorities, SALSA hubs and leaderrank in the directed Rocketfuel Network]{
\includegraphics[width=0.4\textwidth]{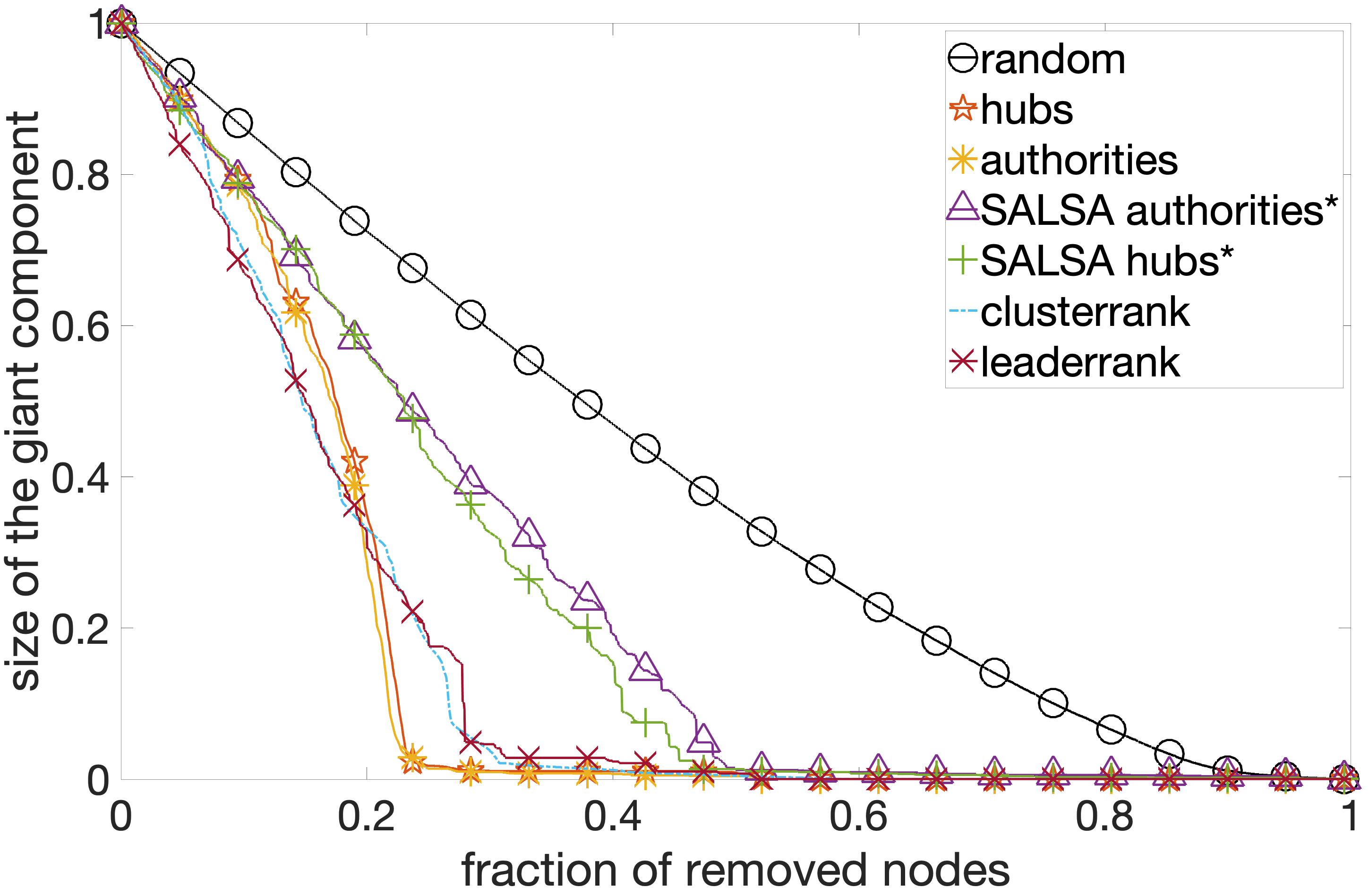}}
\caption{The size of the giant component after removing the initial non-infectious attacker nodes based on the surveyed centrality metrics (39 point centrality metrics tested) in the undirected EU Email Network for (a)-(e) and the directed Rocketfuel Network for (f) where the random node removal is added as a baseline model. The star notation(*) in legend means the result only with a single simulation run due to too high running time. For others without *, 100 simulation runs are used to obtain the shown mean size of the giant component.}
\label{fig:EU-undirected-non-infectious}
\end{figure*}

Fig.~\ref{fig:ia-email-point} shows the size of the giant component in the URV Email Network and UCI Social Network when varying the fraction of removed nodes (i.e., attacked nodes) selected via different point centrality metrics.  Hence, this models a targeted attack based on the given point centrality metric where the attack is not infectious.  From the observation of Fig.~\ref{fig:ia-email-point} (a) -- (f), we found the following: (i) Most targeted attacks are stronger attacks than random attacks (notated as `random' in black), showing a significantly lower size of the giant component; (ii) Betweenness in (a) and GDSP betweenness in (e) show the best performance (i.e., in the sense of reducing the size of the giant component) with the network dissolved after a little more than $4 \backslash 10$ths of the nodes are removed; and (iii) Although most targeted attacks with given point centrality metrics outperform a random attack, the attack with clustering coefficient in (c) performs close to the random attack without showing a higher impact in disconnecting a given network.  We can conjecture the reasons as follows: Since the clustering coefficient measures the number of triangle relationships among a node's adjacent nodes, removing a node with high clustering coefficient still allows neighboring nodes to remain connected. 
%The worst impact is disrupting locally connected networks, not globally connected networks. 
The impact of removing a node is lessened if the selection criteria (or centrality) has a more local, rather than a global, scope.
Therefore, removing a node with high clustering coefficient does not introduce a dramatic effect in reducing the size of the giant component.  In Fig.~\ref{fig:EU-undirected-non-infectious}, we also conducted the same experiment under different network topologies, under the undirected EU Email Network and the directed Rocketfuel Network.  The general trends observed from the results shown in Fig.~\ref{fig:EU-undirected-non-infectious} are highly similar to the results in Fig.~\ref{fig:ia-email-point}.  The key observations are already discussed above while discussing Fig.~\ref{fig:ia-email-point}.

\begin{figure*}
\centering
\includegraphics[width =\textwidth, height=0.6\textwidth]{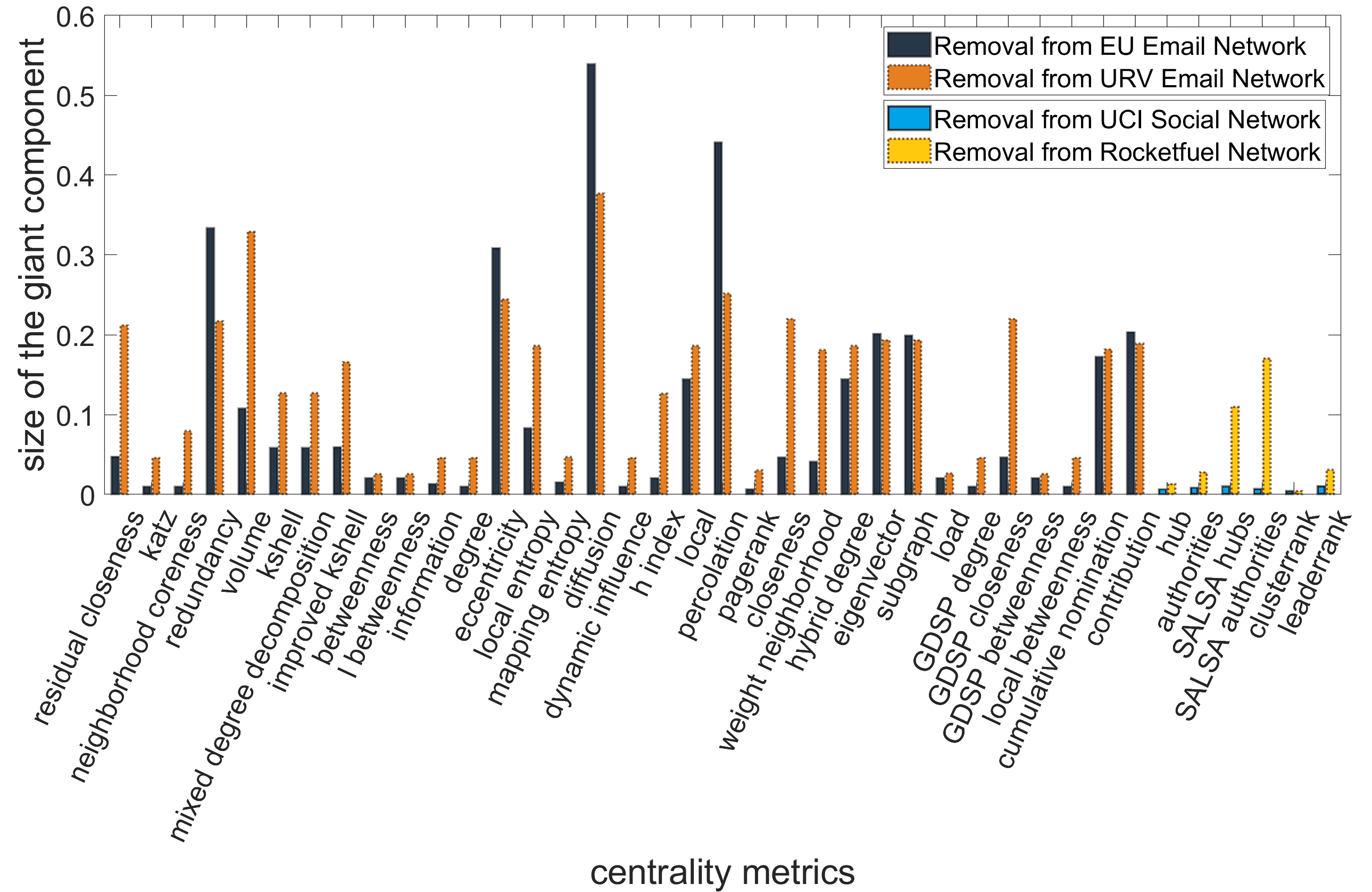}
\caption{The size of the giant component after removing the top 50 percent of the non-infectious attackers selected based on the given point centrality metrics (39 point metrics tested) in both undirected networks (i.e., EU Email Network and URV Email Network) and directed networks (i.e., UCI Social Network and Rocketfuel Network).} 
\label{fig:R2}
\end{figure*}
%\end{comment}

Fig.~\ref{fig:R2} shows the size of the giant component after the top 50 percent of the nodes, ranked based on each point centrality, are removed. Note that this attack is not infectious so an attacked node cannot compromise adjacent nodes.  In undirected networks, most centrality metrics showed a larger size of the giant component in a dense network, which is the EU Email Network. On the other hand, in the URV Email Network, which is a sparse network, we observe a smaller size of the giant component.  Diffusion, percolation, and volume centrality metrics performed relatively poorly perhaps indicating these metrics are less informative for sparser networks.  Except for the clusterrank metric, all metrics evaluated under directed networks performed better (i.e., a smaller size of the giant component from the attacker perspective) under the UCI Social Network than the Rocketfuel Network.  The key observations from Fig.~\ref{fig:R2} are: (i) Katz and dynamic influence centrality metrics show a weaker impact on the size of the giant component, compared to other centrality metrics. This is because both metrics are derived based on eigenvalues and measure the influence of the node based on the influence of its neighbors.  Even if the node itself is removed, the adjacent nodes are connected in the giant component of the network.  Hence, the impact of removing nodes with high Katz or dynamic influence centrality is not stronger than that of removing nodes with high centrality of other types; (ii) The effect of the point centrality on the degradation of the network depends also on the network topology. For example, with volume centrality, node removals in the EU Email Network results in a significantly larger size of the giant component than node removals in the URV Email Network.  In addition, all point centrality metrics tested in the right side of the plot (e.g., from eigenvector centrality to contribution centrality) show a larger size of the giant component for the URV Email Network compared to the EU Email Network; and  (iii) In the metrics evaluated under directed networks, we can clearly see poor performance of authorities, SALSA hubs, and SALSA authorities on a sparse network as the Rocketfuel Network.  This is because an attack only infects in the direction of its directed edges. But these three centrality metrics measure the centrality based on incoming edges, which even prevents the infection from being spread over the network.

\begin{figure*}
\centering
\subfigure[Infectious attacks with degree, closeness, betweenness, pagerank, eigenvector, local entropy and mapping entropy]{
\includegraphics[width=0.4\textwidth]{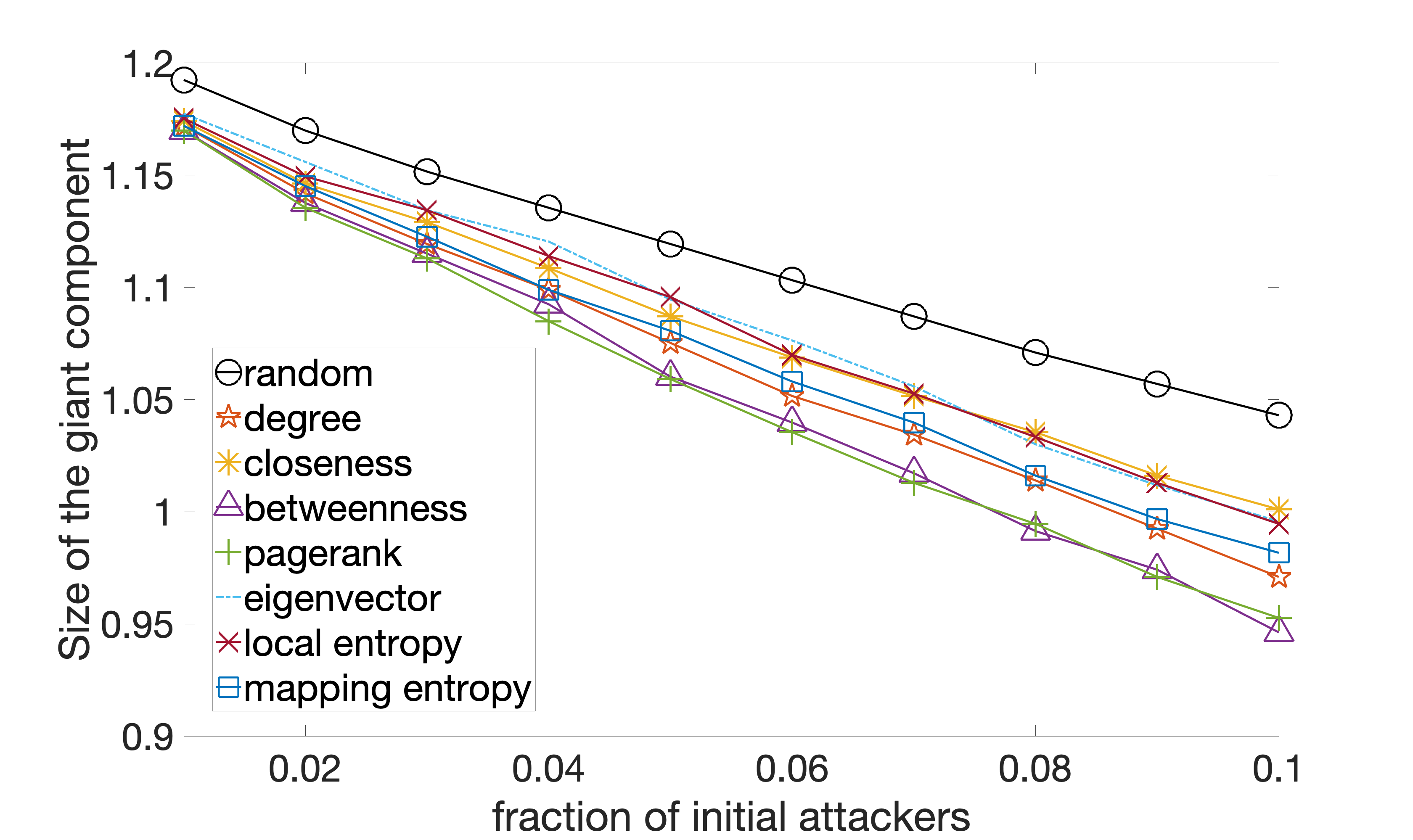}} \hspace{5mm}
\subfigure[Infectious attacks with local betweenness, volume, redundancy, kshell, improved kshell, percolation and hybrid degree]{
\includegraphics[width=0.4\textwidth]{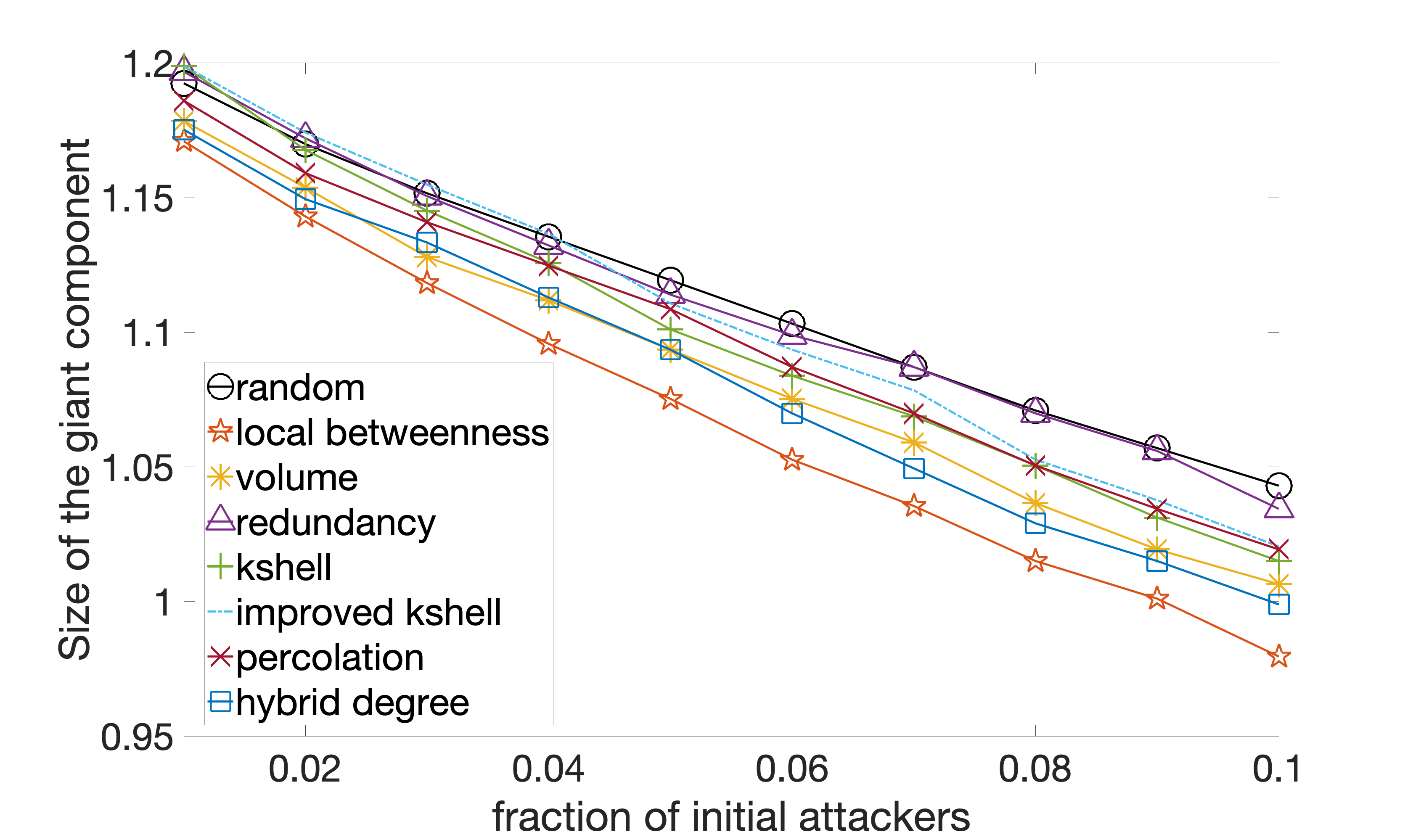}} \hspace{5mm}
\subfigure[Infectious attacks with neighborhood coreness, flow betweenness, katz, deffusion centrality, subgraph and clustering coefficient]{
\includegraphics[width=0.4\textwidth]{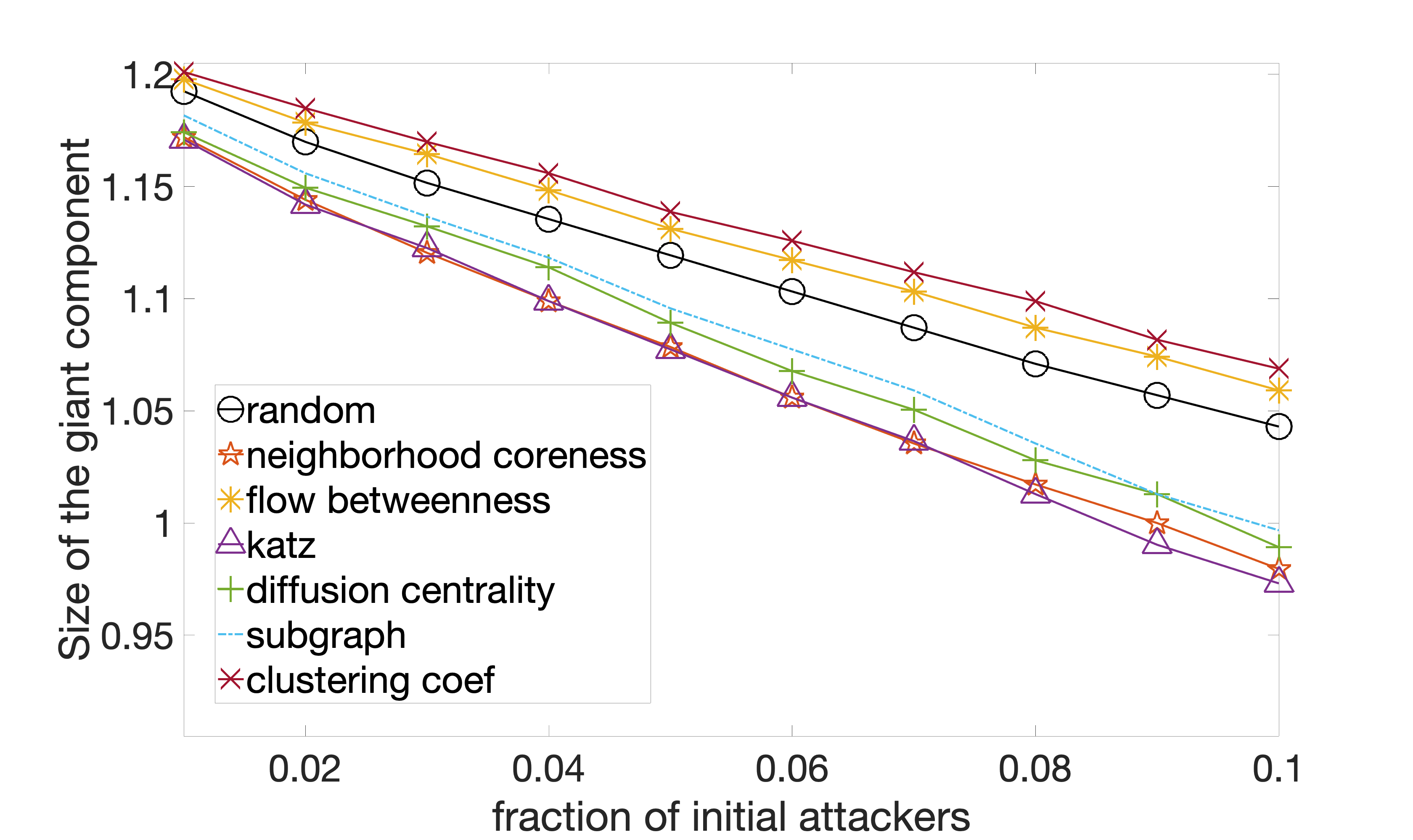}} \hspace{5mm}
\subfigure[Infectious attacks with information centrality, residual closeness, semi local, mixed degree decomposition, dynamic influence and weight neighborhood]{
\includegraphics[width=0.4\textwidth]{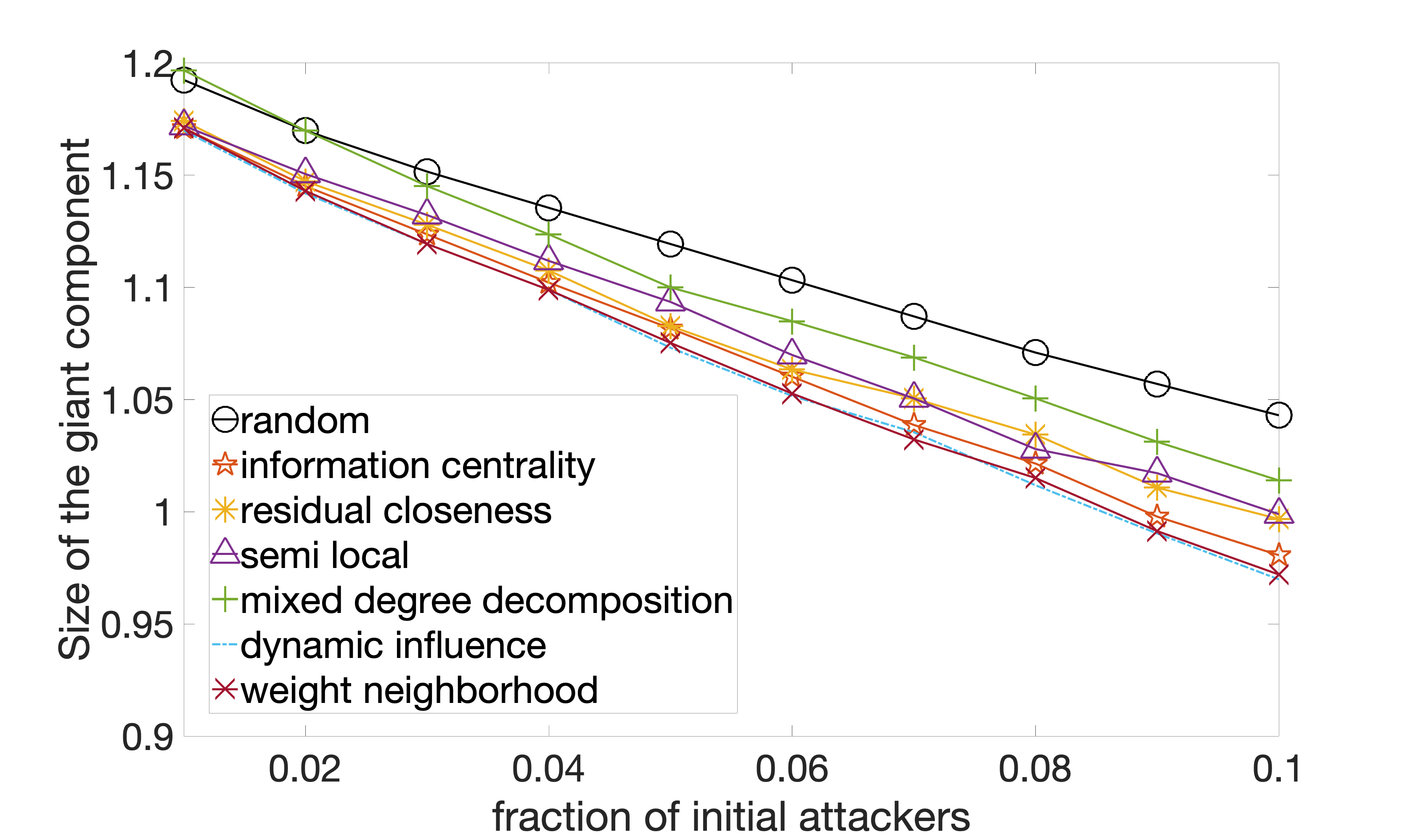}} \hspace{5mm}
\subfigure[Infectious attacks with GDSP degree, GDSP closeness, GDSP betweenness, eccentricity, cumulative nomination, h index, $L$-betweenness and contribution]{
\includegraphics[width=0.4\textwidth]{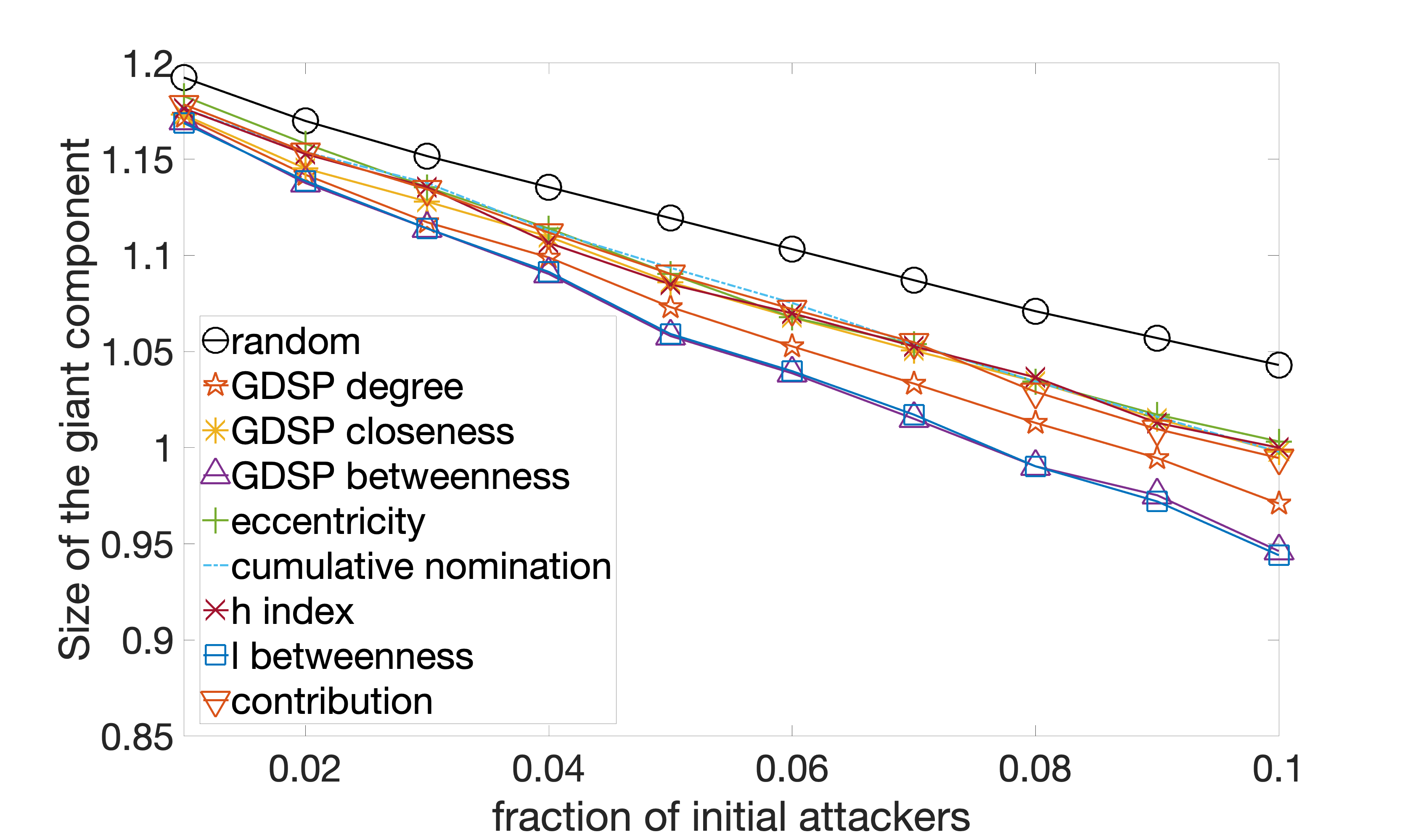}} \hspace{5mm}
\subfigure[Infectious attacks with hubs, authorities, clusterrank, SALSA authorities, SALSA hubs, leaderrank in the (directed) UCI Social Network]{
\includegraphics[width = 0.4 \textwidth]{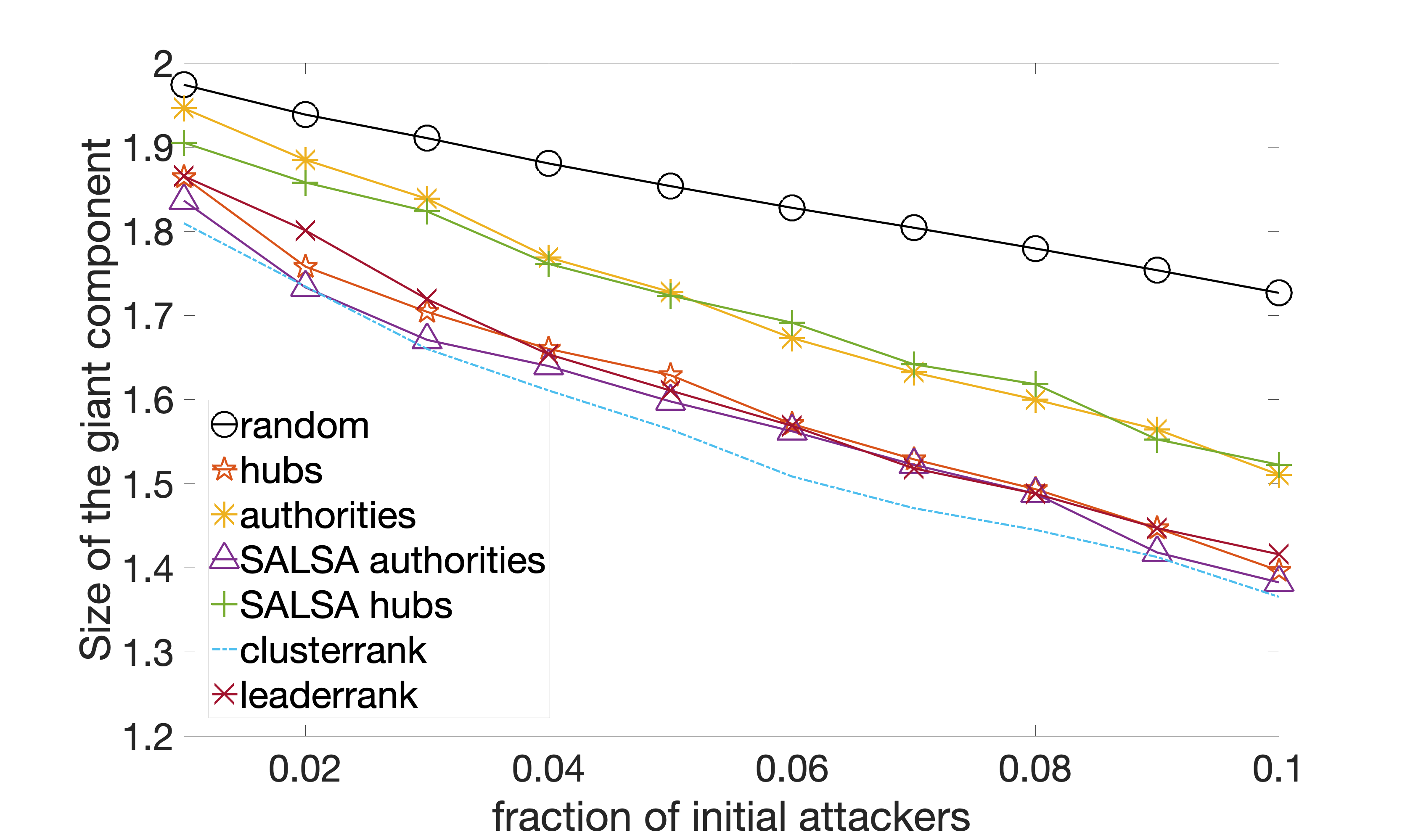}}

\caption{The size of the giant component after removing the initial infectious attacker nodes based on the surveyed point centrality metrics (39 point centrality metrics tested) in the undirected URV Email Network for (a)-(e) and the (directed) UCI Social Network network for (f) where the random node removal is added as a baseline model.}
\label{fig:ia-email-point-infectious}
\end{figure*}
%\end{comment}

\subsubsection{\bf Under Infectious Attacks}
We also evaluated the performance of point centrality metrics surveyed in this work under infectious attacks. As discussed in Section~\ref{subsec:attack-scenarios}, an seeded attacker can infect neighboring nodes with an infection probability $\beta$.  Fig.~\ref{fig:ia-email-point-infectious} shows the size of the giant component under targeted attacks of the URV Email Network and the UCI Social Network for 39 point centrality metrics. 
%The difference is that in Fig.~\ref{fig:ia-email-point-infectious}, the selected attackers infect its adjacent nodes with the infection probability $\beta = 0.05$.  
Here, we varied the fraction of the initial attackers by an increment of 0.01 from 0.01 to 0.1. A node is immune to the attack if the node is attacked but is not infected based on the given infection probability, $\beta$. Note that we report results over a smaller fraction of initial attackers because of the stronger impact of infectious attacks on the size of the giant component.  We observed the following from the results shown in Fig.~\ref{fig:ia-email-point-infectious}. First, overall the decrease of the size of the giant component is linear. Most targeted attacks reduce the size of the giant component compared to random attacks. Second, curiously, three point centrality metrics tested in this work resulted in a comparable or larger size of the giant component than random attacks. These are clustering coefficient, flow betweenness, and redundancy.  For the clustering coefficient, as discussed in Fig.~\ref{fig:ia-email-point} (c), removing a node with high clustering coefficient has a limited effect on its local network due to high connectivity. More generally, when local neighborhoods are well connected, which is the case for nodes with high clustering coefficient, the reduction of the network is tempered.  Similarly, since redundancy captures the overlap of a node's neighborhood with that of other nodes, the network is less likely to be dismantled because the nodes in the neighborhood remain connected. Volume centrality is estimated based on a given hop $h$ which is set to 3 in our work. 
%Due to this hop setting in our experiment, a node's volume centrality is also limited to 3-hop neighbors when estimating a node's local neighbor hood. 
This means that even when a node with high volume centrality is removed, an infectious propagation of the attack may be limited in scope depending on the immunity of the immediate neighbors. Lastly, the performances of betweenness and pagerank in (a) and GDSP betweenness and $L$-betweenness in (e) are impressive compared to other centrality metrics, resulting in a significantly smaller size of the giant component for the undirected URV Email Network. In addition, in the (directed) UCI Social Network, clusterrank, leaderrank, hubs, and SALSA authorities are quite impressive in their performance, resulting in a significantly smaller size of the giant component, compared to other centrality metrics.   

\begin{figure*}
\centering
\subfigure[Infectious attacks with degree, closeness, betweenness, pagerank, eigenvector, local entropy and mapping entropy]{
\includegraphics[width=0.4\textwidth]{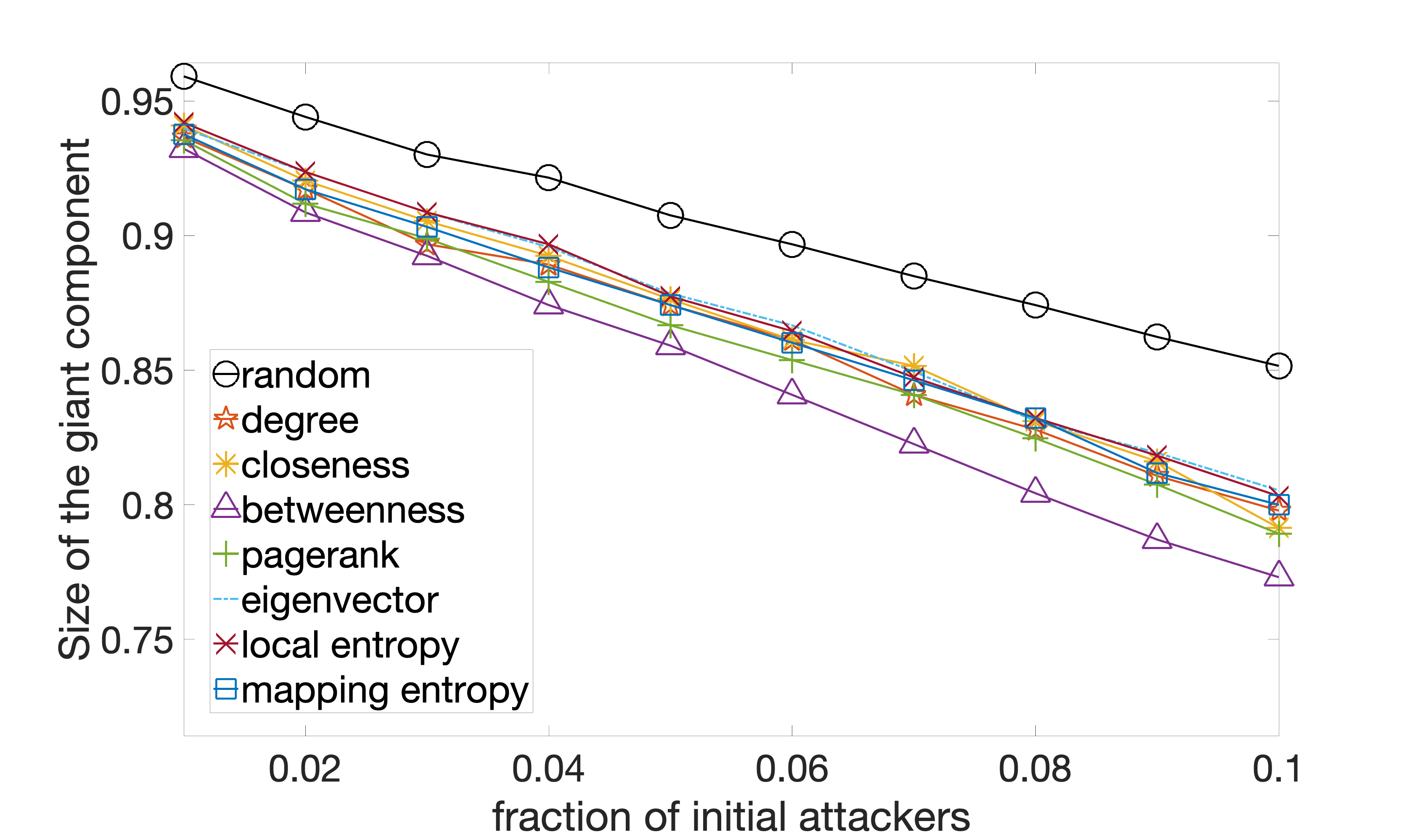}} \hspace{5mm}
\subfigure[Infectious attacks with local betweenness, volume, redundancy, kshell, improved kshell, percolation and hybrid degree]{
\includegraphics[width=0.4\textwidth]{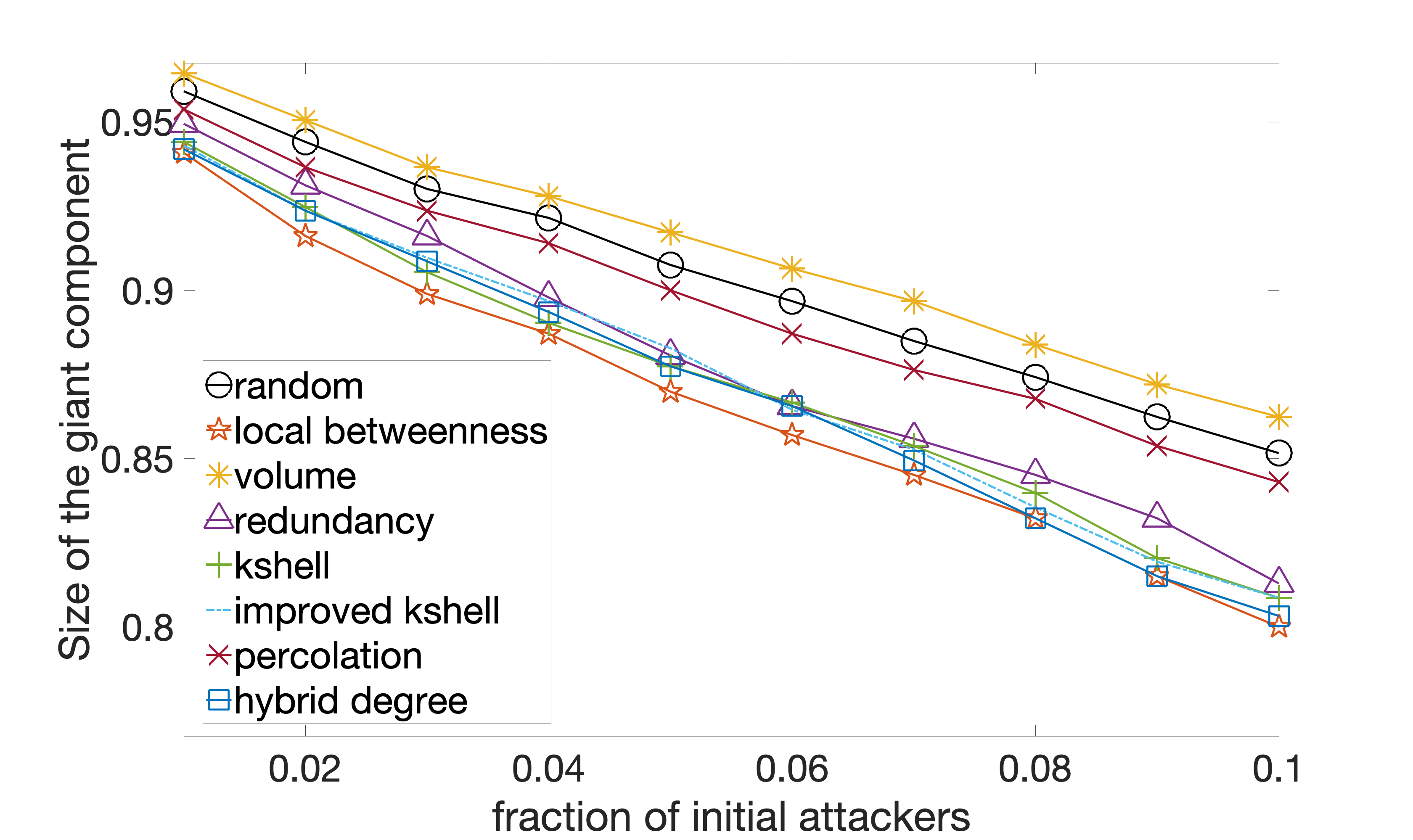}} \hspace{5mm}
\subfigure[Infectious attacks with neighborhood coreness, flow betweenness, katz, deffusion centrality, subgraph and clustering coefficient]{
\includegraphics[width=0.4\textwidth]{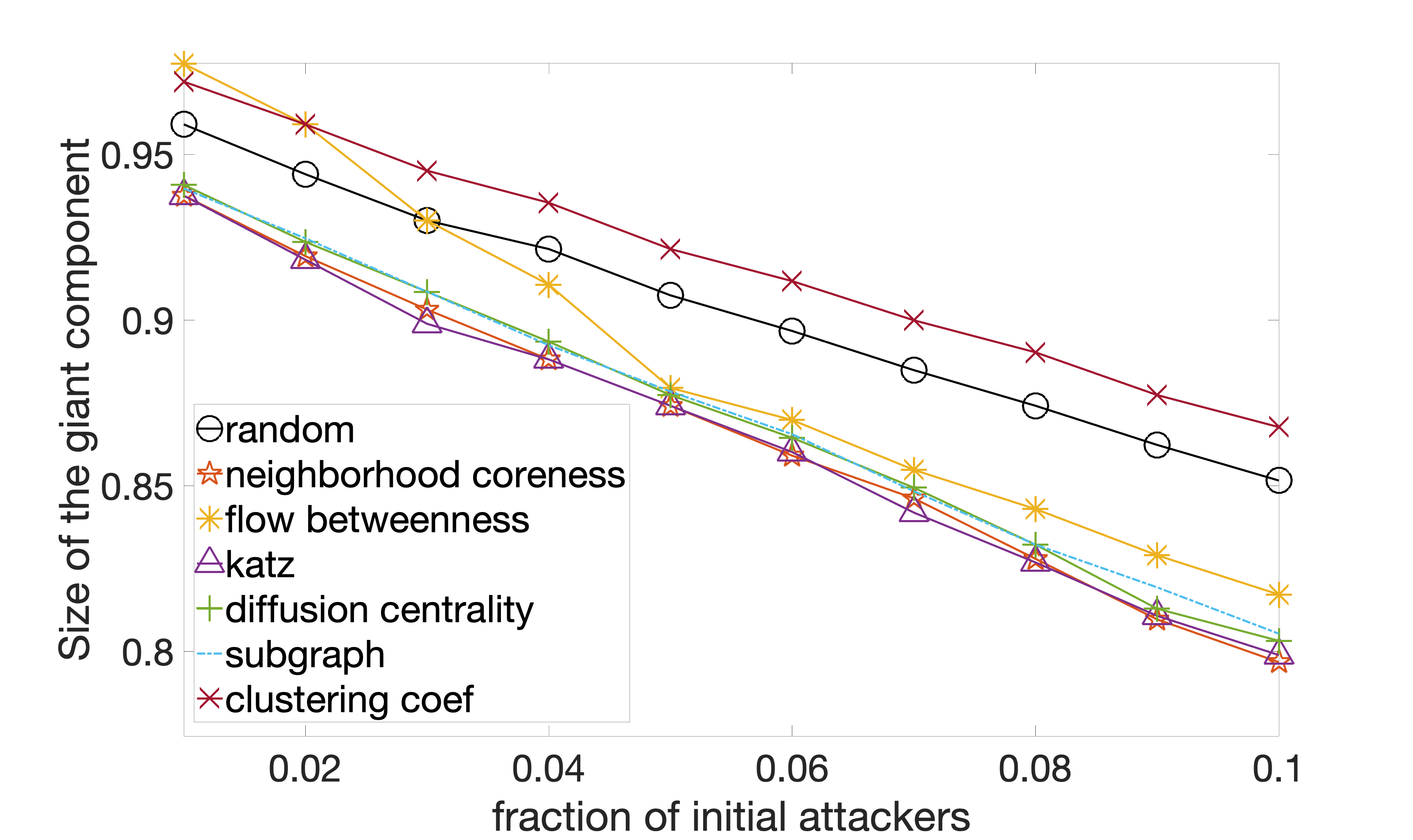}} \hspace{5mm}
\subfigure[Infectious attacks with information centrality, residual closeness, semi local, mixed degree decomposition, dynamic influence and weight neighborhood]{
\includegraphics[width=0.4\textwidth]{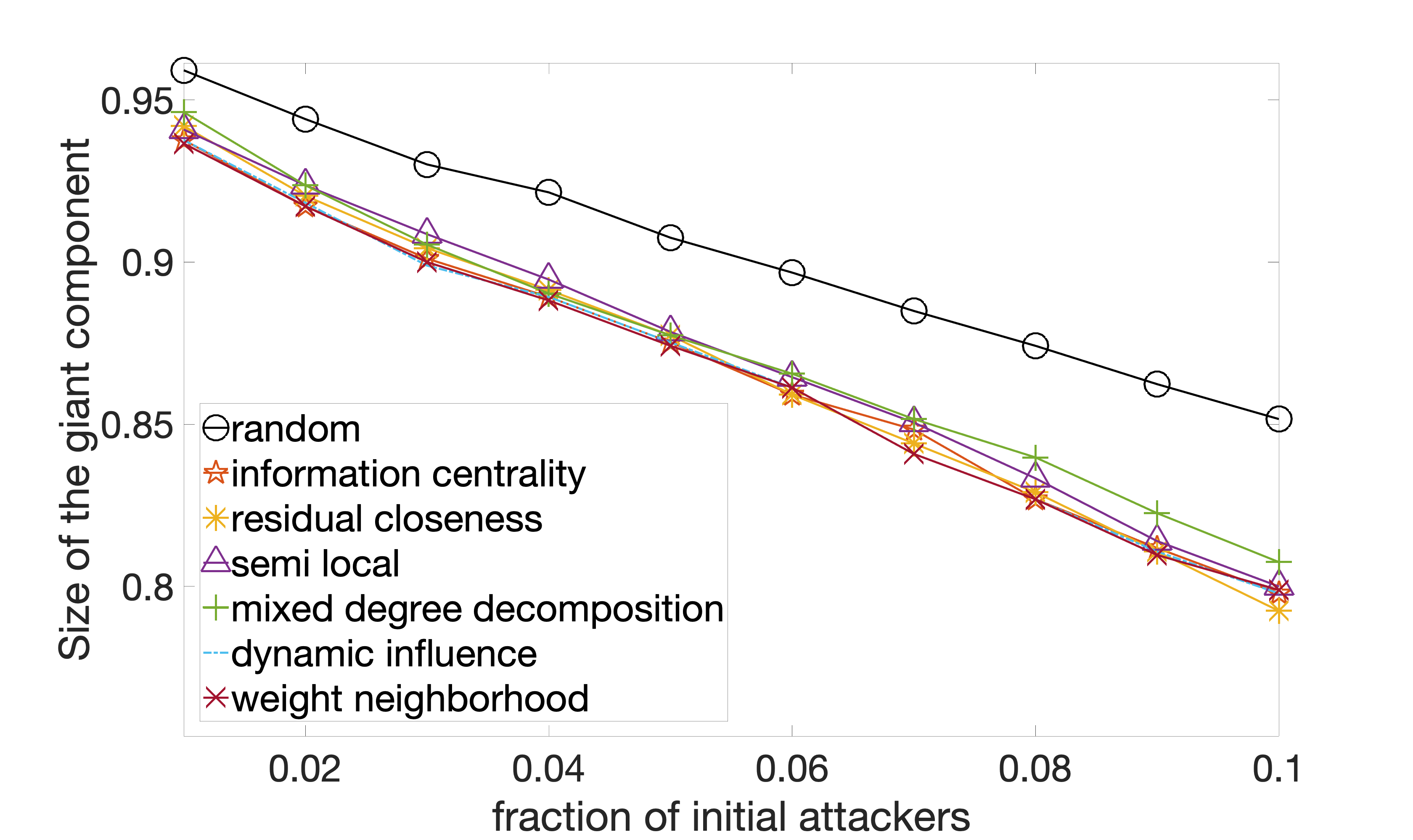}} \hspace{5mm}
\subfigure[Infectious attacks with GDSP degree, GDSP closeness, GDSP betweenness, eccentricity, cumulative nomination, h index, $L$-betweenness and contribution]{
\includegraphics[width=0.4\textwidth]{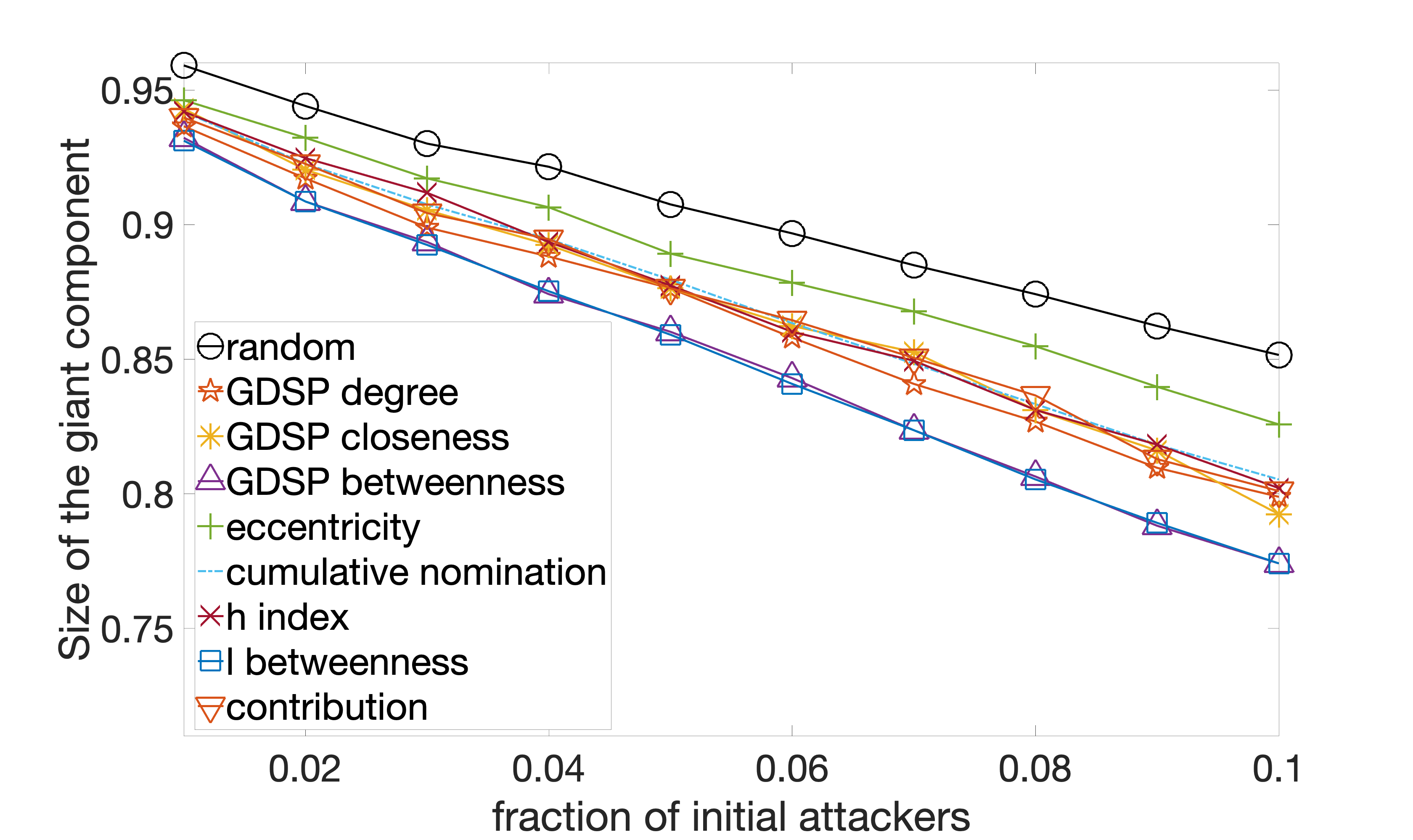}} \hspace{5mm}
\subfigure[Infectious attacks with hubs, authorities, clusterrank, SALSA authorities, SALSA hubs, leaderrank in the directed Rocketfuel Network]{
\includegraphics[width = 0.4 \textwidth]{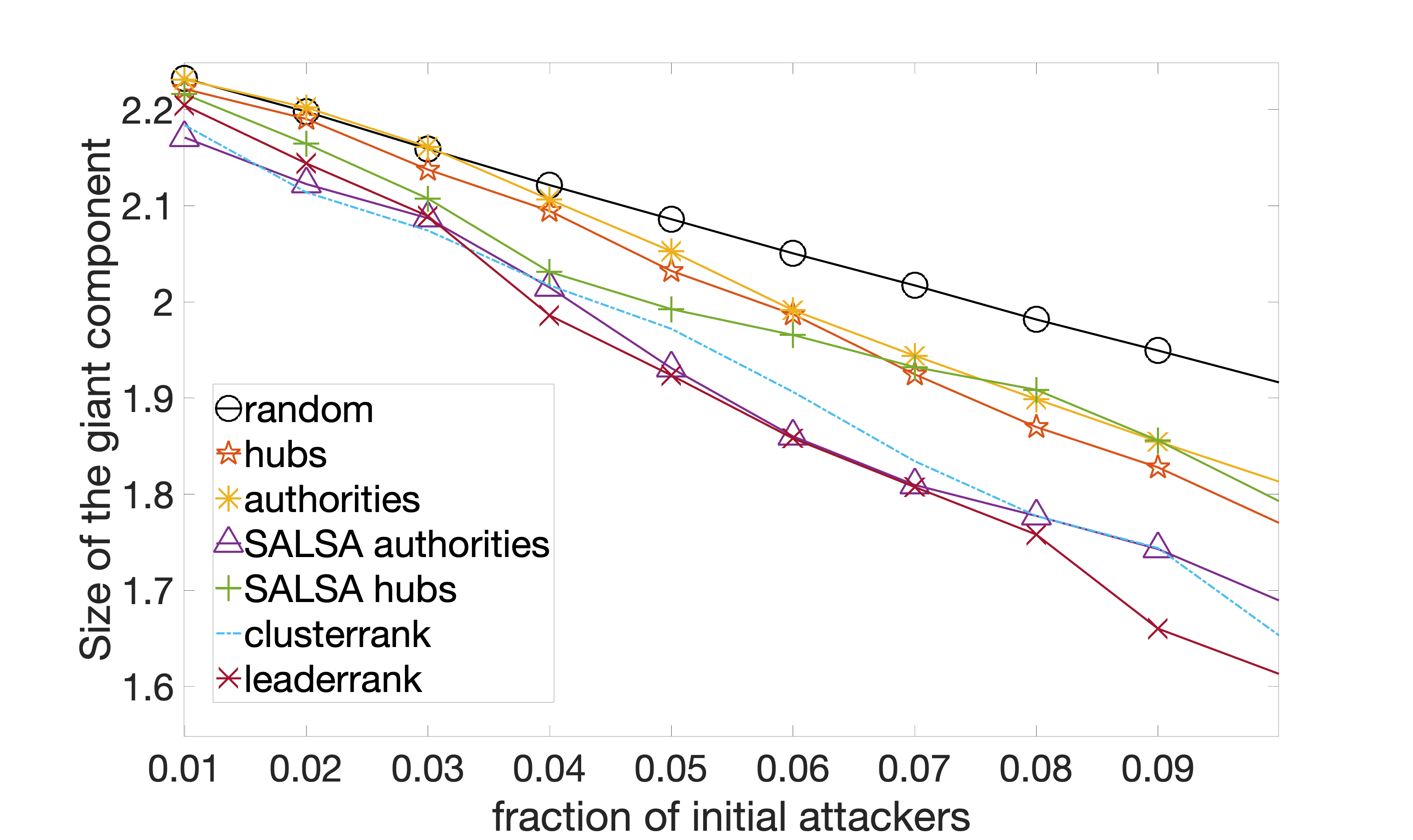}}

\caption{The size of the giant component after removing the initial infectious attacker nodes based on the surveyed point centrality metrics (39 point centrality metrics tested) in the undirected EU Email Network for (a)-(e) and the directed Rocketfuel Network for (f) where the random node removal is added as a baseline model.}
\label{fig:email-Eu-point-infectious}
\end{figure*}
%\end{comment}

Fig.~\ref{fig:email-Eu-point-infectious} shows the size of the giant component under targeted infectious attacks on the EU Email Network and the Rockefuel Network. Again, the infection probability is $\beta = 0.05$, and there are 39 point centrality metrics tested. The overall trends are similar to Fig.~\ref{fig:ia-email-point-infectious}. However, some differences are as follows. First, seeding attackers based on flow betweenness in Fig.~\ref{fig:email-Eu-point-infectious}(c) performs better in the EU Email Network as the fraction of initial infectious attackers increases whereas in the URV Email Network, selection based on flow betweenness performed no better than random selection, as shown in Fig.~\ref{fig:ia-email-point-infectious}(c). Second, volume centrality-based seeding didn't perform as well in the EU Email Network (Fig.~\ref{fig:email-Eu-point-infectious}(b)) compared to the URV Email Network (Fig.~\ref{fig:ia-email-point-infectious}(b)).  This could be because of the reason discussed earlier regarding the clustering coefficient, which also didn't perform better compared to the random attack.  That is, removing a node with high volume centrality may only collapse the local network of the node. This means that under dense networks, the removal of nodes with a highly connected local neighborhood does little to separate the network into smaller components.  Third, the resulting size of the giant component is similar in the EU Email Network for all centrality metrics in Fig.~\ref{fig:email-Eu-point-infectious}(d), while the performances are more distinctive in the URV Email Network, as shown in Fig.~\ref{fig:ia-email-point-infectious}(d) showed distinctive performances. Based on these observations, we can say the network topology really affects the performance of centrality metrics. In particular, the key difference between these two datasets (i.e., the URV Email Network in Fig.~\ref{fig:ia-email-point-infectious} and the EU Email Network in Fig.~\ref{fig:email-Eu-point-infectious}) is that the EU Email Network is a denser network than the URV Email Network.  This can explain why flow betweenness can significantly perform better than random in the EU Email Network, compared to its performance in the URV Email Network.  That is, since a higher network density (i.e., more edges between nodes) can increase the impact of infectious attacks, the flow betweenness-based attacks can take an advantage of the network density to increase its effect in compromising other nodes in the network. In addition, higher network density can also make the performances of targeted attacks less distinctive because the opportunities for infection are more relevant than the marginal benefits of optimizing the selection of initial attackers. 

\begin{figure*}[t]
\centering
\includegraphics[width = \textwidth, height=0.6\textwidth]{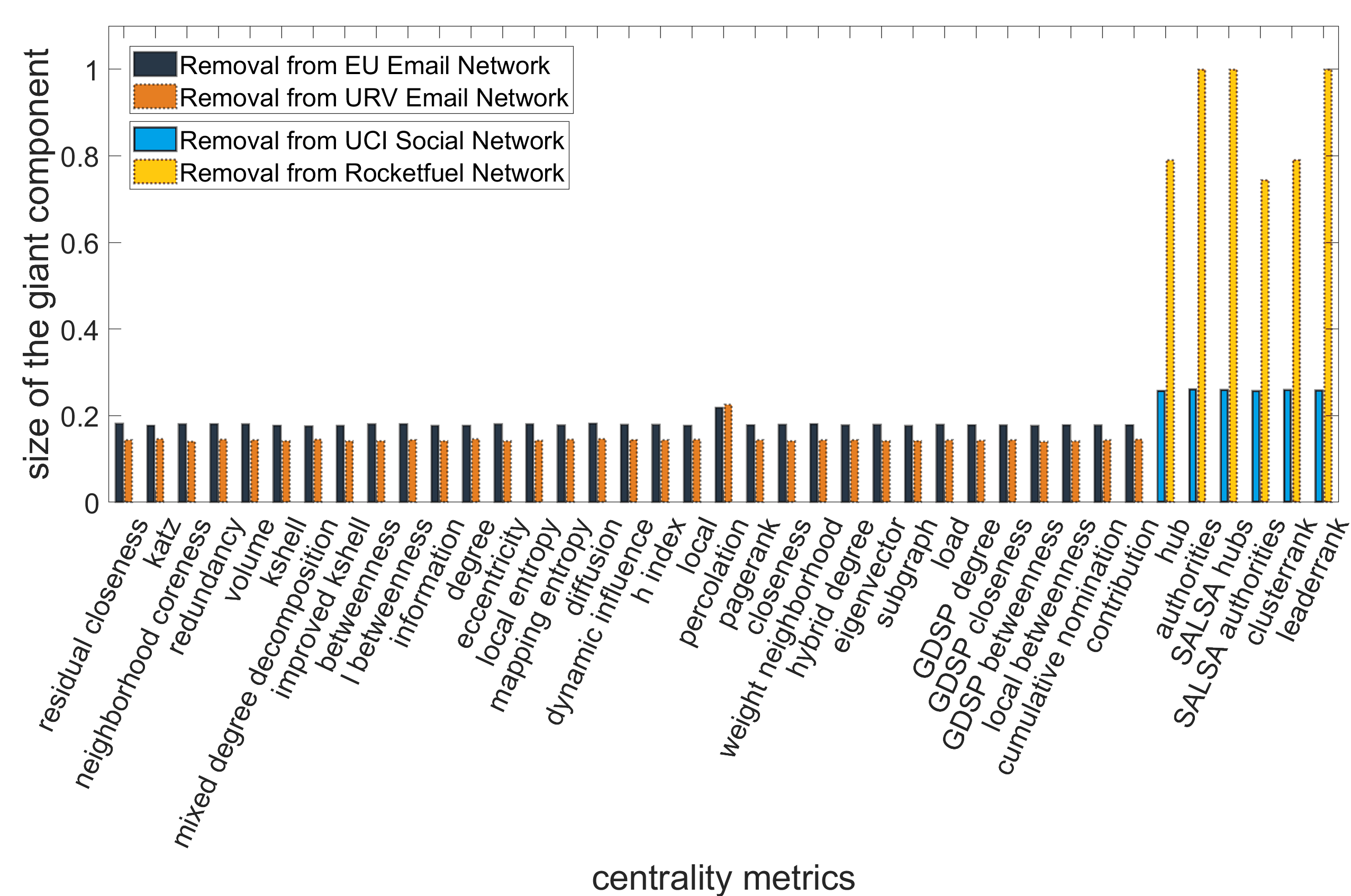}
\caption{The size of the giant component after removing a single top ranked node based on a given centrality metric (39 point centrality metrics tested) in both undirected networks (i.e., EU Email Network and URV Email Network) and directed networks (i.e., UCI Social Network and Rocketfuel Network) where the attack is infectious. }
\label{fig:R5-eu}
\end{figure*}
%\end{comment}

Fig.~\ref{fig:R5-eu} shows the effect of point centrality-based targeted attacks in the undirected networks (EU Email Network, URV Email Network) and directed networks (UCI Social Network, Rocketfuel Network) in terms of the size of the giant component as an indicator of the network resilience when the single top-ranked node based on a given metric is selected as an infectious attacker.  The trends are very similar to Fig.~\ref{fig:R2} in terms of the performance under different networks.  Repeating the trends observed in Fig.~\ref{fig:R2}, the effect of targeted attacks based on point centrality metrics is greater (i.e., smaller size of the giant component) in the sparse URV Email Network than in the dense EU Email Network. It is not surprising that the dense network can absorb the impact of removing nodes and better maintain a connected network. However, interestingly, in directed networks, the sparsity of the directed Rocketfuel Network can mitigate the infection process, leading to a larger size of the giant component while the higher density of the UCI Social Network allows attacks to more easily spread.

\begin{figure*}
\centering
\subfigure[Infectious attacks with degree, closeness, betweenness, pagerank, and eigenvector]{
\includegraphics[width=0.4\textwidth]{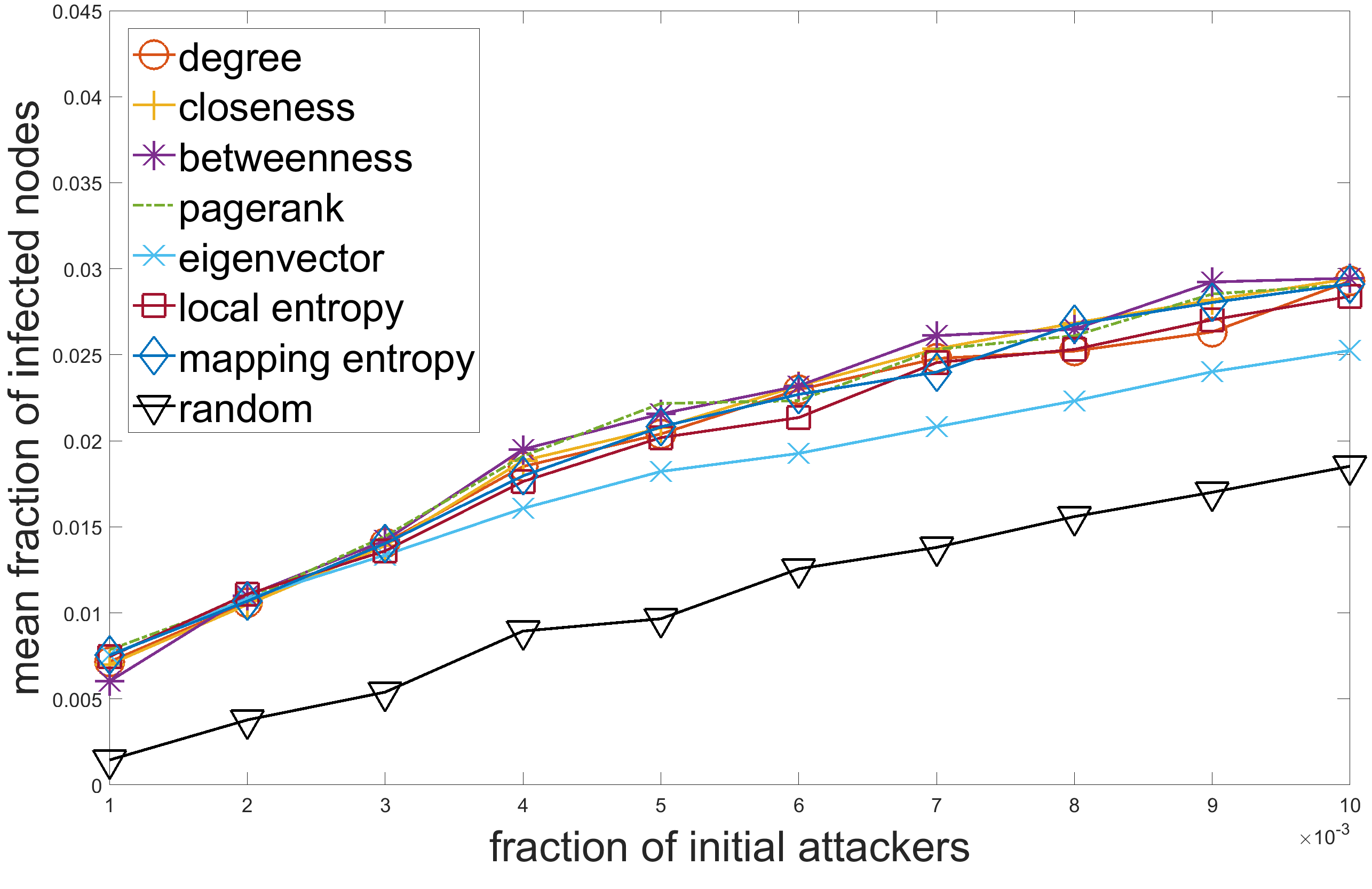}} \hspace{5mm}
\subfigure[Infectious attacks with local betweenness, volume, redundancy, kshell, improved kshell, percolation and hybrid degree]{
\includegraphics[width=0.4\textwidth]{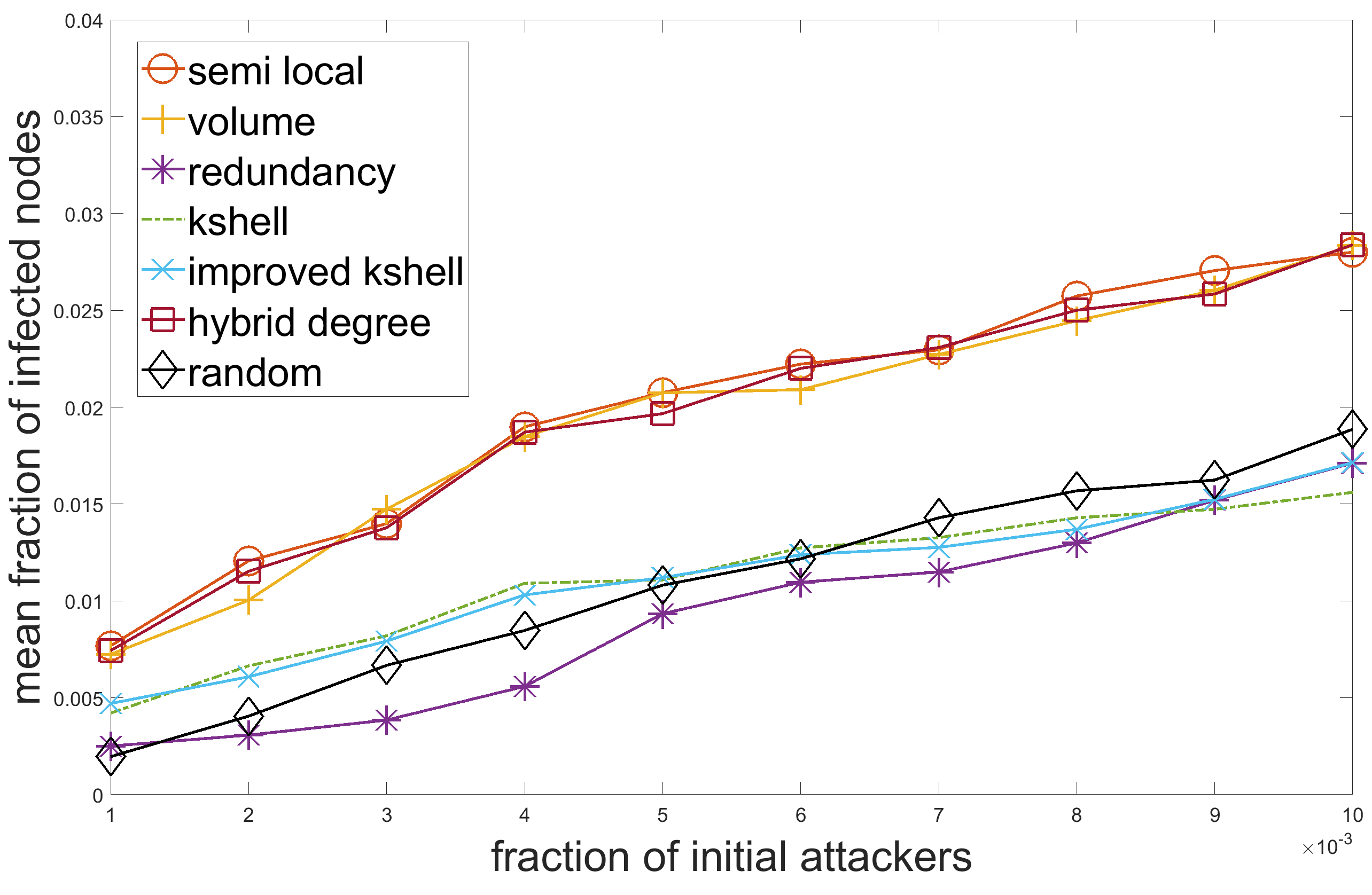}} \hspace{5mm}
\subfigure[Infectious attacks with neighborhood coreness, flow betweenness, katz, diffusion centrality subgraph and clustering coefficient]{
\includegraphics[width=0.4\textwidth]{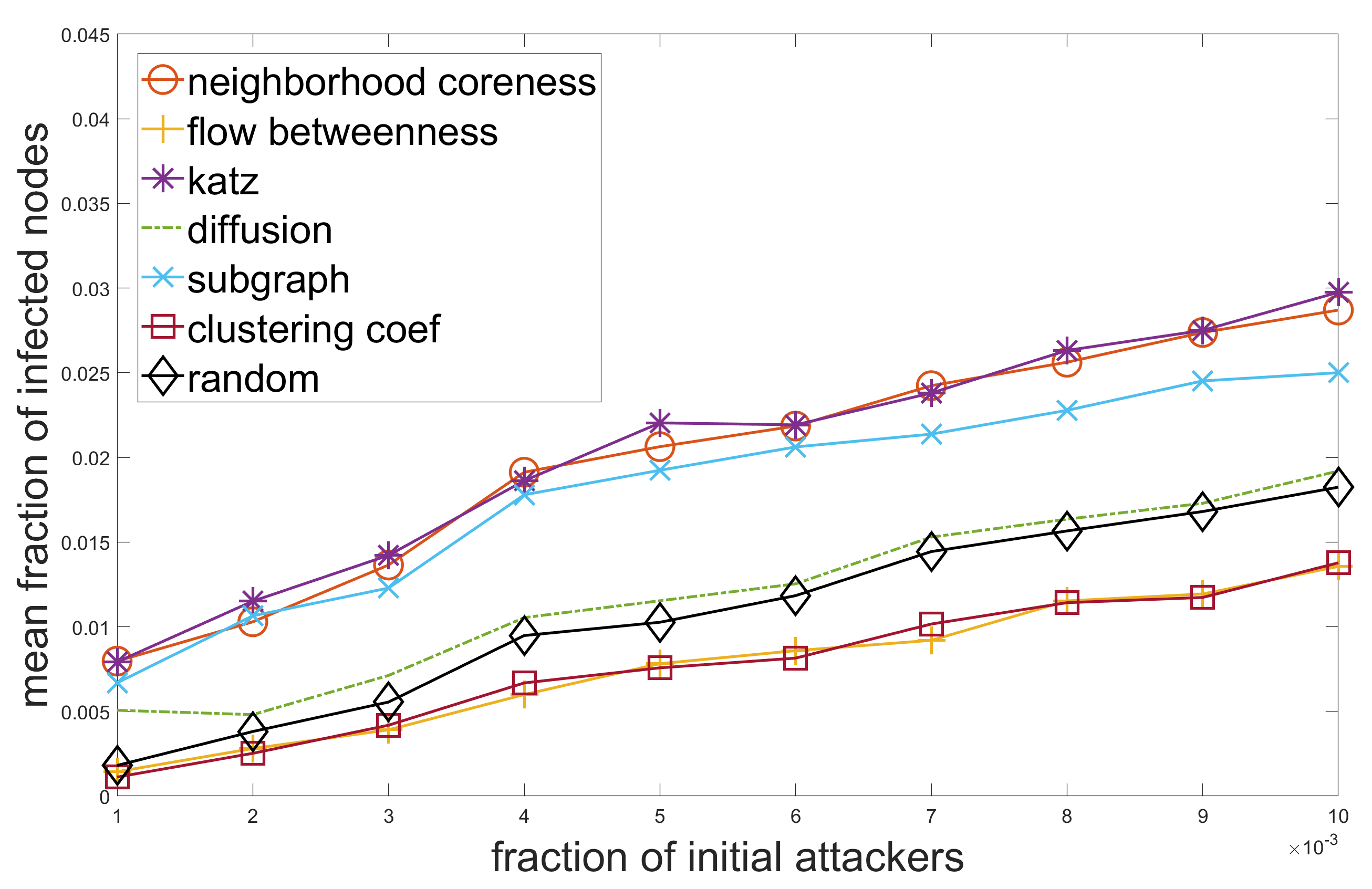}} \hspace{5mm}
\subfigure[Infectious attacks with information centrality, residual closeness, semi local, mixed degree decomposition, dynamic influence and weight neighborhood]{
\includegraphics[width=0.4\textwidth]{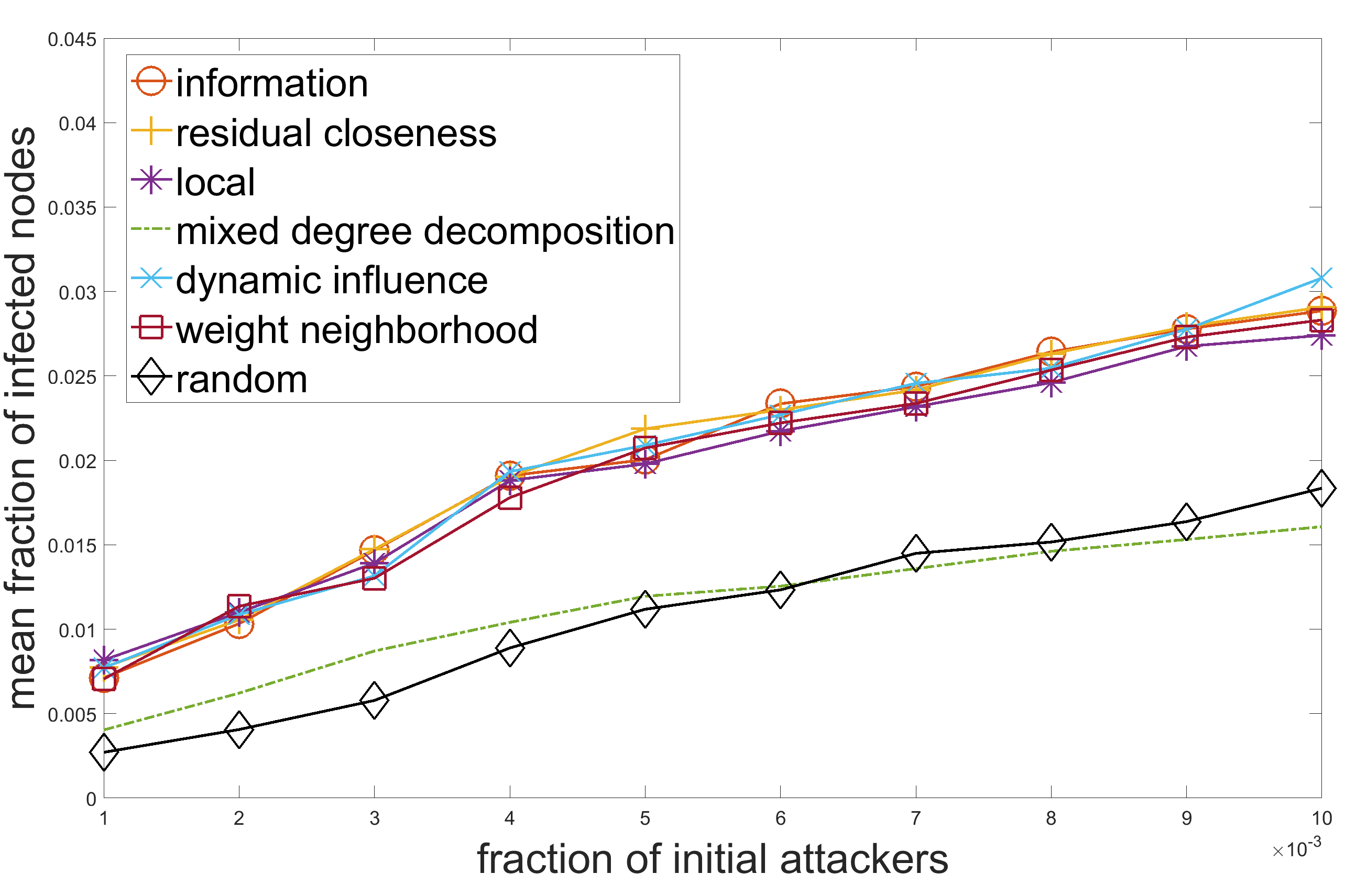}} \hspace{5mm}
\subfigure[Infectious attacks with GDSP degree, GDSP closeness, GDSP be tweenness, eccentricity, cumulative nomination, h index and contribution]{
\includegraphics[width=0.4\textwidth]{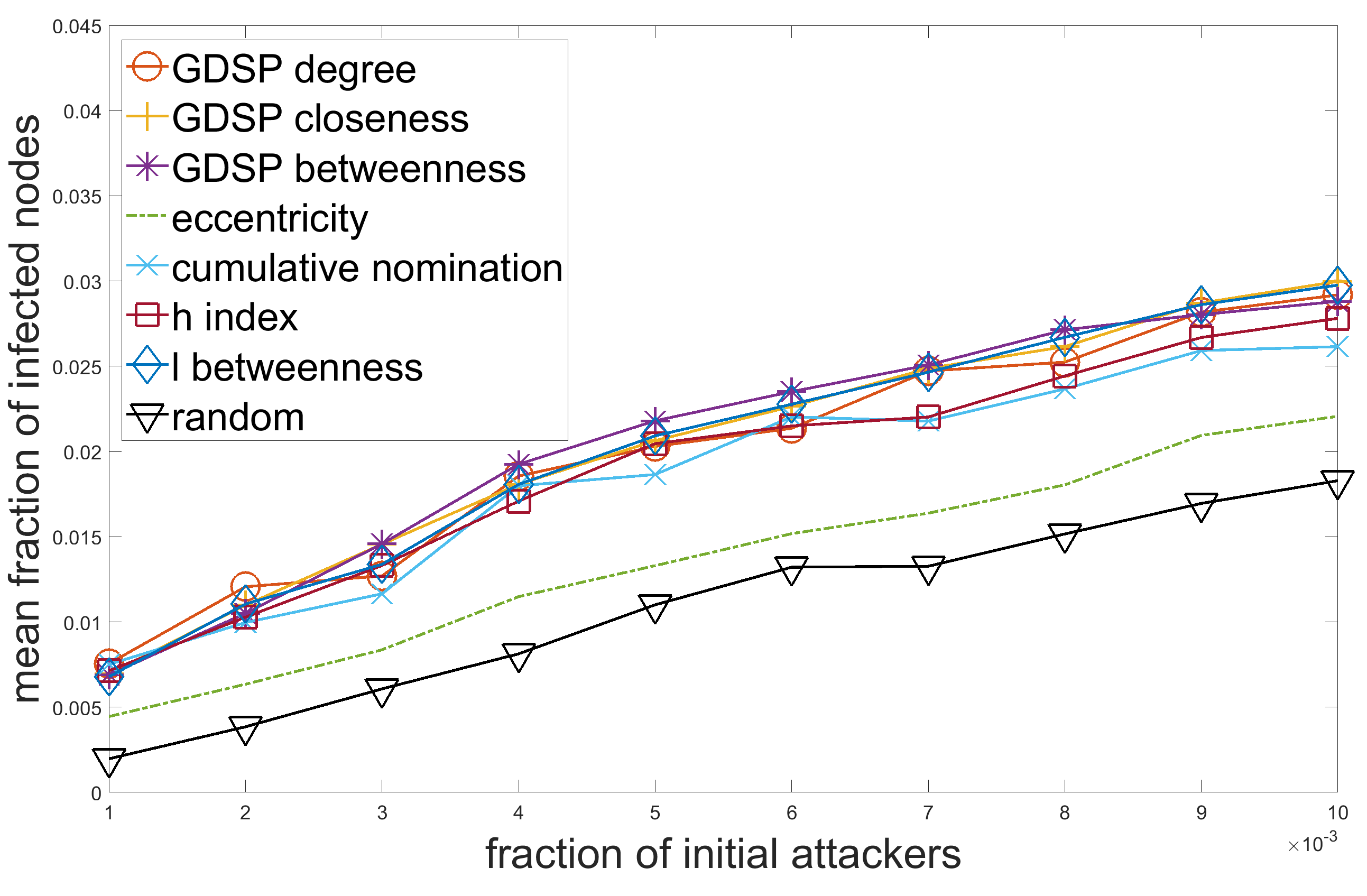}} \hspace{5mm}
\subfigure[Infectious attacks with hubs, authorities, clusterrank, SALSA authorities, SALSA hubs and leaderrank in the directed Rocketfuel Network]{
\includegraphics[width=0.4\textwidth]{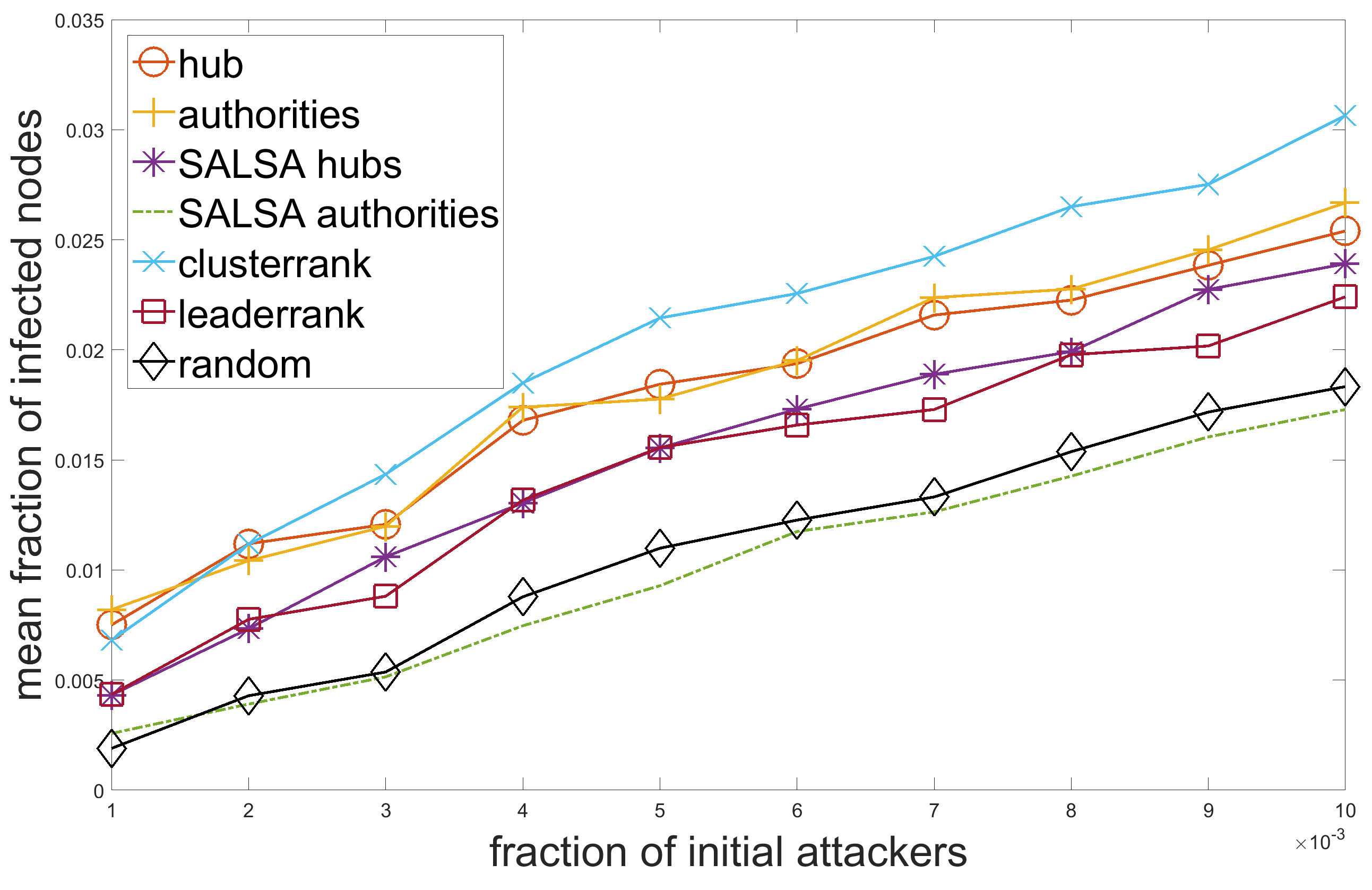}}
\caption{Mean fraction of infected nodes after infectious, initial targeted attackers are selected from 0.001 to 0.01 with the increment of 0.01 based on 38 centrality metrics in the undirected EU Email Network (i.e., (a)-(e)) and in the directed Rocketfuel Network (i.e., (f)).}
\label{fig:point-centrality-infectious-yash-dataset}
\end{figure*}
%\end{comment}

Fig.~\ref{fig:point-centrality-infectious-yash-dataset} shows the mean fraction of nodes infected by a single, initial attacker when the fraction of initial attackers vary from 0.001 to 0.01 with an increment of 0.01 using 38 point centrality metrics to determine the initial selection for the undirected EU Email Network (i.e., Fig.~\ref{fig:point-centrality-infectious-yash-dataset}(a)-(e)) and in the directed Rocketfuel Network (i.e., Fig.~\ref{fig:point-centrality-infectious-yash-dataset}(f)). Most metrics evaluated in this work showed higher rates of infection spread per initial attacker. However, some metrics, such as flow betweenness, clustering coefficient, diffusion centrality, mixed degree decomposition, and SALSA authorities, showed lower rates per initial attacker.  Note that an attack resulting in a smaller size of the giant component does not necessarily mean there are more infected nodes because there may exist many uninfected nodes in smaller components.  Conversely, lower infection rates due to a given centrality-based selection does not imply that the network is resilient to that particular attack. 
%Naturally we can observe a higher fraction of initial attackers introduce a higher fraction of infected nodes per initial attacker. %THIS IS WRONG. IF I INFECT 100% OF NODES, THE INFECTION RATE IS ZERO SINCE THERE ARE NO NODES TO INFECT. SO THE STATEMENT IS TRUE UP TO A POINT, BUT WE DID NOT INVESTIGATE WHERE THE RATE PER SEEDED ATTACKER GOES DOWN.

\subsection{Network Resilience Analysis of Graph Centrality Metrics} \label{subsec:exp-graph-centrality}

We surveyed 14 graph centrality metrics in Section~IV of the main paper. Since the range of each metric varies, we cannot compare their maximum values. However, we can at least investigate whether the value of each metric increases or decreases depending on how many nodes are removed at random and accordingly the size of the giant component. In order to easily observe this, we devised a metric, the {\em relative graph centrality} (RGC) value, which is computed by:
\begin{equation}
\label{eq:rgc}
\mathrm{RGC} = \frac{GC - GC'}{GC},    
\end{equation}
where $GC$ is the value of a given graph centrality (GC) from the original network with the size of the giant component being 1 and $GC'$ is the value of a given GC after removing a certain percentage of nodes being removed at random.  If we observe the RGC value increases under a smaller $S_g$, it implies that the GC value decreases under the smaller $S_g$. On the other hand, if the RGC value decreases under a small $S_g$, this means the GC value increases under a smaller $S_g$.

\subsubsection{\bf Under Non-Infectious Attacks}

\begin{footnotesize}
\begin{table}[t]
\centering
\caption{Relative Graph Centrality (RGC) Values of 10 GC Metrics Under Non-Infectious Attacks in the Undirected Network Datasets (EU Email Network, URV Email Network)}
\label{tab:rgc-gc}
\vspace{-2mm}
\begin{tabular}{|P{2.5cm}|P{1cm}|P{1cm}|P{1cm}|P{1cm}|}
\hline
Dataset & \multicolumn{2}{c|}{EU Email Network} & 
\multicolumn{2}{c|}{URV Email Network} \\
\hline
\% of node removal &  30\% & 70\% & 30\% & 70\%\\
\hline
Size of the giant component & $\sim$0.7 & $\sim$0.3 & $\sim$0.7 & $\sim$0.3\\
\hline
\hline
 distance-based & 0.538 & 0.936 & 0.538 & 0.927\\
\hline
degree-based & 0.396 & 0.756 & 0.423  & 0.830 \\
\hline
$k$-component & 0.109 & 0.477 & 0.105 & 0.34\\
\hline
local assortativity & 0.052 & 0.282 & 0.017  & 0.092 \\
\hline
graph curvature & 0.028 & 0.152 & -0.024 & -0.074 \\
\hline
global clustering & 0.104 & 0.413 & 0.062 & 0.209 \\
\hline
betweenness-based  & -0.376  & -1.427 & -0.041  & -0.0184 \\
\hline
flow betweenness & -0.156 & -0.603 & -0.035 & - 0.146 \\
\hline
closeness-based & -0.105 & -0.146 & 0.016 & -0.005 \\
\hline
degree assortativity & -0.015 & 0.052 & 0.057 & 0.101 \\
\hline
\end{tabular}
\vspace{-2mm}
\end{table}
\end{footnotesize}

For the validation of group selection centrality (GC) metrics, we considered two sets of random attacks with 30\% removal and 70\% removal of nodes in two undirected network datasets (EU Email Network, URV Email Network). Since we considered random attacks in this case to investigate how the GC values are affected under two different scenarios, we observed that the size of the giant component was the similar with approximately 0.3 and 0.7 for the respective cases. Since $k$-plex, $k$-clique, and $k$-core return a set and reciprocity needs to be applied in a directed network, we omitted the discussions of those metrics. In Table~\ref{tab:rgc-gc}, we summarized the RGC values.   

The key observations are as follows: (i) Overall, the size of giant components under different GC metrics is similar because the attacks are random; and (ii) The effects of random attacks on the extent of GC values are different depending on each GC metric.  We found that increasing the number of initial attackers reduces the GC value in the following graph centrality metrics: distance-based GC, degree-based GC, $k$-component, degree assortativity, local assortativity, and global clustering.  On the other hand, we observed greater GC when increasing the number of attackers in the following GC metrics: betweenness-based GC, closeness-based GC, and graph curvature.  The reason of exhibiting the different trends can be explained as follows. If the GC metric measures how the node is locally connected with its close neighbors, then the GC value decreases due to the breakdown of local connections when random attacks are performed. However, if the GC metric estimates how the node is globally connected with other nodes, its value can increase as the normalization of the GC calculation depends on the size of the network. Therefore, we cannot simply rely on whether a network is dense or sparse based on the GC metric because a higher GC metric doesn't always necessarily imply a denser network.

\subsubsection{\bf Under Infectious Attacks}

\begin{footnotesize}
\begin{table}[t]
\centering
\caption{Relative Graph Centrality (RGC) Values of 10 GC Metrics Under Infectious Attacks in the Undirected Network Datasets (EU Email Network, URV Email Network)}
\label{tab:rgc-gc-infectious}
\vspace{-2mm}
\begin{tabular}{|P{2.5cm}|P{1cm}|P{1cm}|P{1cm}|P{1cm}|}
\hline
Dataset & \multicolumn{2}{c|}{EU Email Network} & 
\multicolumn{2}{c|}{URV Email Network} \\
\hline
\% of node removal &  30\% & 70\% & 30\% & 70\%\\
\hline
Size of the giant component  
 & $\sim$0.32 & $\sim$0.12 & $\sim$0.29 & $\sim$0.07\\
\hline
\hline
 distance-based & 0.8833 & 0.9835  & 0.8902 & 0.9929 \\
\hline
degree-based & 0.8165 & 0.9433 & 0.7096 & 0.8761 \\
\hline
$k$-component & 0.302 & 0.5652 & 0.3525 & 0.6977 \\
\hline
local assortativity & 0.0806 & 0.2425 & 0.2206 & 0.7063 \\
\hline
graph curvature & 0.1808 & 0.2345 & 0.2670 & 0.3438\\
\hline
global clustering & 0.2271 & 0.4278 & 0.3780 & 0.6668 \\
\hline
betweenness-based & -0.0049 & -0.2095 & -1.3418 & -1.3036\\
\hline
flow betweenness & -0.1582 & -0.3101 & -0.5204 & -1.0713 \\
\hline
closeness-based & 0.0194 & 0.0426 & -0.1471 & 0.1523 \\
\hline
degree assortativity & 0.0775 & -0.1900 & 0.1608 & 0.3438 \\
\hline
\end{tabular}
\vspace{-2mm}
\end{table}
\end{footnotesize}

Table~\ref{tab:rgc-gc-infectious} shows the RGC values of the graph centrality (GC) metrics when random infectious attacks are performed. Again, the network is seeded with 30\% or 70\% of infected nodes and the results are for the two undirected networks (EU Email Network, URV Email Network).  Due to the infectious nature of this attack, the size of the giant component is observed to be smaller compared to that under non-infectious attacks. But similar to what we observed in Table~\ref{tab:rgc-gc}, some GC metrics (e.g., the top 6 GC metrics in Table~\ref{tab:rgc-gc}) show a similar tendency with decreasing GC under a graph with a smaller size of the giant component. However, other GC metrics (e.g., the bottom 4 GC metrics in Table~\ref{tab:rgc-gc}) do not show a consistent trend. For example, for degree assortativity, the size of GC decreases in the dense EU Email Network while it increases in the sparse URV Email Network. In addition, GC does not always keep increasing or decreasing depending on the size of the giant component even for the same network, as observed in the closeness-based metric.  Therefore, the scale of some GC metrics can be used to predict the size of the giant component.

\subsection{Network Resilience Analysis of Group Selection Centrality Metrics} \label{subsec:exp-group-selection}

%\begin{comment}
\begin{figure*}
    \centering
\subfigure[Under noninfectious attacks in the URV Email Network]{
\includegraphics[width=0.4\textwidth]{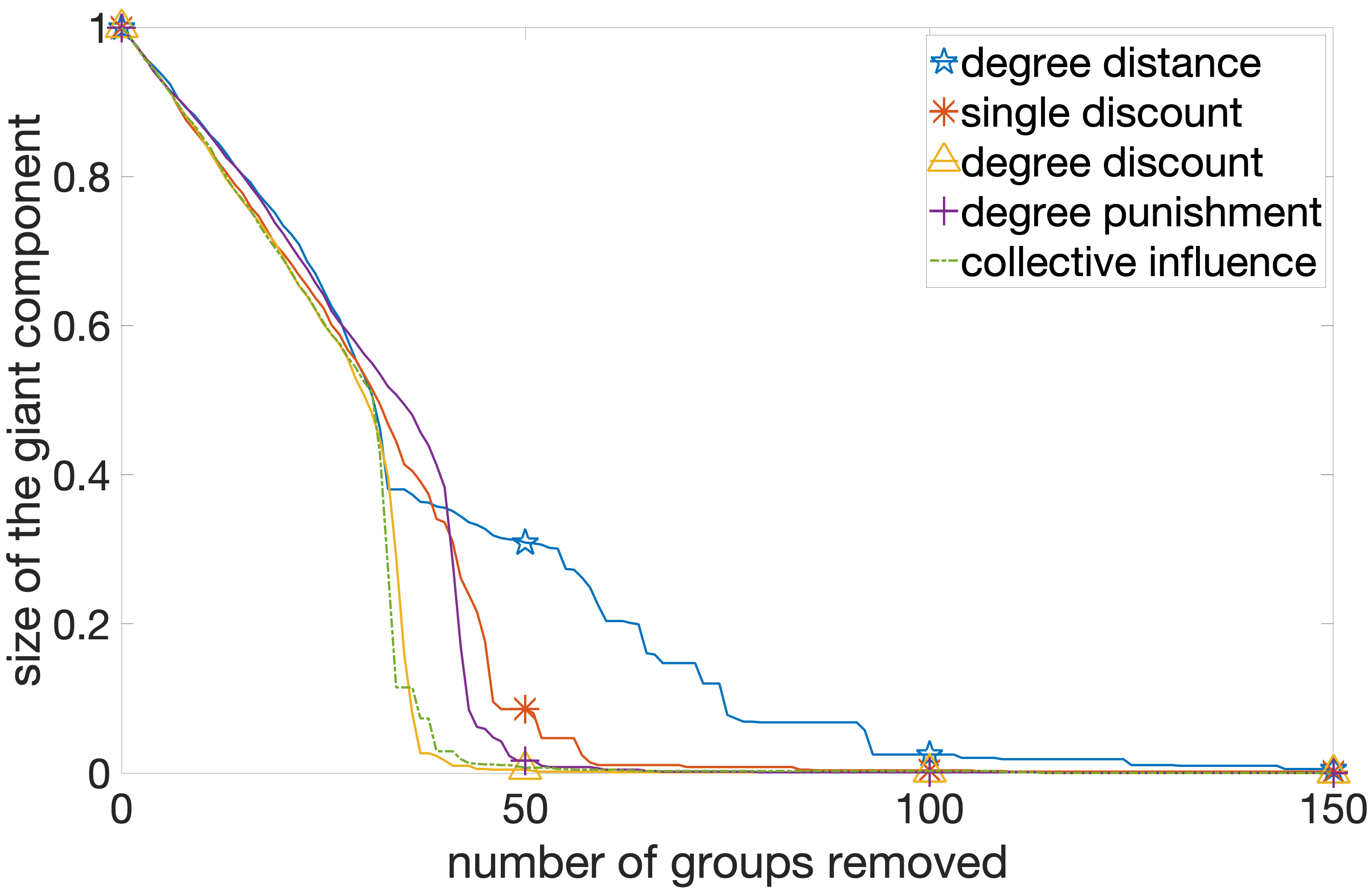}}
\hspace{5mm}
\subfigure[Under noninfectious attacks in the EU Email Network]{
\includegraphics[width=0.4\textwidth]{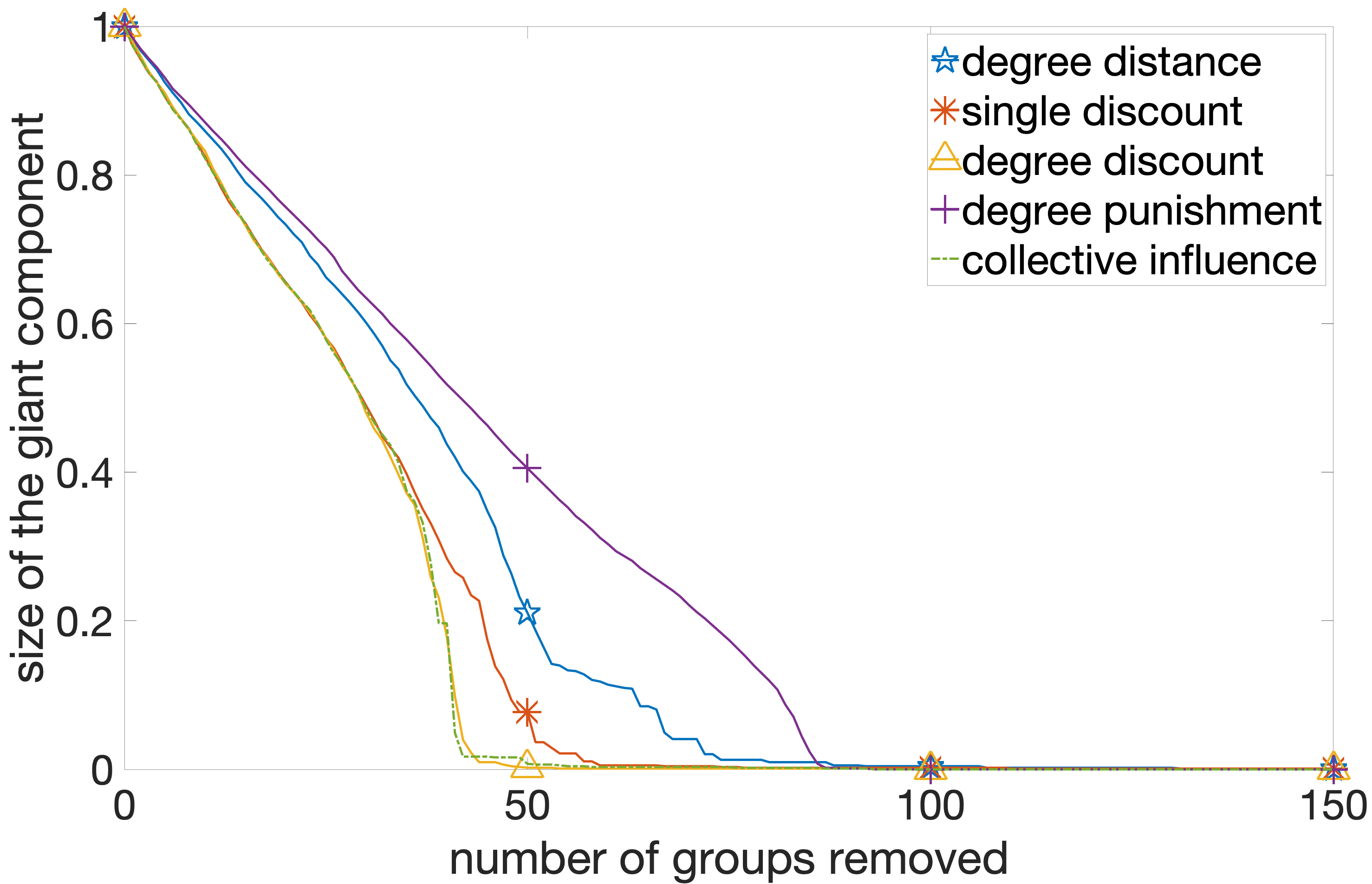}}
\subfigure[Under infectious attacks in the URV Email Network]{
\includegraphics[width=0.4\textwidth]{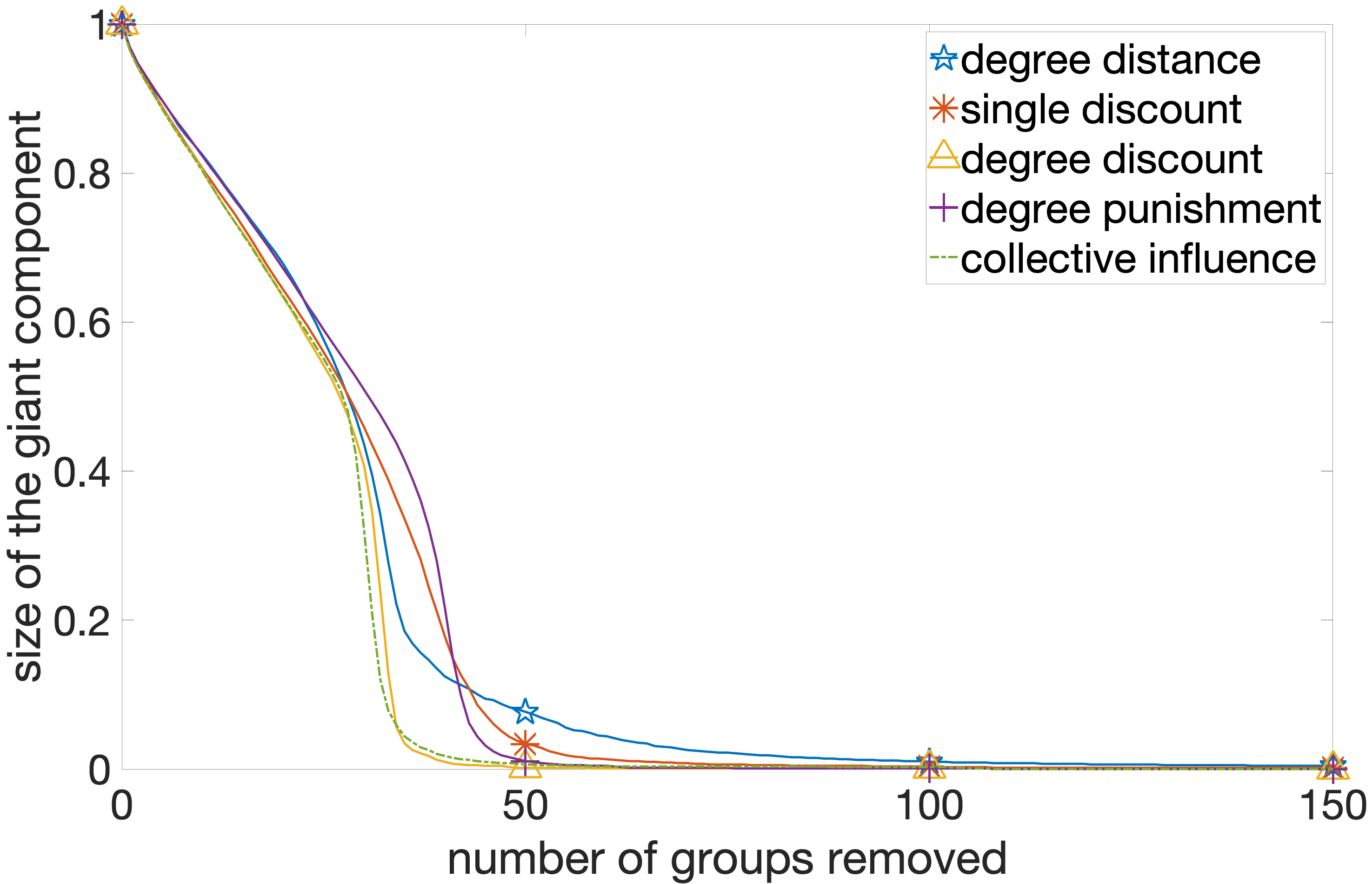}}
\hspace{5mm}
\subfigure[Under infectious attacks in the EU Email Network]{
\includegraphics[width=0.4\textwidth]{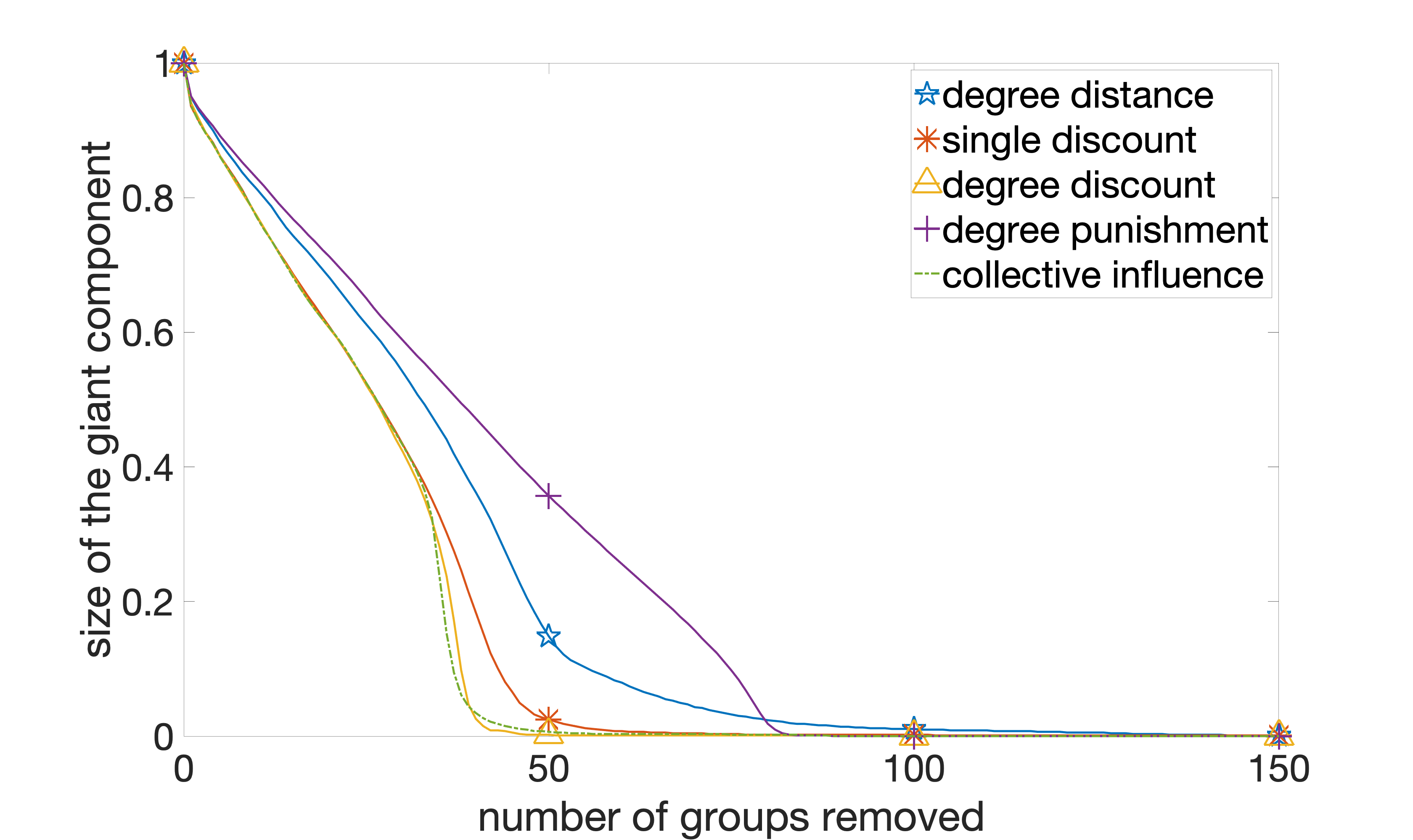}}
\caption{The size of the giant component after removing a set of either non-infectious and infectious initial attackers based a given group selection metrics in the two undirected network datasets (EU Email Network, URV Email Network).}
\label{fig:resilience-group-selection}
\end{figure*}
%\end{comment}

Fig.~\ref{fig:resilience-group-selection} shows sizes of the giant component in both undirected networks (EU Email Network and URV Email Network) as the indicator of network resilience when a set of groups (where a group is defined as 10 nodes) chosen based on a given group selection metric are removed as targeted attacks.  Under non-infectious attacks, each metric's performance is more distinct. In particular, attacks on more dense networks (with more edges) in the EU Email Network are less severe when degree punishment is the selection criteria while attacks on larger networks (with more nodes) are less severe with degree distance.  Under infectious attacks, the results are more interesting. First, for a less dense network like the URV Email Network, the effect of the four metrics on the size of the giant component is similar although the degree discount seems to be the best selection strategy.  However, under the denser network like the EU Email Network, the degree punishment strategy outperforms the others because high network density mitigates the effect of the penalty.  From this observation, we found that under infectious attacks, higher network density can significantly mitigate the effect of the targeted attacks. If a network is not sufficiently dense, regardless of what metric is used to select targets to attack, the network can more easily collapse.  Thus, it is more important to select the right group selection metric for developing more powerful attacks under dense networks than under sparse networks.

\section{Running Time Analysis} \label{sec:running-time}

\begin{figure*}[t]
\centering
\includegraphics[width = \textwidth, height = 0.4\textwidth]{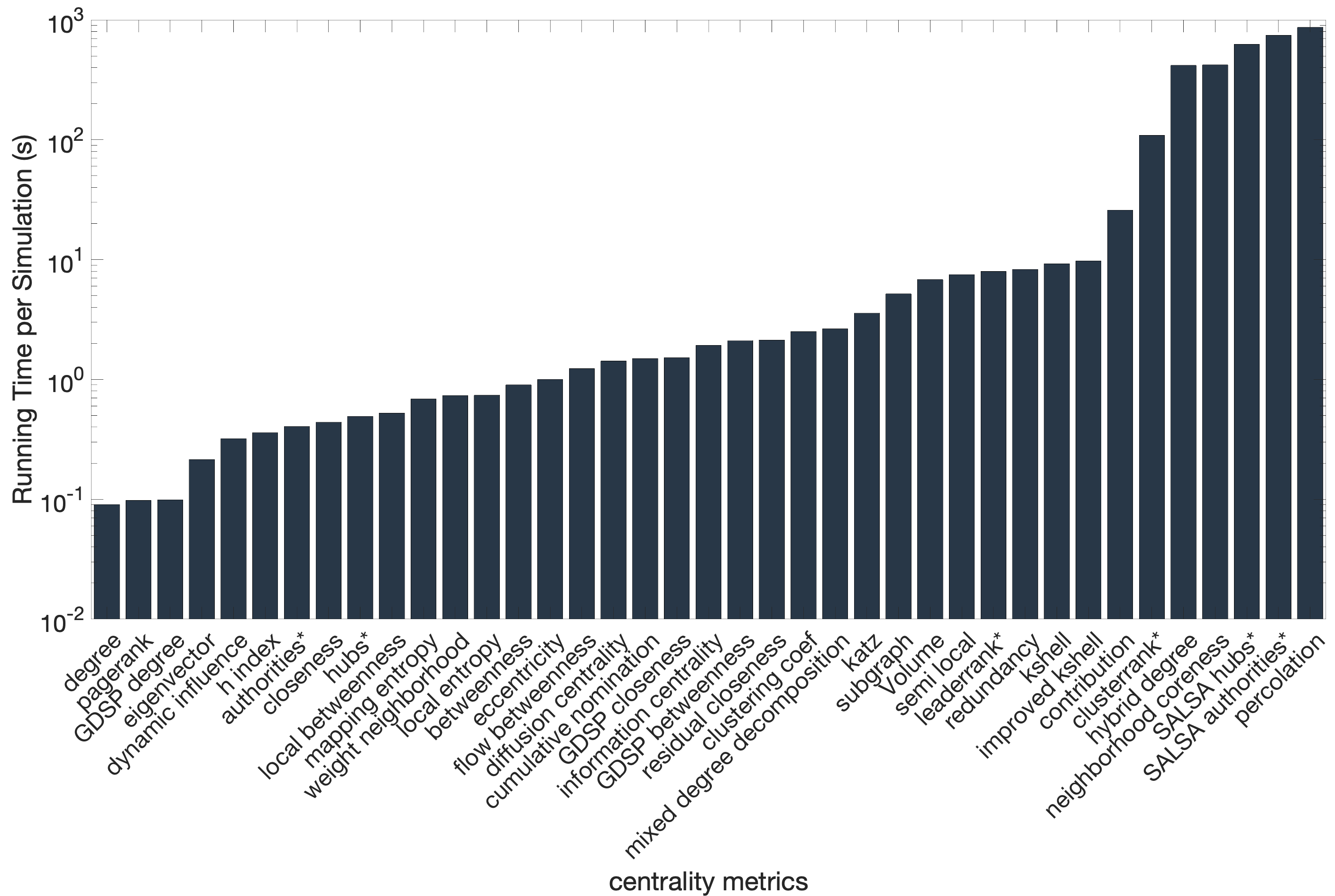}
\caption{Simulation running time (in $\log_{10} \mathrm{sec.}$) of the 39 point centrality metrics in the undirected URV Email Network and the UCI Social Network. Note that centrality metrics that can be only shown in directed networks are indicated with *. }
\label{fig:r8-running-time-pont-1}
\end{figure*}

Fig.~\ref{fig:r8-running-time-pont-1} shows the running time in $\log_{10} \mathrm{sec.}$ to show the efficiency of 39 point centrality metrics surveyed in this work using the undirected URV Email Network and the UCI Social Network.  Degree, pagerank, and GDSP degree exhibit the best efficiency among the point centrality metrics considered in this work.  This is one reason why even though a large volume of centrality metrics have been created in the 2000s and 2010s (see Fig.~2 of the main paper), simple degree-based or similar centrality metrics still dominate in practice due to their efficiency in calculation.  We also observe high running time from contribution centrality to leaderrank centrality in the right side of Fig.~\ref{fig:r8-running-time-pont-1}.  Although these metric offer certain useful features in capturing insightful centrality concepts in terms of power or influence, their high running time may not be attractive particularly in sizable or resource-constrained, distributed environments.  

\begin{figure*}[t]
\centering
\includegraphics[width = \textwidth, height = 0.4\textwidth]{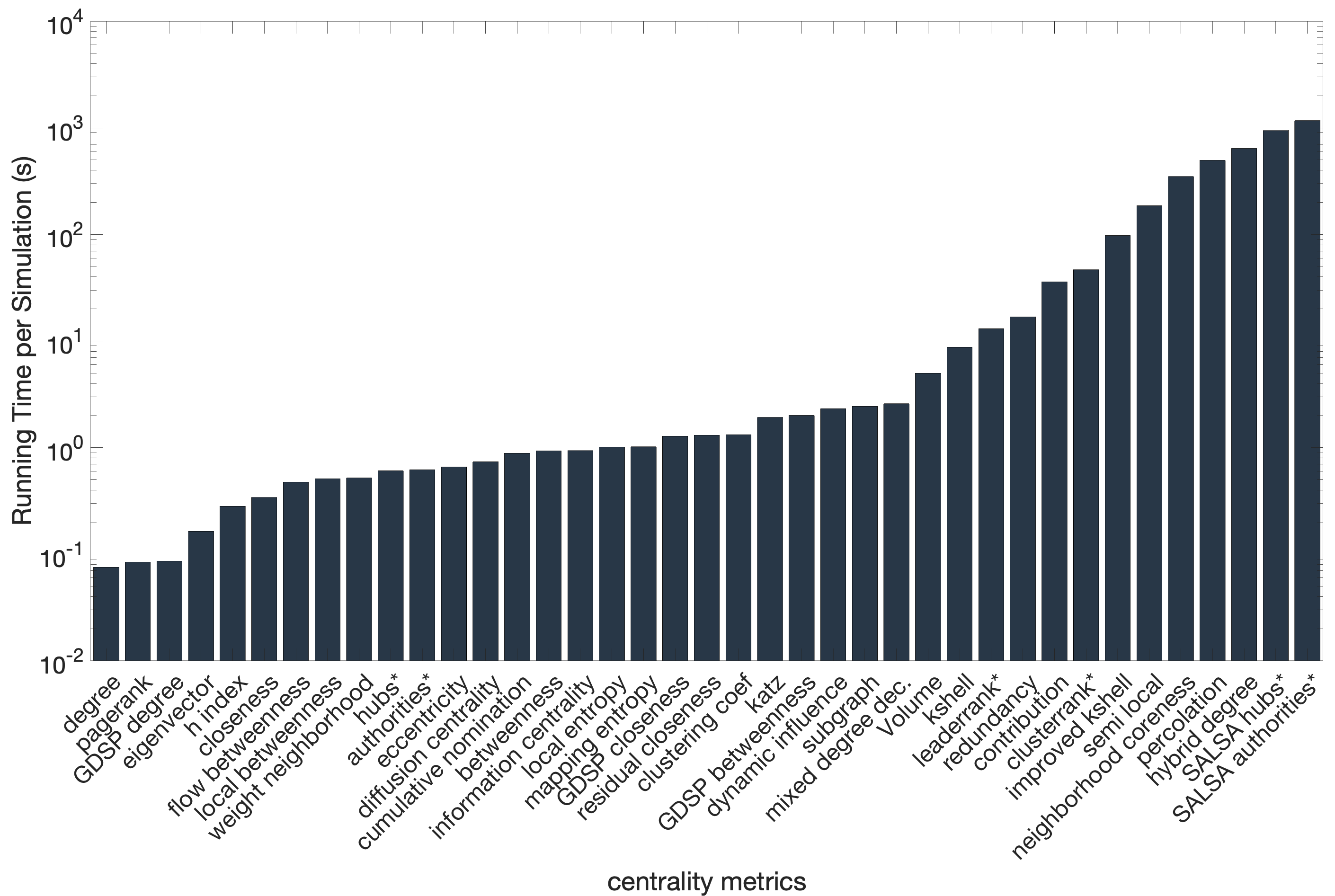}
\caption{Simulation running time (in $\log_{10} \mathrm{sec.}$) of the 39 point centrality metrics in the undirected EU Email Network and the directed Rocketfuel Network. Note that centrality metrics that can be only shown in directed networks are indicated with *. }
\label{fig:r8-2-running-time-pont}
\end{figure*}

We also display the running time analysis of the point centrality metrics using the undirected EU EMail Network and the directed Rocketfuel Network in Fig.~\ref{fig:r8-2-running-time-pont}. 
%This figure shows the running time in $\log_{10} \mathrm{sec.}$ to show the efficiency of 39 point centrality metrics surveyed in this work under the undirected EU Email Network and the directed Rocketfuel Network. 
Comparing the results here with the other networks in Fig.~\ref{fig:r8-running-time-pont-1}, we find there are only slight differences in the performance order. This is because the characteristics of a network dataset affect each centrality metric's running time. However, the trends are similar since the performance order is still dependent on the inherent complexity of each metric.

\begin{figure}[h]
    \centering
    \includegraphics[width =0.5 \textwidth, height = 0.5\textwidth]{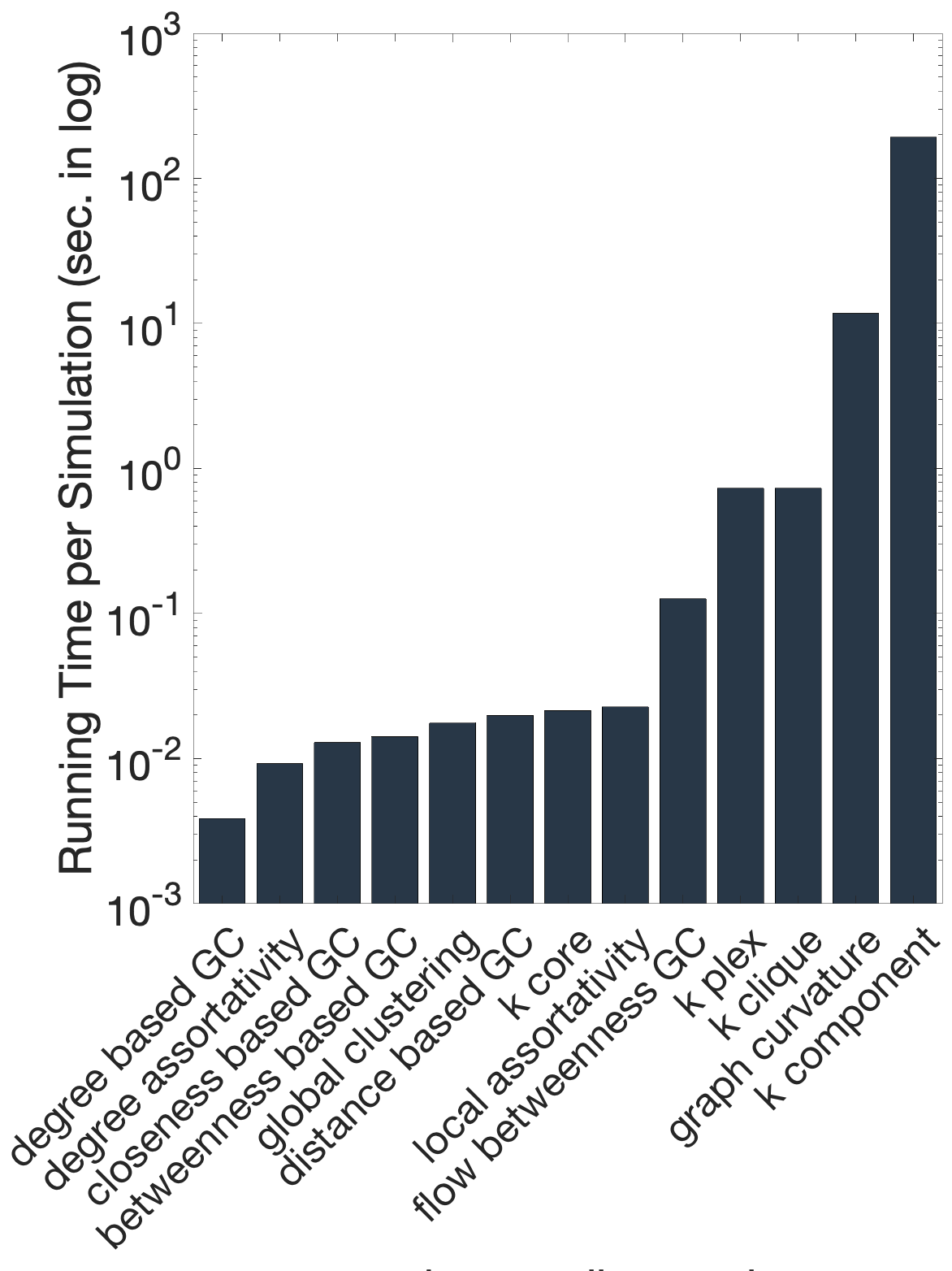}
    \caption{Simulation running time in sec. (in log scale) for 13 graph centrality metrics applied to the URV Email Network dataset.}
    \label{fig:R9_running_time-1}
    \vspace{-3mm}
\end{figure}

\begin{figure}[h]
    \centering
    \includegraphics[width = 0.5\textwidth, height=0.5\textwidth]{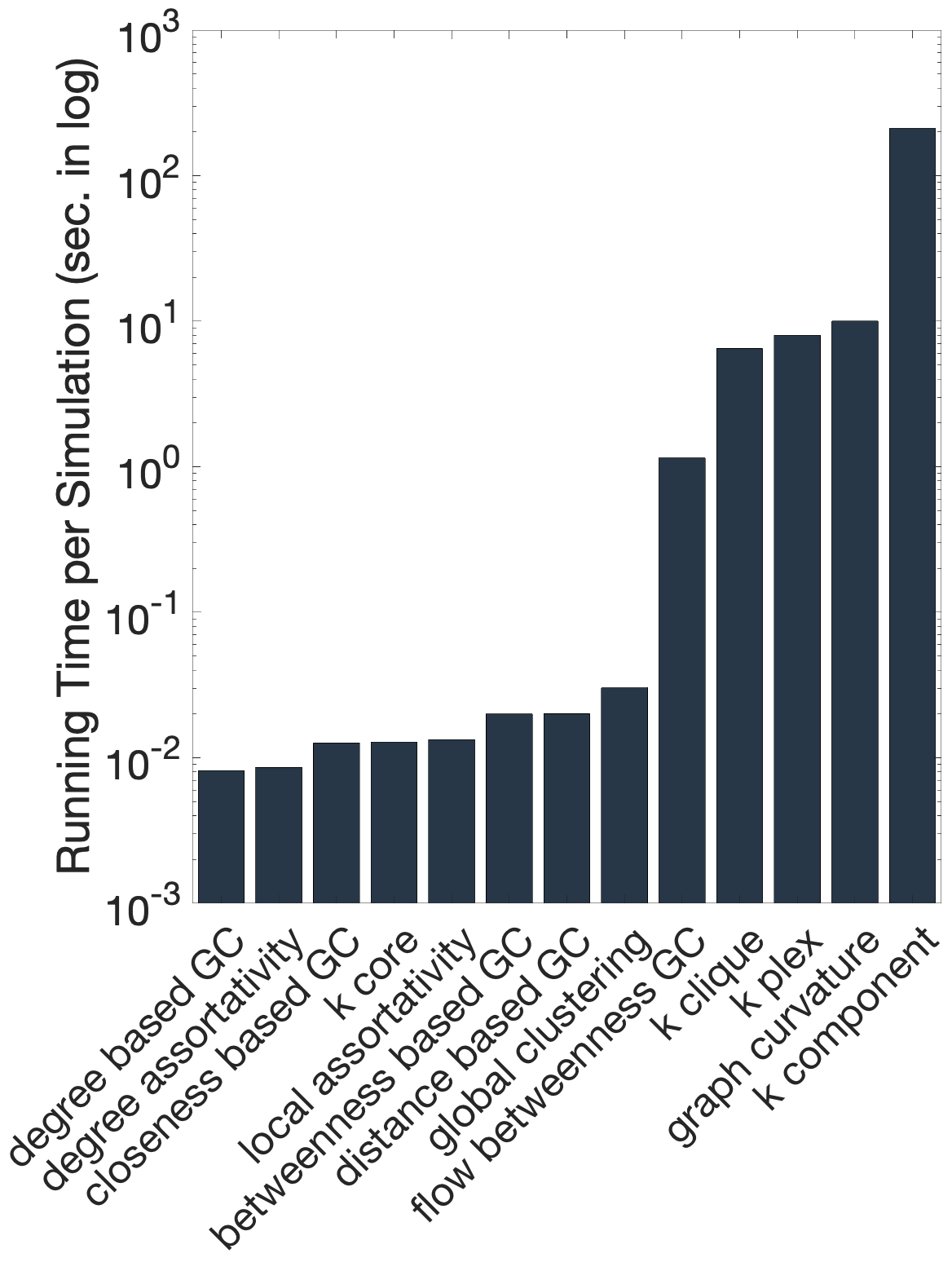}
    \caption{Simulation running time in sec. (in log scale) for 13 graph centrality metrics in the undirected EU Email Network dataset. }
    \label{fig:R9_running_time}
\end{figure}

Fig~\ref{fig:R9_running_time-1} shows the running time of 13 graph centrality (GC) metrics per simulation run on the undirected URV Email Network. We found most $k$-metrics, except $k$-core, are fairly slow while common metrics such as degree-based metrics are faster, which is one reason for their common utilization in various domain applications.  However, it seems there is no clear relationship between algorithmic complexity and the nature of the GC metrics, such as local or global metrics, in the process of their calculation.  Similarly, Fig.~\ref{fig:R9_running_time} shows the running time of 13 graph centrality (GC) metrics per simulation run but using the other undirected EU Email Network dataset. We found a slightly different performance order compared to Fig.~\ref{fig:R9_running_time-1} using the UCI Social Network.  However, the overall trend is similar. As discussed regarding Fig.~\ref{fig:R9_running_time-1}, it seems there is no relationship between algorithmic complexity and local or global centrality nature in the GC metrics.

\begin{figure}[h]
    \centering
    \includegraphics[width = 0.6\textwidth]{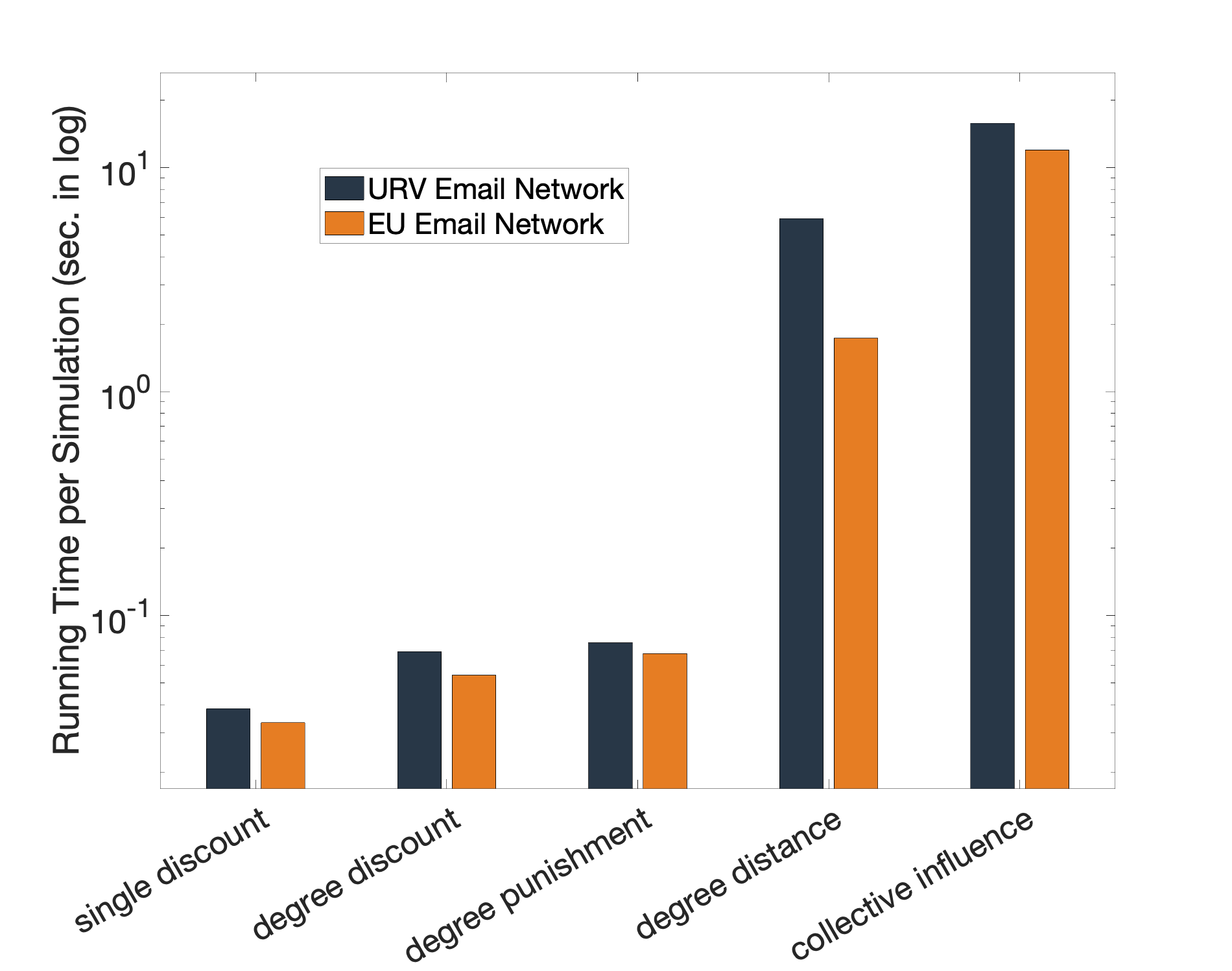}
    \caption{Simulation running time in sec. (in log scale) for 5 group selection metrics in the two undirected network datasets (URV Email Network and EU Email Network). } %We used $d_{td}=4$ for the degree distance metric.
    \label{fig:R10_running_time}
\end{figure}

Fig.~\ref{fig:R10_running_time} shows the running time of the four group selection metrics per simulation round. We found that the degree distance is more expensive than other counterparts that are the enhanced versions to improve the complexity of the degree distance using heuristics.  We also found there is a longer running time for calculating the metrics using the URV Email Network than using the EU Email Network. Even though the URV Email Network has more nodes than the EU Email Network, the EU Email Network has five times higher network density (i.e., more edges) than the URV Email Network. This implies that the complexity of a group selection centrality is more affected by node density rather than network density.

\section{Algorithmic Complexity of Centrality Metrics}

In Tables~\ref{tab:point-complexity-1},~\ref{tab:point-complexity-2}, \ref{tab:graph-complexity}, and \ref{tab:group-selection-compexity}, we summarized asymptotic complexities of all centrality metrics surveyed in this paper.

\begin{footnotesize}
\begin{table*}[htbp!] 
\centering
\caption{Point centrality's meaning, metric, and complexity}
\vspace{-3mm}
\label{tab:point-complexity-1}\begin{tabular}{|P{2cm}|P{6.3cm}|P{1cm}|P{1.7cm}|P{1cm}|}
\hline
{\bf Centrality name} & {\bf Meaning} & {\bf Eq. No.} & {\bf Complexity} & {\bf Ref. No.} \\
\hline
\multicolumn{5}{|c|}{\cb {\bf Local Centrality Metrics}} \\
\hline
Degree & Popularity 
%The number of nodes adjacent (or edges incident) to the node 
& (1) & $O(n+m)$ or $O(n^2)$ & \cite{wasserman1994social, Freeman78}\\
\hline
Semi-local & Popularity + the popularity of the node's neighbors
%Takes into account the number of nearby neighbors of the neighbors of the node
& (2) & 
%$O(n^3)$ 
$O(n \langle k \rangle^2)$ 
& \cite{chen2012identifying} \\
\hline
Hybrid degree & A mixture of degree and a modified semi-local centralities & (3) &  
%$O(n^3)$ 
$O(n \langle k \rangle^2)$ 
& \cite{ma2017identifying} \\
\hline
Volume & Captures the size of a ball of radius $h$ centered at the node
%The sum of degrees of the neighbors within a certain distance from the node 
& (4) & 
%$O(n^{(h+2)})$ 
$O(n \langle k \rangle^{(h+1)})$ 
& \cite{wehmuth2012distributed,Kim12-volume} \\
\hline
Clustering coefficient & Probability of node's neighbors being neighbors of each other
%The likelihood of neighbors of a node to be connected 
& (5) & 
%$O(n^3)$ 
$O(n \langle k \rangle^2)$ 
& \cite{Watts98} \\
\hline
Redundancy & Captures usefulness (social capital) of a link
%The scaled number of links from the node that do not contribute to structural holes 
& (6) & 
%$O(n^3)$ 
$O(n \langle k \rangle^2)$ 
& \cite{Burt95} \\
\hline
Entropy-based measures & Amount of (missing) information in the node's neighborhood system
%The entropy of a node's degree compared with its neighbors 
& (7) & $O(n^2)$ & \cite{nie2016using} \\
\hline
ClusterRank & Clustering-coefficient weighted semi-local centrality
%Incorporates the node degree, the degree of the neighbors, and damps by clustering coefficient 
& (8) & 
%$O(n^3)$ 
$O(n d_{\max}^2 +n^2)$ 
& \cite{chen13cluster} \\
\hline
H-index & Impact (where degree is productivity) of a node's links
%The maximum number $h$ of neighbors with degree at least $h$ 
& (9) & $O(n^2)$ & \cite{korn09lindex} \\
\hline
Curvature & Measure of local geometry near node & (10) & $O(2^n)$ & \cite{knill2012index} \\
\hline
\multicolumn{5}{|c|}{\cb {\bf Iterative Centrality Metrics}} \\
\hline
$k$-shell index or coreness & Hierarchical structure membership of the node in the network & (11) & $O(n+m)$ & \cite{kitsak2010identification} \\
\hline
Mixed degree decomposition & Mixture of $k$-shell and degree & (12) & $O(n+m)$ & \cite{zeng2013ranking} \\
\hline
Neighborhood coreness & Aggregating $k$-shell indices of neighboring nodes & (13) & $O(n^2 + m)$ & \cite{bae2014identifying} \\
\hline
Eigenvector & Importance of neighboring nodes determines node's importance & (14) & $O(n^3)$ & \cite{Bonacich72} \\
\hline
Katz & Similar to eigenvector, with damping effect on distant nodes & (15) & $O(n^3)$ & \cite{Katz53} \\
\hline
Authorities \& Hubs & Eigenvector centrality for directed networks & (16) & $O(n^3)$ & \cite{Kleinberg99} \\
\hline
PageRank & Google's algorithm that adapts Katz centrality, weighting influence by out degree & (17) & $O(n^3)$ & \cite{Brin98theanatomy} \\
\hline
Contribution & Weighted eigenvector centrality using structural dissimilarity & (18) & $O(n^3)$ & \cite{Alvarez-Socorro15} \\
\hline
Diffusion & Models the influence of the spread of information over finite time & (19) & $O(n^3)$ & \cite{Banerjee13} \\
\hline
Subgraph & Incidence of nodes to closed walks weighted by length (motifs) & (20) & $O(n^3)$ & \cite{Estrada05} \\
\hline
%Communicability & Measures all closed paths of all lengths starting and  ending at a same node & \eqref{eq:communicability-c-ij} & $O(n^3)$  & \cite{Estrada08} \\
%\hline
LeaderRank & Parameterless modified (ground node) version of PageRank & (21)& $O(n^3)$ & \cite{Lu11} \\
\hline
Dynamical influence & Incorporates initial dynamic state into the eigenvector concept & (22) & $O(n^3)$ & \cite{Klemm12} \\
\hline
Cumulative nomination & Nomination process that approaches Bonacich centrality & (23) & $O(n^3)$ & \cite{poulin2000nomination} \\
\hline
SALSA & Random walk alternative to hubs \& authorities & (24) & $O(n^3)$ & \cite{lempel2000salsa} \\
\hline
\multicolumn{5}{|c|}{\cb {\bf Global Centrality Metrics}} \\
\hline
Improved method & Improve $k$-shell, and rank the nodes with the same $k$-shell; Used a Binary Search Tree to find distance & (25) & O($n^2  \log n$) & \cite{liu2013ranking} \\
\hline
Betweenness & Measuring the influence of a node as a broker  & (26)& %\textcolor{red}{$O(n^3)$ with Floyd-Warshall, or $O(n^2*log(n)+mn$ if use Johnson's algorithm or Brandes' algorithm and graph is weighted, or $O(mn)$ if use Brandes' algorithm and graph is unweighted \cite{brandes2001faster} } 
& \cite{Freeman77} \\
& when Floyd-Warshall algorithm is used & & $O(n^3)$ & \\
& when Johnson's algorithm or Brandes' algorithm with a weighted graph is used & & $O(n^2 \log n+mn)$ & \\
& when Johnson's algorithm or Brandes' algorithm with a unweighted graph is used & & $O(mn)$ & \\
\hline
\end{tabular}
\end{table*}
% ===== Table Separate Here ===========
\begin{table*}[htbp!] 
\centering
\caption{Point centrality's meaning, metric, and complexity}
\vspace{-3mm}
\label{tab:point-complexity-2}\begin{tabular}{|P{2cm}|P{6.3cm}|P{1cm}|P{1.7cm}|P{1cm}|}
\hline
{\bf Centrality name} & {\bf Meaning} & {\bf Eq. No.} & {\bf Complexity} & {\bf Ref. No.} \\
\hline
\multicolumn{5}{|c|}{\cb {\bf Global Centrality Metrics}} \\
\hline
$L$-betweenness & Increase the efficiency of betweenness centrality by only considering the pair whose distance smaller than $L$  & (27) & same as betweenness & \cite{ercsey2010centrality} \\
\hline
Flow betweenness & The flow level through a node & (28) & $O(m^2n)$ & \cite{Freeman91, newman2005measure} \\
\hline
Random-walk betweenness & Measuring transmit speed to a node with random walk %instead of shortest path 
& (29) & %sparse graph: $O(n^3)$; worst case: 
$O((m+n)n^2)$ & \cite{newman2005measure} \\
\hline
Routing betweenness & Expected number of packet passing through a node  & (30) & $O(n^2m)$ & \cite{Dolev10} \\
\hline
Load & When all nodes send a packet to every other node along with a shortest path, the number of packets passing through a node; used Dijkstra algorithm & (31) & $O(mn)$ & \cite{Goh01, Dolev10} \\
\hline
Closeness &  Reciprocal of distance sum of a node to all other nodes & (32) & $O(mn)$ & \cite{Sabidussi66, saxena2017faster} \\
\hline
Information & Consider all possible paths to decide a node's importance & (33) & $O(n^3)$ & \cite{Stephenson89} \\
\hline
Current-flow betweenness and closeness &  Model information spread over a network as an electric current %; $O(n^2 (n+m))$ for current-flow betweenness, $O(n^3)$ for current-flow closeness 
& (34)-(35) & $O(n^3)$ for $m < n^2$  & \cite{brandes2005centrality} \\
\hline
Residual closeness & Alternative version of closeness metric with a weighting scheme; used Floyd-Warshall algorithm  & (36) & $O(n^3)$  & \cite{dangalchev2006residual} \\
\hline
Spatial & measures the efficiency of the route between two nodes; used Breadth First Search & (37) & $O(n(n+m))$ & \cite{crucitti2006centrality}\\
\hline
AHP-based & Utilize multiple centrality metrics to identify influential nodes; used degree, betweenness, or closeness & (38)
 & $O(n^3)$ & \cite{bian2017identifying} \\
\hline
Generalized degree and shortest paths  & Combine degree, closeness and betweenness metrics with their weighted version & (39) & $O(n^2)$ for degree, $O(n^3)$ for closeness and betweenness & \cite{opsahl2010node} \\
\hline
%Collective influence
%& Finding a minimum node group with the highest influence & \eqref{eq:collectiveinfluence} & $n^2 \times \log(n)$ with max-heap structure & \cite{morone2015influence} \\
%\hline
Weight neighborhood & Measuring the diffusion importance based on benchmark centrality (e.g., degree, betweenness, $k$-shell), given $\phi$ & (40) & $O(n \times m)$  & \cite{wang2017novel} \\
\hline
Percolation & Evaluate the changing of network topology & (41) & $O(n^3)$ & \cite{Piraveenan13} \\
\hline
Eccentricity & Max distance to other nodes; used Floyd-Warshall algorithm & (42) & $O(n^3)$ & \cite{hage1995eccentricity} \\
\hline
\end{tabular}
(Notations: $n$ is the total number of nodes, $m$ is the number of edges, $\langle k \rangle$ is the mean degree of nodes, and $d_{\max}$ is the maximum degree.)
\end{table*}
\end{footnotesize}

\begin{footnotesize}
\begin{table*}[htbp!] 
\centering
\caption{Graph centrality's meaning, metric, and complexity}
\vspace{-3mm}
\label{tab:graph-complexity}\begin{tabular}{|P{2.1cm}|P{6.3cm}|P{1.7cm}|P{1cm}|P{1cm}|}
\hline
{\bf Centrality name} & {\bf Meaning} & {\bf Complexity} & {\bf Metric Eq. No.} & {\bf Ref. No.} \\
\hline
 Distance-based GC & Sum of distances between each vertex and all other vertices using Breadth First Search & $O(n(n+m))$ & (43) & \cite{Freeman78, Shimbel53} \\
\hline
 Degree-based GC & Maximum sum of differences between the largest centrality and all other centralities & $O(n^2)$ & (44) & \cite{Nieminen74} \\
\hline
 Betweeness-based GC & Mean difference between the maximum betweenness and all other betweenness; used Floyd-Warshall algorithm & $O(n^3)$ & (45)& \cite{Freeman77, Freeman78}\\
\hline
 Flow betweenness-based GC & Difference between the highest maximum flow with highest betweenness and maximum flow of all other nodes & $O(n^4)$ & (46) & \cite{Freeman91}\\
\hline
 Closeness-based GC & Mean difference between the maximum closeness metric and all other closeness & $O(n^3) $ & (47) & \cite{Freeman78} \\
\hline
 Reciprocity & Number of bidirectional edges between two nodes over the total number of possible edges in a network & $O(n^2)$ & (48) & \cite{Newman02-virus}\\
\hline
 $k$-component & A maximal subset of nodes where each node can reach the other nodes in the subset based on minimum $k$ paths that are vertex independent & $O(Fn^2)$ & - & \cite{Newman10} \\
\hline
 $k$-clique & A maximal subset of vertices where each vertices of the subset are directly connected to each other  & $O(n^3)$ & - & \cite{Seidman78-plex, Tichy73, Newman10} \\
\hline
 $k$-plex & A maximal subset of $n$ vertices where each vertex is connected to minimum $n-k$ other vertices & $O(n^3)$  & - & \cite{Seidman78-plex} \\
\hline
 $k$-core & A maximum size of the subset where each vertex is connected to minimum $k$ other vertices & $O(n + m)$ & - & \cite{Newman10, Seidman78-plex} \\
\hline
 Global clustering efficient & Mean of a local clustering coefficient of a graph & $O(nd_{max}^2)$ & (49) & \cite{Watts98, Holland71, Newman10} \\
\hline
 Degree assortativity & Linear correlation coefficient between two nodes' excess degree & $O(n^2)$ & (50) & \cite{Newman02-assortativity, Newman03-assortativity, Noldus15}\\
\hline
 Local assortativity & An individual node's assortativity based on the node's degree and its neighbor's degree & $O(n^2)$ & (51) & \cite{Piraveenan10-local-assortativity} \\
\hline
 Graph curvature & Negative curvature of the graph as a whole to identify congestion & $O(n^2(n+m))$ & (52) & \cite{krioukov2010hyperbolic, narayan2011large, gromov1987hyperbolic, jonckheere2007upper}\\
\hline
\end{tabular}
(Notations: $n$ is the total number of nodes, $m$ is the number of edges, $\langle k \rangle$ is the mean degree of nodes, and $d_{\max}$ is the maximum degree.)
\end{table*}
\end{footnotesize}

\begin{footnotesize}
\begin{table*}[htbp!] 
\centering
\caption{Group selection centrality's meaning, metric, and complexity}
\vspace{-3mm}
\label{tab:group-selection-compexity}\begin{tabular}{|P{2.1cm}|P{6.3cm}|P{1.7cm}|P{1cm}|P{1cm}|}
\hline
{\bf Centrality name} & {\bf Meaning} & {\bf Complexity} & {\bf Metric Eq. No.} & {\bf Ref. No.} \\
\hline
DegreeDistance 
& Restrict selected nodes to have a certain distance of separation, unless the number of common neighbors is limited and the influence probability is low
%& Choose the top-degree node and discard all nodes within a distance $d_{td}$ of the selected node from the candidate set 
& $O(n(n+m))$ & (53) & \cite{sheikhahmadi15} \\
\hline
SingleDiscount 
& Node selection determined by maximal degree less the number of neighboring seed nodes
%& Degree Discount heuristic in which each neighbor of a newly selected seed discounts its degree by one 
& $O(n^2)$ & (54) & \cite{chen09} \\
\hline
DegreeDiscount 
& Node selection determined by the degree reduced by the influence probability of neighboring seed nodes
%& Degree of each neighbor of a newly selected seed is discounted based on a small propagation probability $p$ 
& $O(p\log n + m)$ & (55) & \cite{chen09} \\
\hline
DegreePunishment
& Node selection determined by the degree reduced by a punishment from existing on short paths originating from seed nodes 
%& Punishment strategy, more stricter than discount, to restrict the neighbors of the current seed from being selected as the next seed 
& $O(l(n + \langle k \rangle^2))$ & (56) & \cite{wang2016effective} \\
\hline
Collective Influence 
& Node selection determined by hierarchical corona of hubs  
& $O(n \log n)$ & (57)& \cite{morone2015influence,morone16} \\
\hline 
\end{tabular}
(Notations: Given a given network $G$, $n$ is the total number of nodes, $m$ is the number of edges, $\langle k \rangle$ is the mean degree of nodes, $d_{td}$ is the distance threshold, $d_{\max}$ is the maximum degree, $l$ is the number of iterations and $F$ is the time complexity to find the maximum flow between two vertices in a graph $G$.)
\end{table*}
\end{footnotesize}

\end{appendices}

\end{document}